%% file: main.tex
\documentclass[12pt]{article}

\usepackage{amsmath, geometry, amssymb}
\usepackage{authblk, comment, longtable, times}
\usepackage{dsfont, ushort, color, threeparttable, threeparttablex}
\usepackage{subfig, graphicx, booktabs, lscape, mwe}
\usepackage{tabularx, tabulary, rotating, dcolumn, multicol}
\usepackage{setspace, algorithm, algorithmic, subcaption, pdflscape}
\usepackage[sort]{natbib}

\usepackage[font=small]{caption}

\def\doublespacing{\setstretch{2}}

\begin{document}

\title{Retail Investor Horizon and Earnings Announcements}
\author{Domonkos F. Vamossy\thanks{Department of Economics, University of Pittsburgh, d.vamossy@pitt.edu.}}
\date{\today}
\maketitle

This paper moves beyond aggregate measures of retail intensity to explore investment horizon as a distinguishing feature of earnings-related return patterns. Using self-reported holding periods from StockTwits (2010–2021), we observe that separating retail activity into "long-horizon" and "short-horizon" cohorts reveals divergent price anomalies. Long-horizon composition is associated with underreaction, characterized by larger initial reactions and pronounced Post-Earnings Announcement Drift (PEAD), suggesting a slow but persistent convergence toward fundamental value. In contrast, short-horizon activity parallels sentiment-driven overreaction, where elevated pre-event sentiment precedes weaker subsequent performance and price reversals. A zero-cost strategy exploiting this heterogeneity, going long on long-horizon stocks and short on short-horizon stocks, yields risk-adjusted alphas of 0.43\% per month. These findings suggest that accounting for investment horizon helps disentangles the fundamental signal in retail flow from speculative noise.

\bigskip
\medskip
{\bf Keywords}: Investment horizon; dispersed information; NLP; earnings announcements. \\
\bigskip

\textbf{JEL Codes: G14, G41.} \\

\clearpage 
\doublespacing

\input{sections/intro}

\input{sections/measure}

\input{sections/theory}

\input{sections/methodology}

\input{sections/results}

\input{sections/discussion}

\input{sections/conclusion}

\normalsize 

\bibliography{bib}
\bibliographystyle{plainnat}

\clearpage

\input{sections/appendix}

\end{document}

%% file: sections/intro.tex
\section{Introduction}

\begin{quote} "The stock market is a device for transferring money from the impatient to the patient." \end{quote}

\begin{flushright} — Warren Buffett \end{flushright}

Building on Buffett's insight into the virtues of patience, this paper investigates how the investment horizon of investors influences stock price reactions to earnings surprises. How different investor time-horizons shape price discovery is a question of both theoretical and practical importance, as it reveals how trading patience influences stock reactions and informs our understanding of market efficiency.

We leverage social media data from StockTwits, a platform akin to Twitter but tailored for investors, to measure investment horizons among retail investors. By analyzing users' self-reported holding periods, we classify retail investors as either short-term or long-term. For each stock, we tally the unique count of these investors who engage in discussions within the $[-90, -1]$ day window prior to an earnings announcement, calculating the proportion of long-term investors within this group. If this proportion exceeds 50\%, we designate the stock as being predominantly associated with long-term retail investors.

We then examine how investment horizon at the security level affects stock price reactions to earnings surprises. Using earnings data from I/B/E/S and COMPUSTAT, we focus on quarterly earnings announcements for which at least one analyst provides an earnings forecast in the 30 days before the announcement. We define the earnings surprise as the difference between the earnings announcement and the consensus earnings forecast, scaled by the price per share as in \cite{dellavigna:2009}.

Horizon composition meaningfully shapes earnings-announcement returns. Relative to short-horizon names, stocks predominantly followed by long-horizon investors exhibit stronger immediate reactions (about $+0.21$ pp over days 0–1 in fully controlled specifications) and a larger, more persistent post-announcement drift. Long-horizon stocks earn an additional $+2.08$ pp over days 2–75, resulting in a total outperformance of $+2.31$ pp over the full 0–75 day window. These differentials are robust to firm fixed and year-quarter effects and matching. A simple monthly long–short drift strategy formed within earnings-surprise bins delivers economically large, factor-adjusted alphas, approximately 30 basis points per month for the overall portfolio and nearly 40 basis points when restricting to the top three surprise bins.

The existing literature highlights several channels explaining the role of short-horizon institutional investors in stock price behavior, including demand pressure or demand shock mechanisms \cite{gompers:2001, hotchkiss:2003, cella:2013}, informational advantages \cite{ke:2006, yan:2009}, and corporate governance influences \cite{gaspar:2005}. However, these mechanisms may not fully capture the dynamics of stock price reactions to earnings surprises in a landscape increasingly shaped by retail investors.

Recent research indicates that retail investors' behavior can significantly impact stock prices, especially around high-news events \cite{barber2008retail, vamossy2021investor,
vamossy2023social, barber2023resolving, friedman2024retail}. Retail investors' horizons vary widely; some pursue longer-term ``buy and hold" strategies, while others are inclined toward frequent trading in response to recent news or social sentiment. This behavioral divergence may reflect underlying factors that influence stock price reactions, and the holding patterns of retail investors could serve as a proxy for these unobserved variables.

This paper connects and extends several strands of the literature. Prior work highlights how short-horizon institutional ownership can influence prices via demand pressure and shocks (\cite{gompers:2001,hotchkiss:2003,cella:2013}), information advantages (\cite{ke:2006,yan:2009}), and governance channels (\cite{gaspar:2005}). At the same time, a growing literature shows that retail trading shapes prices around high-news events and in attention-driven settings (\cite{barber2008retail,barber2023resolving,friedman2024retail}). We add a retail-horizon perspective to classic work on horizon heterogeneity and trading by institutions (\cite{lakonishok:1992,grinblatt:1995,wermers:1999,nofsinger:1999,gompers:2001,grinblatt:2000,cohen:2002,hotchkiss:2003}), showing that the composition of retail horizons is a powerful cross-sectional predictor of both the immediate and delayed components of earnings-related returns.

Our interpretation emphasizes differences in information processing and optimism bias. Using textual analysis and sentiment extraction, we find that short-horizon investors rely heavily on technical trading cues (e.g., \textit{break}, \textit{target}) and exhibit higher pre-announcement sentiment volatility. In contrast, long-horizon investors focus on fundamental concepts (e.g., \textit{EPS}, \textit{financials}). Mechanism tests using above-median pre-EA social-media valence support an optimism-bias channel: upbeat chatter predicts subsequent giveback, especially in short-horizon stocks, while a long-horizon presence attenuates this reversal. We formalize these predictions in a simple model in Section~\ref{sec:theory} and test them empirically.

The remainder of the paper is organized as follows. Section 2 introduces the measure for the horizon of retail investors and provides descriptive statistics highlighting the differences between the long-term and short-term securities. Section 3 presents a simple model, and Section 4 presents the methodology. Section 5 discusses the main results. Section 6 offers additional findings and probes into mechanisms. Section 7 concludes.

%% file: sections/measure.tex
\section{A measure of retail investors' investment horizon: StockTwits Holding Period}

We use data from StockTwits, a popular social networking platform for investors similar to Twitter, where users share stock opinions using ``cashtags" linked to specific company ticker symbols (e.g., \$AMZN). Users can tag their sentiment (bullish, bearish, or unclassified), and we can also access a sentiment score computed by StockTwits. The platform allows the tracking of likes per message, user identifiers, and user attributes like follower count.

Most users self-report their investment philosophy along two axes: Approach (technical, fundamental, momentum, value, growth, global macro) and Holding Period (day trader, swing trader, position trader, long-term investor). They also categorize their experience level as novice, intermediate, or professional. This enables us to classify firms based on the predominance of short-term or long-term traders among their investors. Specifically, we measure each firm’s retail‐investor horizon by counting unique investors, classified as short-term (day or swing traders) or long-term (position or buy-and-hold investors), who post about the stock between 90 and 1 calendar days before an earnings announcement. We then calculate the share of long-term investors in the total sample; firms with a long-term share above 50 \% are designated as predominantly long-term.

\subsection{Sample Construction}

Table~\ref{tab:sample_restrictions} summarizes the construction of the analysis sample. We begin with 180{,}556 quarterly earnings announcements for 5{,}419 tickers in I/B/E/S. After merging these events on ticker and announcement dates with CRSP to obtain prices, returns, and firm characteristics, 142{,}025 announcements remain. We then apply a series of filters to ensure well-defined earnings surprises: (i) the stock price five days before the announcement must exceed \$5, (ii) SUE must be non-missing, and (iii) the pre-announcement price must be at least as large as the absolute values of both the forecasted and actual EPS. These restrictions remove penny stocks and mechanically invalid denominators, yielding 130{,}496 announcements. We exclude firm--quarters with multiple earnings announcements and those lacking abnormal returns over the \([-30,90]\) window, yielding a final sample of 123{,}880 firm--quarter observations.

We next merge these announcements to StockTwits message activity via ticker symbols. Across the full unrestricted data, there are 51{,}065{,}435 posts by 693{,}965 users mentioning 5{,}143 tickers within the $[-90,-1]$ window preceding earnings releases. Restricting to users who self-report an investment horizon, required for classifying investors as short-term or long-term, yields 20{,}605{,}979 posts from 132{,}958 users covering 5{,}082 tickers and 104{,}919 earnings announcements. These restrictions ensure that our horizon measure is based solely on users with explicitly stated trading styles. 

\input{tables/sample_restrictions}

These steps show that our merge between StockTwits and I/B/E/S retains the large majority of earnings announcements, ensuring that the final sample remains broad and representative of the underlying universe of public firms. Appendix~C compares the characteristics of the I/B/E/S-only and matched samples to verify this representativeness.

\subsection{Descriptive Statistics}

Table \ref{tab:summary_stats_stocktwits} reports descriptive statistics, including the mean, standard deviation, and key percentiles (1st, 25th, 50th, 75th, and 99th), along with sample sizes. Panel A presents the full sample (N = 104,919); Panel B splits observations by retail-investor horizon (53,354 Long-Term vs. 51,565 Short-Term announcements); and, following \cite{dellavigna:2009}, Panel C groups earnings surprises into eleven quantiles, reporting each bin’s average surprise and count. Earnings surprises that are negative are allocated to quintiles 1 through 5, followed by zero surprises (quintile 6), and positive surprises (quintiles 7 through 11). The cutoffs for these groups are determined separately each quarter. Given that positive surprises are roughly twice as prevalent as negative ones, quintiles 7 through 11 contain approximately twice as many instances as quintiles 1 through 5. Within each group, we distinguish between Long-Term and Short-Term announcements. We plot the earnings surprise distribution, the day-of-the-week distribution, and the earnings surprise histogram across quantiles for Long-Term and Short-Term announcements in Figure \ref{fig:figure4abc}.

Firms with a higher share of long-term investors tend to have significantly lower market capitalization and volatility, lower institutional turnover, and lower analyst coverage compared to those dominated by short-term investors. The observation that stocks classified as long-term based on retail activity also exhibit lower churn among institutions suggests a consistency between retail and institutional investment behaviors. Additionally, long-term announcements appear more frequently later in the sample period (mean year 2016.8 vs. 2014.8). Regarding social media activity, we observe evidence consistent with an optimism bias among short-horizon investors. Stocks followed by short-term users exhibit significantly higher pre-announcement valence (0.537 vs. 0.512, $t$-stat $= -53.40$) and higher dispersion of opinion (SD Valence of 0.153 vs. 0.072) compared to long-term stocks. Finally, Panel C indicates that short-term announcements are disproportionately represented in the extreme tails (quantiles 1 and 11) of the earnings surprise distribution. To mitigate bias from unobserved firm characteristics, we include regressions that incorporate firm fixed effects.

\input{tables/table_1}
\input{figures/figure_1}

\clearpage 

\subsection{Measure Stability}

Table \ref{tab:stability} evaluates the stability of our investor-horizon classification. Panel A examines within-event persistence by tracking whether a stock that is classified as Long-Term or Short-Term in the \([-90,-1]\) window retains that status in the subsequent 
\([0,45]\) and \([46,90]\) windows. Long-Term stocks remain Long-Term 72.2\% of the time in \([0,45]\) and 65.65\% in \([46,90]\). Short-Term classifications are even more persistent, with 77.09\% of stocks staying Short-Term in \([0,45]\) and 77.08\% in \([46,90]\). These high levels of within-event stability indicate that the horizon measure does not fluctuate mechanically around the announcement. 

\input{tables/table_2}

Panel B assesses persistence across earnings announcements at the firm level. Among firms initially classified as Long-Term, 70.1\% of their announcements are also labeled Long-Term, while 29.9\% fall into the Short-Term category. Conversely, firms initially classified as Short-Term remain in that category for 69.0\% of their announcements. Moreover, 11.27\% of firms are classified as Long-Term at every announcement, 15.10\% are always Short-Term, and the remaining 73.63\% switch at least once. Although the measure shows substantial persistence, there is also meaningful switching across announcements. Hence, when we include firm fixed effects in our regressions to control for time‐invariant heterogeneity, this level of persistence leaves ample within‐stock variation for identification.

%% file: tables/sample_restrictions.tex
\begin{table}[htbp]
\centering
\footnotesize
\begin{threeparttable}
\caption{Sample Restrictions}
\label{tab:sample_restrictions}
\begin{tabular}{lrrrr}
\toprule
Restriction & \# of Posts & \# of Users & \# of Tickers & \# of Announcements \\
\midrule
\underline{I/B/E/S} & & & & \\
Raw I/B/E/S announcements
  & -- & -- & 5{,}419 & 180{,}556 \\
Stock price available (CRSP)
  & -- & -- & 5{,}419 & 142{,}025 \\
Size, SUE, and Forecast filters
  & -- & -- & 5{,}419 & 130{,}496 \\
Dropping multiple announcements per quarter
  & -- & -- & 5{,}419 & 128{,}982 \\
  & -- & -- & 5{,}383 & 123{,}880 \\
\midrule 
\underline{StockTwits} & & & & \\
Posts within [$-90,-1$] of announcement                       & 51{,}065{,}435 & 693{,}965 & 5{,}143 & 117{,}389 \\
Self-reported holding period                                  & 20{,}605{,}979 & 132{,}958 & 5{,}082 & 104{,}919 \\
\bottomrule
\end{tabular}
\begin{tablenotes}
\footnotesize
\item \textit{Notes}: This table summarizes the construction of the final sample used in the analysis. We cover 2010 - 2021 June.
\end{tablenotes}
\end{threeparttable}
\end{table}

%% file: tables/table_1.tex
\begin{scriptsize}
\setlength{\tabcolsep}{1pt}
\renewcommand{\arraystretch}{0.6}
\setlength{\LTleft}{0pt}
\setlength{\LTright}{0pt}

\begin{longtable}{l*{12}{c}} 
\caption{Summary Statistics}
\label{tab:summary_stats_stocktwits} \\
\toprule
\multicolumn{13}{l}{\textbf{A. CRSP/Compustat-StockTwits-I/B/E/S Sample Statistics}} \\
\midrule
& Mean & Std. Dev. & 1\% & 25\% & 50\% & 75\% & 99\% & & & & & \\
\midrule
\endfirsthead

\multicolumn{13}{c}{\textit{(Continued)}} \\
\toprule
\multicolumn{13}{l}{\textbf{B. Differences Between Announcements for Long-Term and Short-Term Stocks}} \\
\midrule
& Full sample & Long-Term & Short-Term & Norm. diff. & t-stat & & & & & & & \\
& (N = 104,919) & (N = 53,354) & (N = 51,565) & & & & & & & & & \\
\midrule
\endhead

\midrule
\multicolumn{13}{r}{\textit{(Continued on next page)}} \\
\endfoot

\bottomrule
\multicolumn{13}{p{\dimexpr\textwidth-2\tabcolsep\relax}}{\textit{Notes}: Panel A provides descriptive statistics for market, I/B/E/S, and social media variables. Market variables include size (in millions), volatility, institutional ownership, institutional turnover, and abnormal short interest. Institutional ownership is the fraction of shares held by 13F firms at quarter-end $t$, while turnover is the lagged churn rate calculated from holding changes between $t-2$ and $t-1$. Volatility is computed from 182 days before the announcement up to the day before. I/B/E/S variables include earnings surprise, number of analysts, forecast dispersion, and buy-and-hold abnormal returns at various horizons. Social media variables include posting activity by user type, as well as the mean and standard deviation of sentiment (valence). Valence is extracted following \cite{vamossy2023emtract} over the window $[-30, -1]$ relative to the announcement. Panel B presents differences between announcements for long-term and short-term stocks, including normalized differences and t-statistics. Earnings surprise is the difference between actual and median forecast scaled by the stock price five trading days before the announcement. The sample spans 2010–2021.} \\
\endlastfoot

\multicolumn{13}{l}{\textbf{Market Variables}} \\
Market Cap              & 12,965.8    & 90,845.5    & 72.4     & 621.7    & 1,845.9    & 6,054.3    & 180,277.3   & & & & & \\
Volatility              & 0.0257     & 0.0274     & 0.0083   & 0.0154   & 0.0214    & 0.0312    & 0.0812     & & & & & \\
Turnover                & 0.2334     & 0.2661     & 0.0000   & 0.1181   & 0.1765    & 0.2600    & 2.0000     & & & & & \\
Institutional Ownership & 0.7003     & 0.2817     & 0.0000   & 0.5599   & 0.7878    & 0.9196    & 1.0000     & & & & & \\
Abnormal Short Interest & 0.0001     & 0.0147     & -0.0417  & -0.0031  & 0.0000    & 0.0023    & 0.0476     & & & & & \\
Past 3-Month Return     & 0.0269     & 0.2281     & -0.4921  & -0.0822  & 0.0178    & 0.1180    & 0.7181     & & & & & \\
\midrule
\multicolumn{13}{l}{\textbf{Earnings-Announcement and Analyst Forecast Variables}} \\
Earnings Surprise       & 0.0001 & 0.0252 & -0.0552 & -0.0008 & 0.0005 & 0.0024 & 0.0435 & & & & & \\
Number of Analysts      & 9.0    & 7.1    & 1       & 4       & 7      & 13     & 31     & & & & & \\
Forecast Dispersion     & 0.0819 & 0.4131 & 0.0000  & 0.0100  & 0.0300 & 0.0600 & 0.8800 & & & & & \\
BHAR [-30,-1]           & -0.0068& 0.1215 & -0.3728 & -0.0674 & -0.0069& 0.0511 & 0.4218 & & & & & \\
BHAR [0,1]              & -0.0006& 0.0767 & -0.2389 & -0.0376 & -0.0005& 0.0370 & 0.2338 & & & & & \\
BHAR [2,75]             & -0.0256& 0.2285 & -0.7392 & -0.1407 & -0.0224& 0.0899 & 0.7222 & & & & & \\
\midrule
\multicolumn{13}{l}{\textbf{Social Media Information}} \\
Distinct Long-Term Users  & 14.16  & 52.06   & 0      & 3      & 6      & 11      & 143.82   & & & & & \\
Distinct Short-Term Users & 20.82  & 86.20   & 0      & 2      & 5      & 14      & 255      & & & & & \\
Long-Term Posts      & 88.76  & 843.27  & 0      & 9      & 23     & 47      & 1{,}060.82 & & & & & \\
Short-Term Posts     & 107.64 & 1{,}087.90 & 0    & 3      & 10     & 36      & 1{,}434.64 & & & & & \\
Valence [$-30,-1$]   & & & & & & \multicolumn{7}{c}{} \\ 
\quad Mean                 & 0.5241	&0.0767&	0.3420&	0.4960&	0.5020&	0.5474&	0.8171 & & & & & \\
\quad Std. Dev.           & 0.1118&	0.1008&		0.0000&	0.0053&	0.1063&	0.2021&	0.3244 & & & & & \\
\midrule
Observations & 104,919
 & & & & & & & & & & & \\
\midrule\midrule

\multicolumn{13}{l}{\textbf{B. Differences Between Announcements for Long-Term and Short-Term Stocks}} \\
\midrule
& Full sample & Long-Term & Short-Term & Norm. diff. & t-stat & & & & & & & \\
& (N = 104,919) & (N = 53,354) & (N = 51,565) & & & & & & & & & \\
\midrule
Market Cap              & 12{,}965.8 & 7{,}639.1  & 18{,}477.2 & -0.12 & -19.35 & \multicolumn{7}{c}{} \\
Volatility              & 0.026      & 0.023      & 0.029      & -0.22 & -35.35 & \multicolumn{7}{c}{} \\
Turnover                & 0.233      & 0.198      & 0.270      & -0.27 & -43.98 & \multicolumn{7}{c}{} \\
Institutional Ownership & 0.700      & 0.694      & 0.706      & -0.04 &  -6.95 & \multicolumn{7}{c}{} \\
Past 3-Month Return     & 0.027      & 0.024      & 0.029      & -0.02 &  -3.63 & \multicolumn{7}{c}{} \\
Abnormal Short Interest & 0.000      & -0.000     & 0.000      & -0.03 &  -5.12 & \multicolumn{7}{c}{} \\
Valence [$-30,-1$]   & & & & & & \multicolumn{7}{c}{} \\ 
\quad Mean & 0.524      & 0.512      & 0.537      & -0.33 & -53.40 & \multicolumn{7}{c}{} \\
\quad Std. Dev.     & 0.112      & 0.072      & 0.153      & -0.81 & -143.84& \multicolumn{7}{c}{} \\
Year                    & 2015.799   & 2016.784   & 2014.780   &  0.63 & 107.50 & \multicolumn{7}{c}{} \\
Earnings Surprise (SUE) & 0.000      & -0.000     & 0.000      & -0.01 &  -2.07 & \multicolumn{7}{c}{} \\
Number of Analysts      & 8.958      & 6.706      & 11.288     & -0.65 & -110.89& \multicolumn{7}{c}{} \\
Forecast Dispersion     & 0.082      & 0.064      & 0.101      & -0.09 & -14.44 & \multicolumn{7}{c}{} \\
BHAR [0, 1]             & -0.001     & 0.001      & -0.003     &  0.05 &   8.01 & \multicolumn{7}{c}{} \\
BHAR [2, 75]            & -0.026     & -0.012     & -0.040     &  0.12 &  19.90 & \multicolumn{7}{c}{} \\
\midrule\midrule

\multicolumn{13}{l}{\textbf{C. Average Surprise by Earnings Surprise Quantile}} \\
\midrule
& Quantile & 1 & 2 & 3 & 4 & 5 & 6 & 7 & 8 & 9 & 10 & 11 \\
\midrule
Long-Term  & Average & -0.036 & -0.006 & -0.003 & -0.001 & -0.000 & 0.000 & 0.000 & 0.001 & 0.002 & 0.004 & 0.018 \\
           & N       & 3{,}222 & 3{,}680 & 3{,}841 & 3{,}879 & 3{,}661 & 3{,}995 & 5{,}865 & 6{,}248 & 6{,}458 & 6{,}517 & 5{,}988 \\

Short-Term & Average & -0.043 & -0.006 & -0.003 & -0.001 & -0.000 & 0.000 & 0.000 & 0.001 & 0.002 & 0.004 & 0.023 \\
           & N       & 3{,}542 & 3{,}054 & 2{,}895 & 2{,}855 & 3{,}093 & 3{,}807 & 6{,}832 & 6{,}418 & 6{,}219 & 6{,}149 & 6{,}701 \\

\end{longtable}
\end{scriptsize}

%% file: figures/figure_1.tex
\begin{figure*}[t!]
    \centering
    \begin{minipage}[b]{\linewidth}
        \centering
        \includegraphics[width=0.6\linewidth]{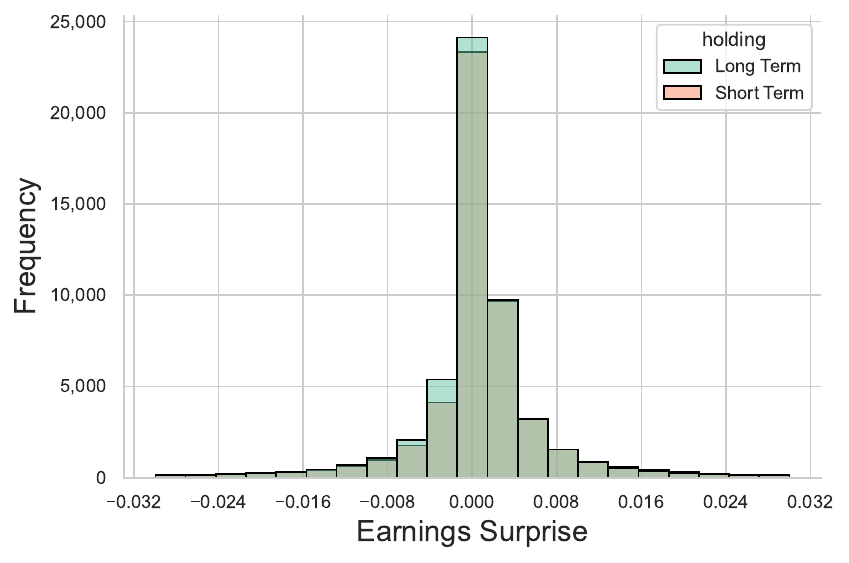}
        \caption*{(a) Earnings Surprise Distribution}
        \label{fig:2a}
    \end{minipage}
    \begin{minipage}[b]{\linewidth}
        \centering
        \includegraphics[width=0.6\linewidth]{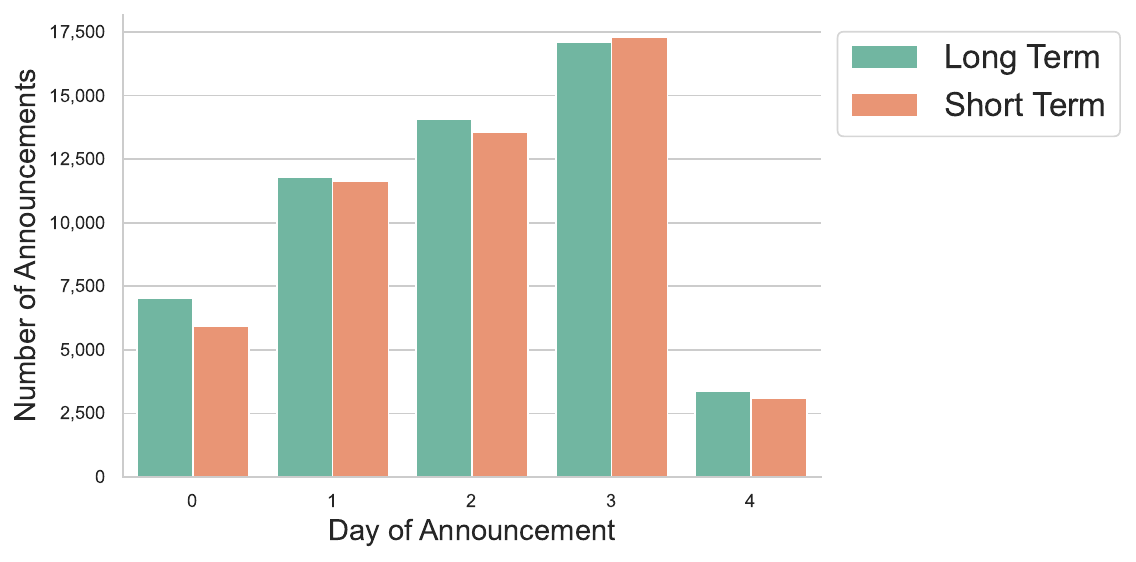}
        \caption*{(b) Day of the Week Distribution}
        \label{fig:2b}
    \end{minipage}
    \begin{minipage}[b]{\linewidth}
        \centering
        \includegraphics[width=0.6\linewidth]{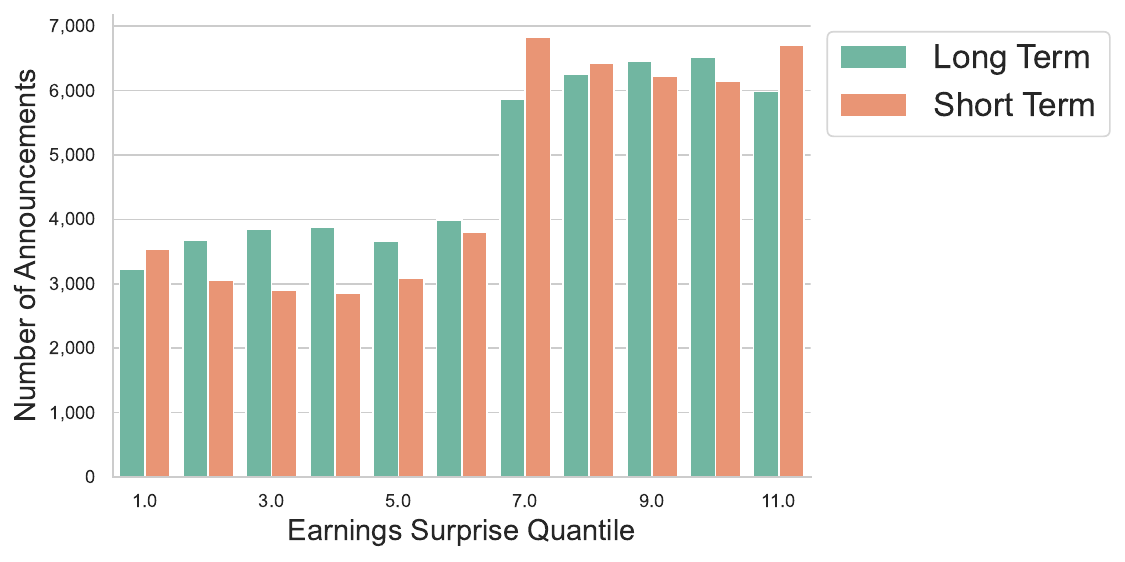}
        \caption*{(c) Earnings Surprise Histogram for Quantiles}
        \label{fig:2c}
    \end{minipage}
    \caption{Earnings Announcements by Retail Investor Horizons}
    \label{fig:figure4abc}
    \captionsetup{justification=justified, font=scriptsize}
    \caption*{Notes: The earnings surprise for an announcement is defined as the difference between the actual earnings for the quarter, as recorded by I/B/E/S, and the median analyst forecast featured in the I/B/E/S detail file within the 30 days preceding the quarterly earnings announcement. This is then scaled by the stock price 5 trading days before the announcement. Earnings announcements are divided into 11 groups: Quintiles 1 to 5 consist of five quintiles of negative earnings surprises, while quintiles 7 to 11 comprise five quintiles of positive earnings surprises. Quintile 6 includes all announcements where the earnings surprise is zero. The breakpoints are determined quarterly.}
\end{figure*}

%% file: tables/table_2.tex
\begin{table}[htbp]
\centering
\scriptsize 
\begin{threeparttable}
\caption{Investor Horizon Measure Stability}\label{tab:stability}
\begin{tabular}{@{}p{2in}p{1in}p{1in}@{}} \hline 
& & \\
\multicolumn{3}{l}{\underline{Panel A: Switching Behavior}} \\ 
& & \\
 \multicolumn{3}{c}{Event Time} \\
{[-90, -1]} & {[0, 45]} & {[46, 90]} \\
Long Term                    & 72.2\%     & 65.65\%    \\
Short Term                   & 77.09\%     & 77.08\%    \\
& & \\ 
\multicolumn{3}{l}{\underline{Panel B: Persistence}} \\ 
& \multicolumn{2}{c}{Percent of time classified as} \\
$\downarrow$ Investor Horizon at $t=t'$               & Long-Term & Short-Term \\
Long Term                    & 70.1\%   & 29.9\%    \\
Short Term                   & 31.0\%   & 69.0\%    \\ \hline \hline 
\end{tabular}
\begin{tablenotes}

\scriptsize
\item \textit{Notes}: This table evaluates the stability of our measure. Panel A reports, for each announcement, the share of stocks that retain their initial ``Long-Term" or ``Short-Term” label when moving from the \([-90,-1]\)-day pre-announcement window to the \([0,45]\) and \([46,90]\) windows around the event, demonstrating the within-event stability of our horizon measure. Panel B shows, for each firm, the average fraction of announcements on which its stock maintains the same initial classification, highlighting across-announcement persistence.
\end{tablenotes}
\end{threeparttable}
\end{table}

%% file: sections/theory.tex
\section{A Model of Earnings Announcement Reactions and Investor Horizons}\label{sec:theory}

This section develops a simple rational-expectations model in which investors differ in their trading horizons. The framework is designed to capture how short- and long-term investors process earnings news differently and how these differences can translate into heterogeneous price reactions around earnings announcements. The model provides clear comparative statics that generate horizon-dependent patterns in immediate and delayed returns, guiding the empirical analysis that follows.

\subsection*{Setup}

We consider a market over three periods, indexed by $t=1,2,3$, with two assets: a \textbf{risk-free asset} yielding a zero interest rate, normalized to have a constant price of $1$; and a \textbf{risky asset} with an uncertain liquidation value $\theta$ that is publicly revealed at date $t=3$. The prices of the risky asset at dates $t=1$ and $t=2$ are denoted by $p_1$ and $p_2$, respectively. The risky asset has a fixed supply $\bar{K}$. 

There are two types of investors with different investment horizons: \textbf{Short-term investors} ($S$): They plan to hold the risky asset from date $t=1$ to date $t=2$ and may re-balance their portfolios at date $t=2$ based on new information. \textbf{Long-term investors} ($L$): They plan to hold the risky asset from date $t=1$ to date $t=3$ and do not trade at date $t=2$. All investors are risk-averse, have mean-variance preferences, and are price-takers.

\subsection*{Information Structure}

At date $t=1$, investors receive private signals about the future values of the risky asset.

\paragraph{Short-term Investors}

Each short-term investor $i \in S$ observes a private signal about the upcoming earnings announcement $e$ at date $t=2$: 
\begin{equation} z_i^S = e + \epsilon_i^S, 
\end{equation} 
where $\epsilon_i^S \sim \mathcal{N}(0, \sigma_S^2)$ is an idiosyncratic noise term.

\paragraph{Long-term Investors}

Each long-term investor $j \in L$ observes a private signal about the fundamental value $\theta$ at date $t=3$: 
\begin{equation} 
z_j^L = \theta + \epsilon_j^L, 
\end{equation} where $\epsilon_j^L \sim \mathcal{N}(0, \sigma_L^2)$ is an idiosyncratic noise term.

\paragraph{Relationship between $e$ and $\theta$}

The earnings announcement $e$ is informative about the fundamental value $\theta$, but may deviate from it due to short-term fluctuations: \begin{equation} 
e = \theta + \eta, 
\end{equation}
where $\eta \sim \mathcal{N}(\mu_{\eta}, \sigma_{\eta}^2)$ represents transitory components or noise in earnings that are not reflective of long-term fundamentals. We assume that $\theta$, $\eta$, $\epsilon_i^S$, and $\epsilon_j^L$ are independent random variables.

\subsection*{Investor Expectations and Biases}

We assume that short-term investors' prior belief about $e$ has a positive bias: \begin{equation} \mathbb{E}^S[e] = \hat{e}_S = \hat{e} + b, \quad b > 0, \end{equation} where $\hat{e}$ is the consensus expectation of $e$, and $b$ represents the optimism bias of short-term investors.\footnote{This lines up with our empirical findings, i.e., short-term investors have systematically more optimistic expectations about the earnings announcement $e$ compared to long-term investors and the consensus.} Long-term investors and the consensus have unbiased expectations: \begin{equation} \mathbb{E}^L[e] = \hat{e}, \quad \mathbb{E}^L[\theta] = \hat{\theta}. \end{equation}

\subsection*{Payoffs and Utility Functions}

All investors have mean-variance utility functions defined over their wealth at their respective investment horizons.

\paragraph{Short-term Investors}

A short-term investor $i$ plans to liquidate at date $t=2$. Her utility function at date $t=1$ is: 
\begin{equation} U_i^S = \mathbb{E}_i^S[W_{i,2}] - \frac{\gamma}{2} \text{Var}_i^S[W_{i,2}], 
\end{equation} 
where $W_{i,2} = k_i^S p_2 + (1 - k_i^S p_1)$ is the wealth at date $t=2$, $k_i^S$ is the number of shares held from $t=1$ to $t=2$, and $\gamma$ is the coefficient of absolute risk aversion.

\paragraph{Long-term Investors}

A long-term investor $j$ plans to hold until date $t=3$. Her utility function at date $t=1$ is: 
\begin{equation} U_j^L = \mathbb{E}_j^L[W_{j,3}] - \frac{\gamma}{2} \text{Var}_j^L[W_{j,3}], 
\end{equation} where $W_{j,3} = k_j^L \theta + (1 - k_j^L p_1)$ is the wealth at date $t=3$, and $k_j^L$ is the number of shares held from $t=1$ to $t=3$.

\subsection*{Optimization and Equilibrium at Date $t=1$}

Investors choose their holdings of the risky asset at date $t=1$ to maximize their expected utility.

\paragraph{Short-term Investors}

Maximizing $U_i^S$ with respect to $k_i^S$ yields: \begin{equation} k_i^S = \frac{\mathbb{E}_i^S[p_2] - p_1}{\gamma \text{Var}_i^S[p_2]}, \label{eq
} \end{equation} where $\mathbb{E}_i^S[p_2]$ is investor $i$'s expectation of the price at date $t=2$, conditional on their information at date $t=1$. Short-term investors form expectations about $p_2$ based on their biased expectations of $e$ and the anticipated market reaction. Since they are optimistic about $e$, they expect a higher $p_2$: 
\begin{equation} \mathbb{E}_i^S[p_2] = \mathbb{E}_i^S[p_2 | z_i^S, \hat{e}_S] = \mathbb{E}[p_2 | e = \hat{e}_S + z_i^S - \hat{e}_S], 
\end{equation} 
where $z_i^S - \hat{e}_S$ represents the deviation of their private signal from their biased prior.

\paragraph{Long-term Investors}

Maximizing $U_j^L$ with respect to $k_j^L$ yields: \begin{equation} k_j^L = \frac{\mathbb{E}_j^L[\theta] - p_1}{\gamma \text{Var}_j^L[\theta]}. \label{eq
} \end{equation} Long-term investors form expectations about $\theta$ based on their unbiased private signals: 
\begin{equation} \mathbb{E}j^L[\theta] = \mathbb{E}[\theta | z_j^L] = \hat{\theta} + \frac{\sigma{\theta}^2}{\sigma_{\theta}^2 + \sigma_L^2} (z_j^L - \hat{\theta}). 
\end{equation}

\paragraph{Market Clearing at Date $t=1$}

The market-clearing condition at date $t=1$ is: 
\begin{equation} \int_{i \in S} k_i^S , di + \int_{j \in L} k_j^L , dj = \bar{K}. \label{eq
} 
\end{equation}

\paragraph{Price Formation at Date $t=1$}

The equilibrium price $p_1$ adjusts so that the market clears. Due to the optimism bias of short-term investors, their aggregate demand is higher, putting upward pressure on $p_1$.

\subsection*{Events at Date $t=2$}

At date $t=2$, the earnings announcement $e$ is publicly revealed. Short-term investors update their beliefs about the fundamental value $\theta$ based on $e$.

\paragraph{Short-term Investors}

Upon observing $e$, which may be lower than their optimistic expectation $\hat{e}_S$, short-term investors adjust their expectations: 
\begin{equation} \mathbb{E}_i^{S,2}[\theta] = \hat{\theta} + \frac{\sigma{\theta}^2}{\sigma{\theta}^2 + \sigma_{\eta}^2} (e - \hat{\theta}). 
\end{equation}

Given that $e$ may be lower than their biased expectation $\hat{e}_S$, short-term investors may find $e$ disappointing, leading to a downward revision of their expectations of $\theta$.

\paragraph{Trading Behavior at Date $t=2$}

Short-term investors decide whether to hold or sell the risky asset based on their updated expectations. If $e$ is lower than their optimistic expectation, they are likely to sell, exerting downward pressure on $p_2$.

\paragraph{Price Reaction at Date $t=2$}

The price change from $p_1$ to $p_2$ reflects the aggregate trading behavior of investors:

\begin{equation} p_2 - p_1 = \Delta p = \frac{1}{\lambda} \left( \int_{i \in S} \Delta k_i^S , di \right), 
\end{equation} 
where $\lambda$ is a price impact parameter, and $\Delta k_i^S$ is the change in holdings by short-term investor $i$. Due to the short-term investors' optimistic bias, the negative surprise (when $e < \hat{e}_S$) leads to larger selling pressure, causing a more negative immediate price reaction.

\subsection*{Simulation}

To illustrate the theoretical model and support our analytical results, we conduct a simulation replicating the market dynamics between short-term and long-term investors with different expectations. Setting \( b = 0.5 \) and \( \gamma = 2.0 \), with standard deviations \( \sigma_\theta = 0.25 \), \( \sigma_\eta = 0.5 \), \( \sigma_S = 1.0 \), \( \sigma_L = 1.0 \), we observe that short-term investors exhibit lower abnormal returns across all deciles due to optimism bias (see Figure \ref{fig:model_simulation}). Additional details on the simulation setup and methodology are provided in Appendix B.

\input{figures/figure_8}

\subsection*{Implications}

The model delivers clear predictions for how investor horizons shape price reactions to earnings news. Short-horizon investors, who over-extrapolate recent information, tend to enter the announcement with inflated expectations. As a result, negative surprises trigger sharp selling pressure, while even positive announcements may generate muted responses when they fall short of these elevated priors. Long-horizon investors, by contrast, form expectations around fundamental value and therefore react more moderately in the immediate window. Because prices partially reflect short-term optimism before the announcement, stocks held predominantly by short-horizon investors are more prone to post-announcement reversals. In contrast, stocks followed by long-horizon investors, which begin closer to fundamental value, experience a gradual upward adjustment as information is incorporated. The model therefore rationalizes asymmetric immediate reactions and horizon-dependent drift around earnings events.

%% file: figures/figure_8.tex
\begin{figure*}[t!]
    \centering
    \includegraphics[scale=0.55]{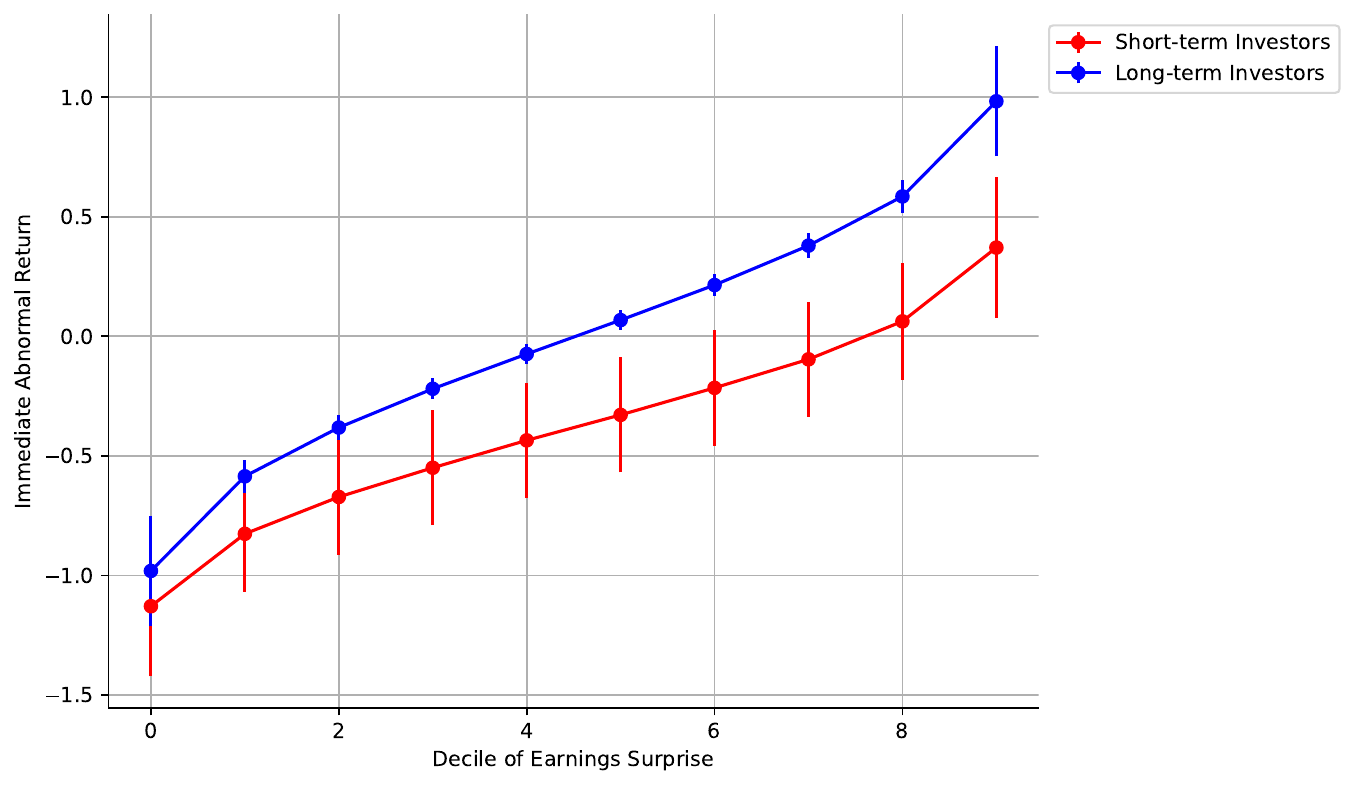}
    \caption{Model Simulation}\label{fig:model_simulation}
    \captionsetup{justification=justified, font=scriptsize}
    \caption*{Notes: This figure illustrates the results of the model simulation examining abnormal returns for stocks predominantly held by short-term versus long-term investors following earnings announcements. The simulation is based on key parameters: optimism bias \( b = 0.5 \) for short-term investors and risk aversion coefficient \( \gamma = 2.0 \). Deciles are formed based on earnings surprises, with mean abnormal returns and standard deviations plotted for each investor type. Error bars represent the standard deviations within each decile.}
\end{figure*}

%% file: sections/methodology.tex
\section{Methodology}

We closely follow the approach of \cite{dellavigna:2009}. The earnings surprise is measured using the actual historical quarterly earnings per share and the consensus forecast of quarterly earnings per share from I/B/E/S, scaled by the historical price per share from the Center for Research in Security Prices (CRSP):

\begin{equation}
s_{t,k} = \frac{e_{t,k} - \hat e_{t,k}}{p_{t-5,k}},
\end{equation} 
where $e_{t,k}$ is the earnings per share announcement of company $k$ released on day $t$; $\hat{e}_{t,k}$ is the related consensus analyst forecast, defined as the median forecast among all analysts who made a forecast in the last 30 trading days prior to the earnings announcement date $t$; and $p_{t-5,k}$ is the price per share for firm $k$ five days preceding the announcement. An interpretation of the surprise measure is the unexpected profits as a share of the total market value of the company. For example, an earnings surprise of $s_{t,k} = 0.01$ means that the company earned unexpected profits equal to one percent of its market value.

To explore the stock reaction to earnings surprises, each announcement date is matched with stock returns, market capitalization, and trading volume from CRSP. The cumulative abnormal return (CAR) and buy-and-hold abnormal return (BHAR) is constructed for different event time windows as follows. Let \(R_{j,k}\) be the raw return on stock \(k\) and \(R_{M,j}\) the return on the value-weighted CRSP market index on trading day \(j\); \(r_{f,j}\) is the one-month Treasury bill rate (continuously compounded) on the same day. Define excess returns \(r_{j,k}=R_{j,k}-r_{f,j}\) and \(r_{M,j}=R_{M,j}-r_{f,j}\). For each announcement we estimate the market model over the
pre-event window \(W_e=[\,t-300,\;t-46\,]\):

\[
r_{j,k}=\alpha_{k}+\beta_{k}\,r_{M,j}+\varepsilon_{j,k},
\qquad j\in W_e.
\tag{A1}
\]

The daily abnormal return is
\[
AR_{j,k}
  = r_{j,k} - \hat{\alpha}_{k} - \hat{\beta}_{k}\,r_{M,j}.
\tag{A2}
\]

For an event window \([h,H]\) (in trading days relative to the announcement date \(t\)) the cumulative abnormal return is
\begin{equation}
\label{eq:cardef}%
CAR_{k}^{h,H}
  =\sum_{j=t+h}^{t+H}AR_{j,k}.
\tag{A3}
\end{equation}

The buy-and-hold abnormal return is calculated on raw (gross) returns:
\begin{equation}
\label{eq:bhardef}%
BHAR_{k}^{h,H}
  =\prod_{j=t+h}^{t+H}\!(1+R_{j,k})
   -\;
   \prod_{j=t+h}^{t+H}\Bigl[1+r_{f,j}+\hat{\beta}_{k}\,r_{M,j}\Bigr]
\tag{A4}
\end{equation}

If the expected return is modeled with the Fama–French three factors, Carhart momentum, or the five-factor extension, replace the term \(\hat{\beta}_{k}\,r_{M,j}\) in (A2)–(A4) with \(\displaystyle\sum_{q=1}^{K}\hat{\beta}_{k}^{(q)}\,f_{j}^{(q)}\), where \(f_{j}^{(q)}\in\{SMB,HML,UMD,RMW,CMA\}\) and \(K\in\{3,4,5\}\); the \(\hat{\beta}^{(q)}_{k}\) are estimated over \(W_e\).

%% file: sections/results.tex
\section{Primary Findings}

In this section, we examine how stock returns respond to earnings surprises over different time horizons. Specifically, we compare the responsiveness for announcements followed by long-term retail investors to those followed by short-term investors. We present visual evidence and carry out regression analyses to support our findings.

\subsection{Immediate Reaction}

Figure \ref{fig:figure1ab}(a) breaks down the immediate abnormal return following each announcement by whether long-term or short-term investors dominate the ensuing discussion. Across every earnings-surprise quintile, stocks with a heavier long-term investor presence see noticeably stronger initial price reactions than those led by shorter-horizon traders. To formally test this observation, we estimate the following regression model:

\begin{equation}
BHAR^{0,1}_{t,k}=\alpha + \beta \times \mathbb{I}_{t,k, \text{Long-term} = 1} + \sum_{j=2}^{11} \Big[\delta_j \times \mathbb{I}_{t,k, \text{Earnings Quantile} = j}\Big] + \gamma^{k} \times X_{t,k}+\lambda_{t} +\eta_{k} +\epsilon_{t,k}, \label{car_reg}
\end{equation}

In Equation (\ref{car_reg}), $BHAR^{0,1}_{t,k}$ denotes the buy-and-hold abnormal return for stock $k$ at announcement $t$ from the close of the day before the earnings announcement to the close of the day after the announcement (i.e., days 0 to 1). The term $\mathbb{I}_{t,k, \text{Long-term} = 1}$ is an indicator variable equal to 1 if the announcement is classified as long-term, and 0 if short-term. The variables $\mathbb{I}_{t,k, \text{Earnings Quantile} = j}$ are indicator variables for earnings surprise quantiles $j$, where $j = 2, \ldots, 11$. The vector $X_{t,k}$ is a set of control variables which include the size of the security, the dispersion of analysts forecasts, the number of analysts, the buy-and-hold abnormal returns [-30, -1], valence and standard deviation of valence over [-30, -1], institutional ownership \& turnover, volatility, and abnormal short interest. The term $\lambda_t$ represents time fixed effects (e.g., quarter and year dummies), and $\epsilon_{t,k}$ is the error term. To address the concern that companies followed by long-term investors may also have unobservable features that differ from companies followed by short-term investors, we control for firm fixed effects in some regression specifications ($\eta_{k}$). The standard errors account for heteroskedasticity as well as the correlation of errors across securities announcing on the same day by clustering observations by firm and day of the announcement.  

The coefficient $\beta$ captures the additional immediate return response associated with long-term announcements compared to short-term announcements. A positive and significant $\beta$ indicates that long-term announcements have higher immediate abnormal returns, controlling for other factors.

Table \ref{table:2_stocktwits} reports estimates of the differential price reaction between stocks predominantly followed by long- versus short-horizon retail investors. In the baseline specification without controls or fixed effects (Column 1), the coefficient on the long-horizon indicator is 0.53 percentage points (s.e.\ 0.05 pp; $p<0.01$), indicating a meaningfully stronger two-day response for long-horizon stocks. Conditioning on the full vector of firm- and announcement-level covariates together with firm and year–quarter fixed effects (Column 2) reduces the estimate to 0.21 pp (s.e.\ 0.06 pp; $p<0.01$). A propensity-score-matching specification (Column 3) delivers a slightly smaller estimate of 0.16 pp (s.e.\ 0.09 pp; $p<0.10$).

The cross-sectional heterogeneity by news type is striking. For negative earnings surprises (Columns 4–6), the long-horizon coefficient becomes small and statistically indistinguishable from zero once controls and fixed effects are included (Column 5). In contrast, for positive earnings surprises (Columns 7–9), the fully saturated specification (Column 8) yields a sizeable and precisely estimated effect of 0.28 pp (s.e.\ 0.08 pp; $p<0.01$). Thus, the differential reaction between long- and short-horizon stocks is concentrated in good-news environments, consistent with long-horizon investors reacting more strongly to favorable information.

Panel~A of Table~\ref{table:3_stocktwits} estimates the immediate price reaction separately by earnings-surprise group. Across the five negative-surprise groups, the long-horizon coefficient is uniformly small and statistically indistinguishable from zero: estimates range from 0.07~bp in group $-5$ to 19~bp in group $-1$, with standard errors between 24–35~bp. The no-surprise category exhibits a modest and statistically significant premium of 0.56~pp (s.e.\ 0.28~pp; $p<0.10$). On the positive side, the differential reaction strengthens. The coefficient is 0.22~pp for group $+1$ (s.e.\ 0.15~pp), becomes statistically significant in groups $+2$ and $+4$ (0.39~pp and 0.53~pp, respectively), and remains positive in group $+5$ (0.43~pp; s.e.\ 0.22~pp; $p<0.10$). Overall, significant long–short differences emerge only in the no-surprise and positive-surprise bins, and they increase monotonically through the middle of the positive tail.

Investor horizon predicts how quickly, and how strongly, prices incorporate earnings information. The effect is highly asymmetric: we find no significant difference in the immediate price reaction between long- and short-horizon stocks following negative news. In contrast, prices of long-horizon securities rise more on good news and react more favorably when there is no surprise. These patterns, particularly the monotonic increase in the differential premium across positive-surprise quantiles, suggest that long-horizon investors trade more aggressively around favorable information events, generating larger and more precisely estimated immediate abnormal returns for the stocks they follow.

\subsection{Delayed Reaction}

We next turn to the delayed reaction, measured as the buy-and-hold abnormal return from day 2 to day 75. Figure \ref{fig:figure1ab}(c) plots the average drift by surprise decile and shows that the long-horizon line lies above the short-horizon line in every bin, with the gap widening in the extreme tails. 

To assess post-announcement drift, we re-estimate Equation~(\ref{car_reg}) using the buy-and-hold abnormal return from days 2 to 75 as the dependent variable. Panel~B of Table~\ref{table:2_stocktwits} reports the results. In the specification without controls or fixed effects (Column~1), long-horizon stocks earn a 2.83~pp higher return over the 2--75 day window (s.e.~0.22~pp; $p<0.01$). Conditioning on the full set of announcement- and firm-level controls, together with firm and year--quarter fixed effects (Column~2), reduces the estimate to 2.08~pp (s.e.~0.21~pp; $p<0.01$). The propensity-score–weighted specification (Column~3) yields a closely related estimate of 1.61~pp (s.e.~0.28~pp; $p<0.01$). Thus, between 1.5 and 2.8 percentage points of additional drift accrues to long-horizon stocks, even after extensive conditioning.

The pattern is similar when the sample is split by news type. For negative surprises (Column~5), the fully saturated coefficient is 1.81~pp (s.e.~0.39~pp; $p<0.01$). For positive surprises (Column~8), the corresponding estimate is 2.18~pp (s.e.~0.27~pp; $p<0.01$). Hence, the drift differential is not concentrated exclusively in good-news events.

Panel~B of Table~\ref{table:3_stocktwits} examines drift by earnings-surprise category, re-estimating Column~2 specification of Table~\ref{table:2_stocktwits}. Long-horizon coefficients are positive in most bins and statistically significant in the moderate-to-extreme surprise groups on both sides of the distribution. Effects range from 1.47~pp in the most negative bin to 3.17~pp in the $+4$ bin, with statistical significance in more than half of the categories. The delayed-return premium for long-horizon stocks therefore appears broad-based, spanning both negative and positive surprise environments and strengthening in the tails of the distribution.

These results show that the horizon differential documented in the first two trading days does not dissipate; it widens materially over the subsequent quarter. Across specifications, long-horizon stocks earn an additional 2--3~pp of abnormal return relative to short-horizon stocks over days 2--75, and this drift emerges in both negative- and positive-surprise environments. The by-bin estimates in Panel~B of Table~\ref{table:3_stocktwits} further indicate that the premium is broad-based: coefficients are uniformly positive and become largest and most precisely estimated in the moderate-to-extreme surprise categories. Overall, the evidence points to a systematic and economically meaningful post-announcement drift in the stocks followed by long-horizon investors, consistent with prices adjusting more slowly to earnings news for these firms.

\subsection{Total Reaction}

Panel~C of Table~\ref{table:2_stocktwits} aggregates the immediate and delayed windows by regressing the buy-and-hold abnormal return from day~0 to day~75 on the long-term indicator. In the specification without controls or fixed effects (Column~1), securities followed predominantly by long-horizon investors earn 3.41~pp more than those followed by short-horizon investors (s.e.\ 0.23~pp; $p<0.01$). Conditioning on the full set of covariates $X_{t,k}$ together with firm and year--quarter fixed effects (Column~2) reduces the estimate to 2.31~pp (s.e.\ 0.23~pp; $p<0.01$). The propensity-score--weighted specification (Column~3) delivers 1.81~pp(s.e.\ 0.31~pp; $p<0.01$). When the sample is partitioned by the sign of the earnings surprise, the long-term coefficient remains positive and significant: 1.90~pp for negative surprises (Column~5; s.e.\ 0.42~pp; $p<0.01$) and 2.56~pp for positive surprises (Column~8; s.e.\ 0.29~pp; $p<0.01$). Thus, long-horizon stocks outperform short-horizon stocks over the full 0--75 day window in every specification and for both good- and bad-news announcements.

Panel~C of Table~\ref{table:3_stocktwits} shows that the long-term coefficient is positive in every earnings-surprise category, with magnitudes ranging from 1.40~pp in the most negative bin to 3.90~pp in the \(+4\) bin. Eight of the eleven coefficients are statistically significant at conventional levels, including the \(-4\), \(-3\), \(-1\), no-surprise, \(+2\), \(+3\), \(+4\), and \(+5\) groups. The strongest effects appear in the middle-to-upper part of the positive distribution, where estimates reach 3.06--3.90~pp. Even in the negative-surprise bins, the point estimates remain uniformly positive. 

These estimates indicate that the horizon effect is not confined to the first two trading days but continues to accumulate over the full post–announcement window. Across specifications, securities followed predominantly by long–horizon retail investors earn between 1.8 and 3.4 percentage points more than those followed by short–horizon investors over days 0–75, an economically meaningful magnitude for a single quarterly earnings event. The persistence and breadth of this differential, appearing after both negative and positive surprises, echoes the classic post–earnings–announcement–drift (PEAD) pattern of slow information incorporation. In our setting, the drift is systematically stronger among firms with a longer–horizon retail investor base. One interpretation is that investor horizons shape the speed at which beliefs about future cash flows are incorporated into prices: long–horizon investors may update more gradually, generating a sustained price adjustment, whereas short–horizon investors react more immediately, leaving less scope for drift. The result is a robust and economically significant horizon–based wedge in post–announcement price dynamics.\footnote{Appendix A presents comprehensive robustness checks, including alternative risk factor models (three-, four-, and five-factor specifications), sensitivity to different post-announcement event windows, variations in propensity score matching parameters, and tests for the impact of winsorization.}

\subsection{Performance of Drift}

A zero-cost portfolio that is long the long-term group and short the short-term group does not require immediate trading after the announcement. Panel~(a) of Figure~\ref{fig:figure3} shows that the cumulative buy-and-hold abnormal return for this portfolio drifts only mildly during the first two trading weeks, but then accelerates sharply. Between trading days 20 and 75 the strategy earns an additional three percentage points, indicating that substantial abnormal returns remain available well after the initial price response. The evidence points to a repricing process that continues over several weeks rather than concluding in the immediate aftermath of the earnings release.

We next examine whether this drift varies with the market’s prevailing tone. Using the emotion extraction procedure of (\cite{vamossy2023emtract}) applied to social-media posts from the \([-90,-1]\) window, we classify firms as \emph{bullish} or \emph{bearish} based on the sign of their average valence. Panel~(b) of Figure~\ref{fig:figure3} replicates the long-minus-short portfolio within each sentiment group. The drift appears exclusively among bullish firms, with little to no effect for bearish firms. This pattern suggests that favorable sentiment amplifies the delayed price adjustment, consistent with horizon-driven underreaction being stronger when the tone of investor discourse is positive.

\subsection{Trading Strategy}

We implement a simple monthly zero-cost strategy to gauge investability and the economic magnitude of the horizon channel effect. The strategy is designed to isolate the drift component by using announcement information from month \(t-1\) to predict returns in month \(t\). At the end of month \(t-1\), all stocks with an earnings announcement are identified and assigned to one of eleven earnings-surprise bins. Within each bin, we form a zero-cost portfolio that goes long stocks predominantly followed by Long-Term retail investors and short stocks predominantly followed by Short-Term retail investors. Each bin-level portfolio is \emph{equal-weighted across firms}, and the aggregate strategy return is simply the equal-weighted return across all stocks in all bins. The portfolio is rebalanced monthly using the latest earnings announcements and updated horizon classifications. We then analyze the factor-adjusted alphas of the aggregate LT--ST portfolio, the portfolio restricted to the top three surprise bins, and the results for each bin separately.

Table \ref{table:factor_horizon} presents Fama–French Five-Factor regressions for the horizon-sorted long–short drift portfolios. The portfolio across all bins (Column 1) displays a clear defensive profile, loading negatively on the market factor and positively on both value (HML) and profitability (RMW). These exposures indicate that long-term-investor stocks resemble more profitable and value-oriented firms, whereas short-term-investor stocks lean toward growth and lower profitability. Despite these tilts, the portfolio delivers a statistically significant alpha of roughly 30 bps per month, implying economically meaningful abnormal returns beyond standard factor compensation. The Top-3-bin portfolio (Column 2) shows nearly identical factor structure but produces an even larger and more precisely estimated alpha of nearly 40 bps per month, suggesting that concentrating weights on the bins associated with the strongest drift amplifies the signal.

Factor loadings vary substantially across the earnings-surprise bins (Columns 3–13). Many bins load positively on value and profitability, reinforcing the idea that long-term investors systematically hold higher-quality, more value-like firms. Several bins also exhibit negative market betas, reflecting a defensive component that persists throughout the cross-section. CMA loadings alternate in sign, indicating that horizon-based spreads do not line up cleanly with investment intensity across bins. Alphas at the bin level are mostly positive but imprecise, expected given the much higher noise in narrow long–short portfolios. Two bins stand out: Bin 6 and Bin 9, which generate statistically significant and economically large alphas. In contrast, the remaining bins have estimates clustered near zero with wide confidence intervals, indicating considerable heterogeneity in how investor horizons interact with earnings news.

These results show that horizon-sorted long–short drift portfolios earn positive risk-adjusted returns that are not absorbed by standard factor exposures. Long-horizon stocks resemble high-quality value firms, yet the residual alpha remains significant even after controlling for these characteristics. The performance is strongest in the bins where disagreement between long- and short-horizon investors is most pronounced, which aligns with the earlier finding that long-horizon investors react more strongly to favorable earnings news while the delayed adjustment appears in both good- and bad-news environments. Aggregating across bins, or concentrating on those with the largest ex post drift, reduces idiosyncratic noise and reveals a robust and economically meaningful return premium associated with long-term investor orientation.

\clearpage 

\input{figures/figure_2}
\input{figures/figure_4}
\input{tables/table_3}
\input{tables/table_4}
\input{tables/trading_strategy}

%% file: figures/figure_2.tex
\begin{figure*}[t!]
    \centering
    \begin{minipage}[b]{0.48\linewidth}
        \centering
        \includegraphics[scale=0.32]{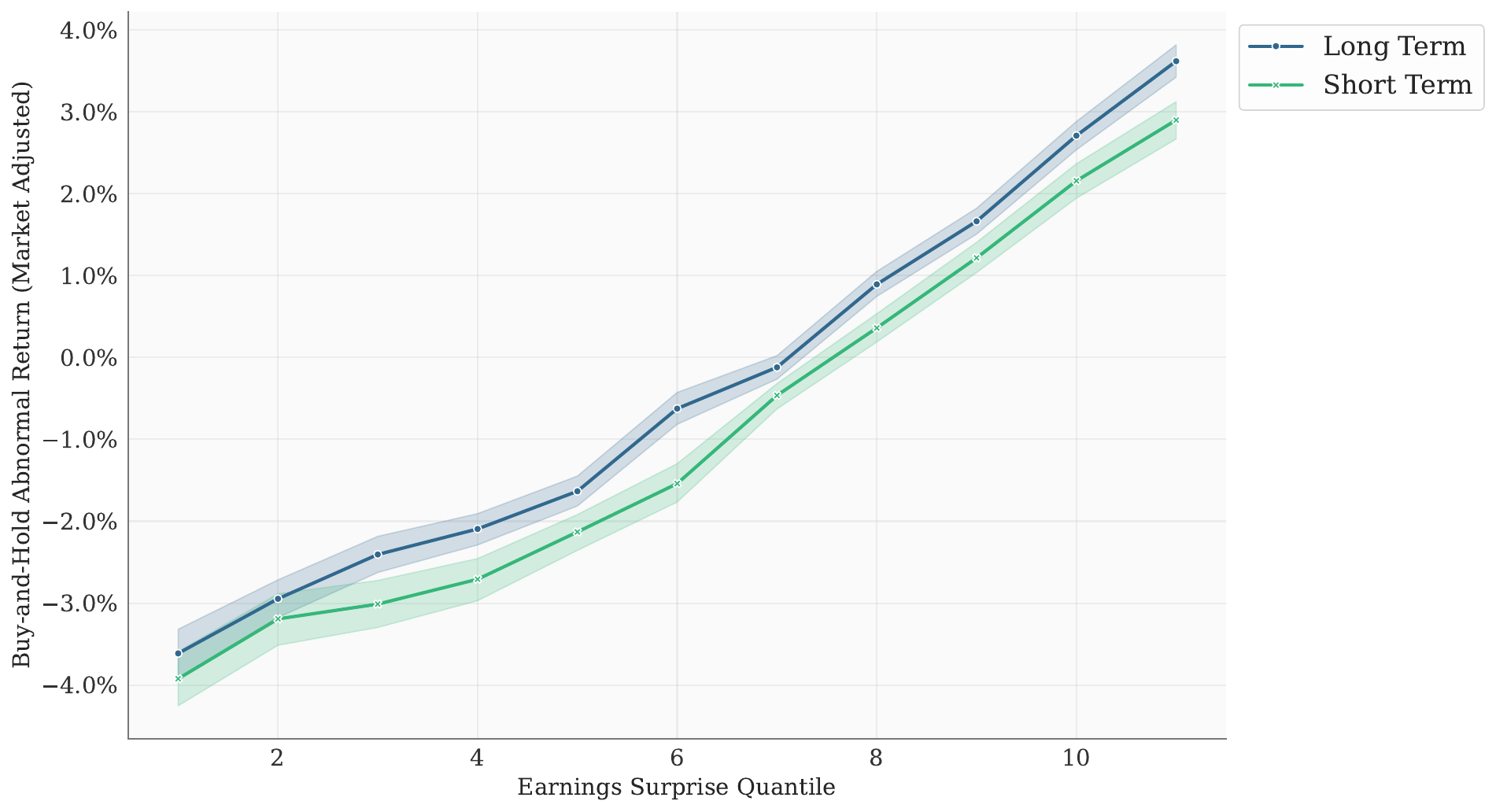}
        \caption*{(a) Response to Earnings Surprise from 0 to 1}
        \label{fig:dellavigna_1a}
    \end{minipage}
    \begin{minipage}[b]{0.48\linewidth}
        \centering
        \includegraphics[scale=0.32]{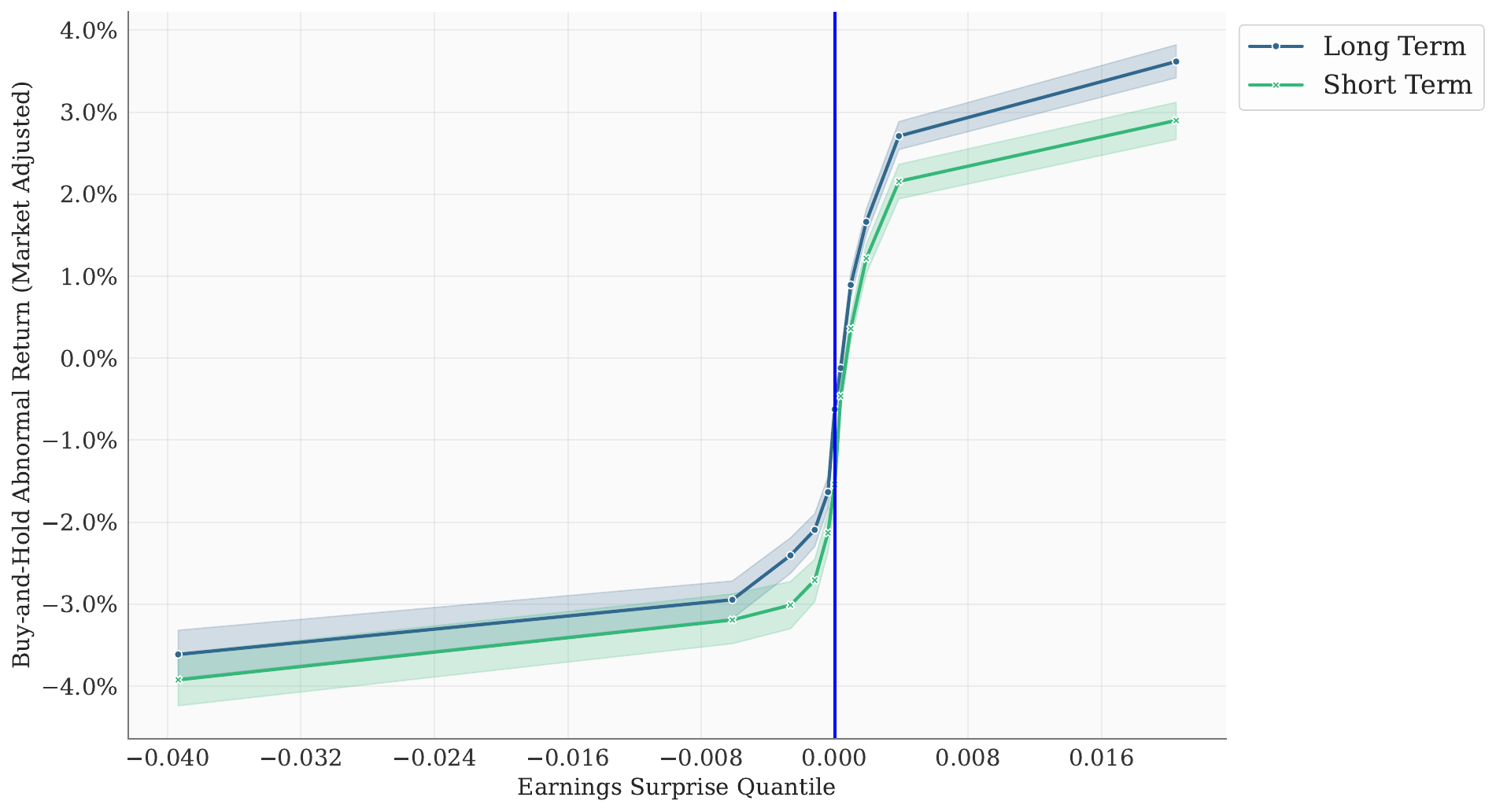}
        \caption*{(b) Nonlinear Earnings Response From 0 to 1}
        \label{fig:dellavigna_1d}
    \end{minipage}
    \begin{minipage}[b]{\linewidth}
        \centering
        \includegraphics[scale=0.35]{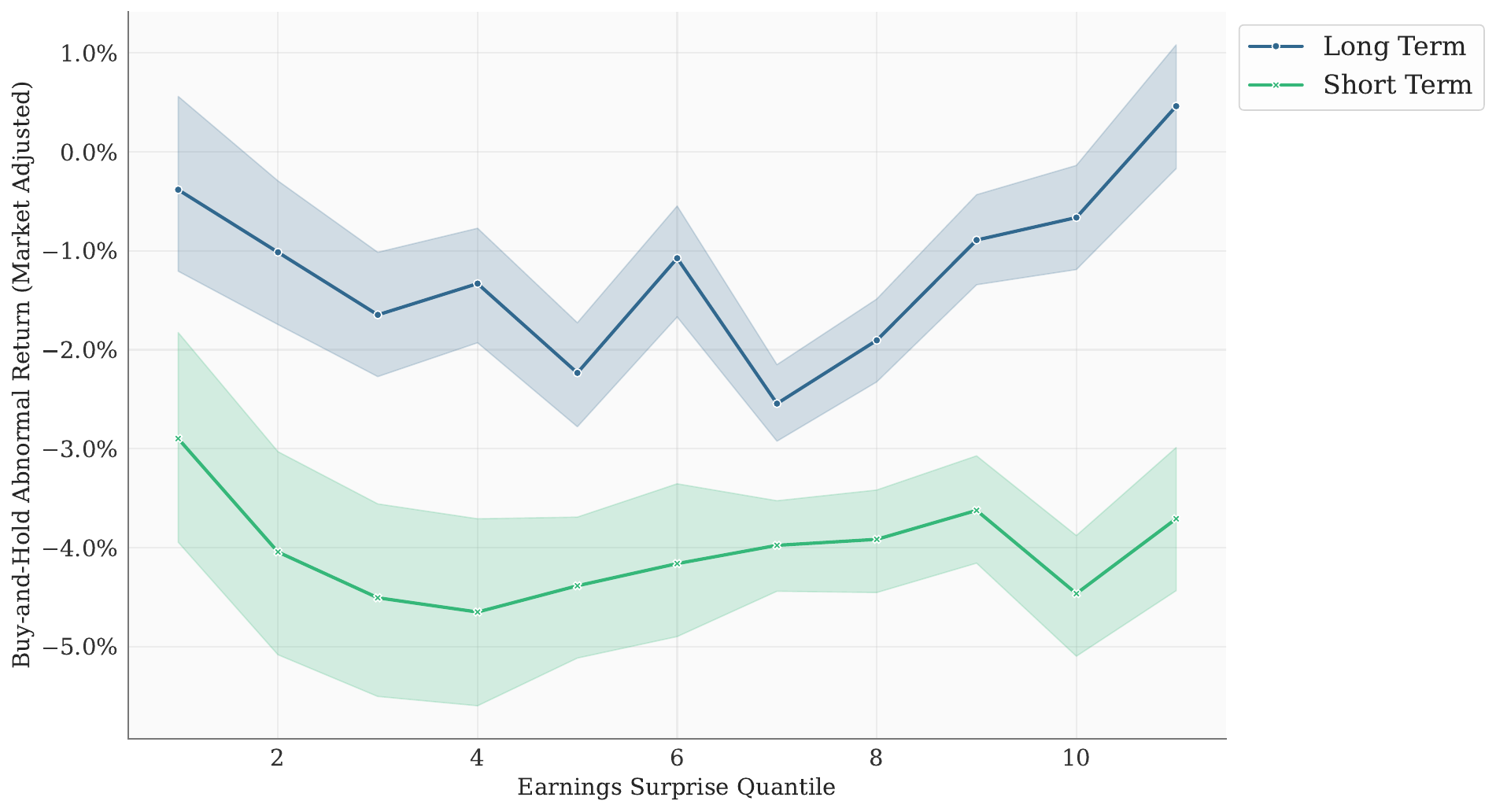}
        \caption*{(c) Response to Earnings Surprise from 2 to 75}
        \label{fig:dellavigna_1b}
    \end{minipage}
    \caption{Immediate (a-b) and delayed (c) response to earnings surprises}\label{fig:figure1ab}
    \captionsetup{justification=justified, font=scriptsize}
    \caption*{Notes: Figure \ref{fig:figure1ab} illustrates the mean buy-and-hold abnormal return reactions. From January 2010 until June 2021, stocks in CRSP are matched with quarterly earnings announcements in I/B/E/S. In the timeline of the event, the day of the announcement is designated as day 0. The buy-and-hold abnormal return for each stock is the raw buy-and-hold return adjusted based on the estimated beta from the market model. The earnings surprise for an announcement is defined as the difference between the actual earnings for the quarter, as recorded by I/B/E/S, and the median analyst forecast featured in the I/B/E/S detail file within the 30 days preceding the quarterly earnings announcement. This is then scaled by the stock price 5 trading days before the announcement. Earnings announcements are divided into 11 groups: Quintiles 1 to 5 consist of five quintiles of negative earnings surprises, while quintiles 7 to 11 comprise five quintiles of positive earnings surprises. Quintile 6 includes all announcements where the earnings surprise is zero. The breakpoints are determined quarterly.}
\end{figure*}

%% file: figures/figure_4.tex
\begin{figure*}[t!]
    \centering
    \begin{minipage}[b]{\linewidth}
        \centering
        \includegraphics[scale=0.4]{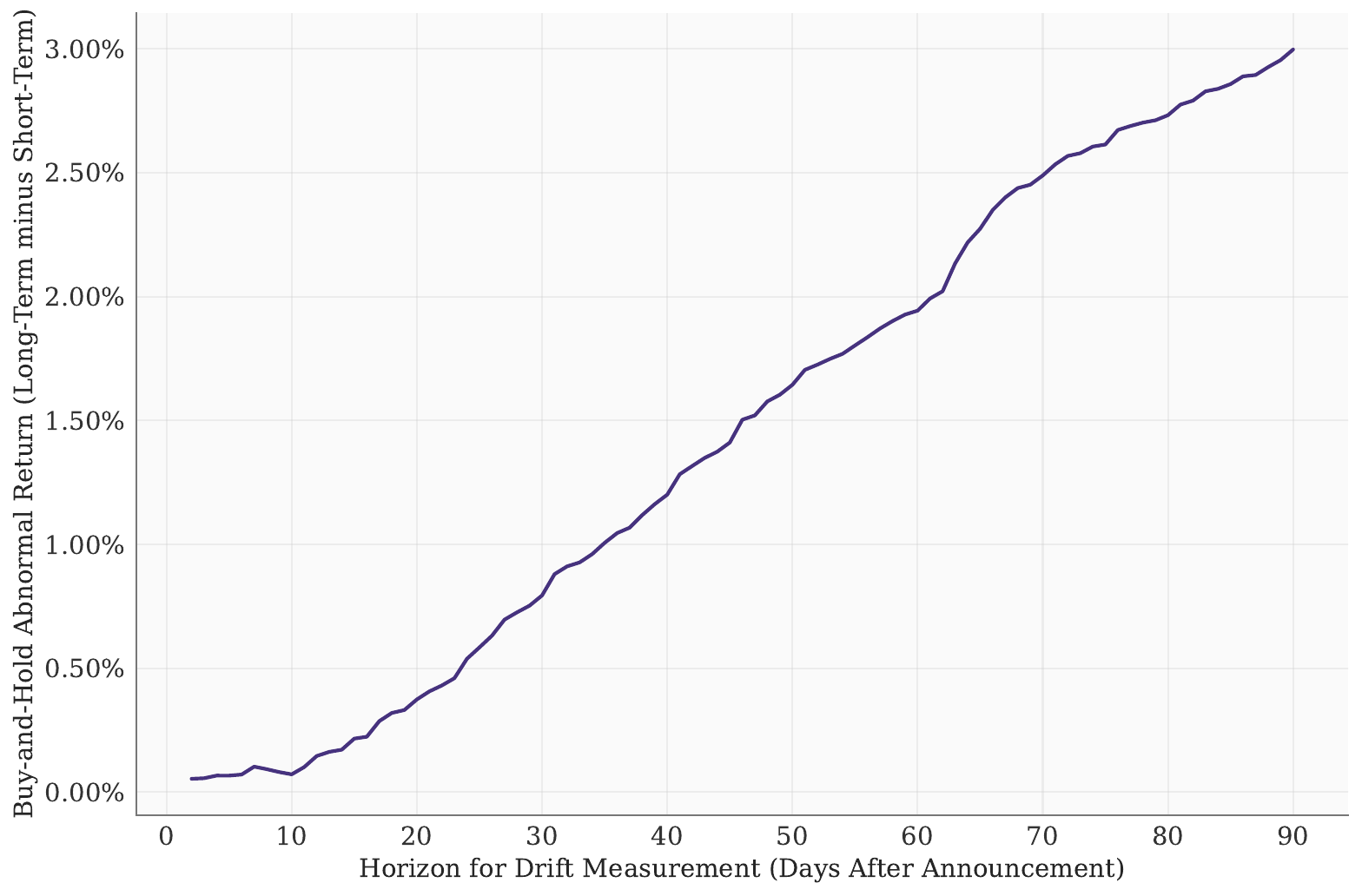}
        \caption*{(a) Performance of drift at different horizons}
        \label{fig:figure3_a}
    \end{minipage}
 \smallskip   
    \begin{minipage}[b]{0.65\linewidth}
        \centering
        \includegraphics[scale=0.4]{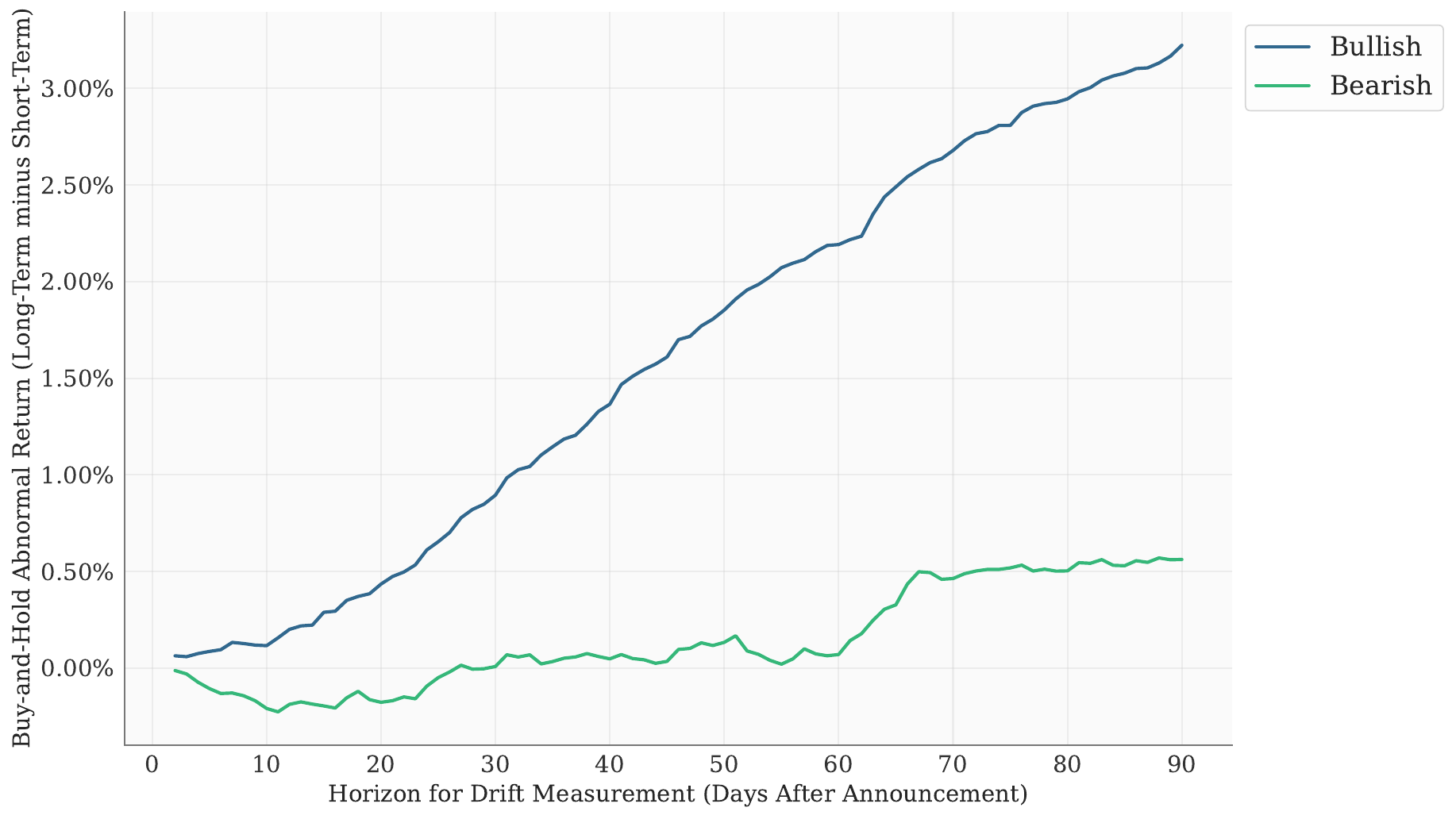}
        \caption*{(b) Performance of drift at different horizons by emotions}
        \label{fig:figure3_b}
    \end{minipage}
    
    \caption{Post-earnings announcement drift at different horizons, overall (a) and by emotions (b)}
    \label{fig:figure3}
    \captionsetup{justification=justified, font=scriptsize}
    \caption*{Notes: Figure \ref{fig:figure3} illustrates the mean buy-and-hold abnormal return reactions. From January 2010 until June 2021, stocks in CRSP are matched with quarterly earnings declarations in I/B/E/S. In the timeline of the event, the day of the announcement is designated as day 0. The buy-and-hold abnormal return for each stock is the raw buy-and-hold return adjusted based on the estimated beta from the market model. The measure for post-earnings announcement drift for horizon \(h\) is the average buy-and-hold abnormal return from day 2 to day \(h\) for long-term announcements minus the average buy-and-hold abnormal return from day 2 to day \(h\) for short-term announcements. Panel (b) estimates this \(\Delta\mathrm{BHAR}_h\) separately for the subsample of stocks with predominantly bullish pre-announcement sentiment and for the subsample with predominantly bearish sentiment in the \([-90,-1]\) window.}
\end{figure*}

%% file: tables/table_3.tex
\begin{landscape}
\begin{table}[htbp]
\scriptsize
\centering
\begin{threeparttable}
\caption{Response to Earnings Surprise}\label{table:2_stocktwits}
\def\sym#1{\ifmmode^{#1}\else\(^{#1}\)\fi}
\begin{tabular*}{\hsize}{@{\hskip\tabcolsep\extracolsep\fill}l*{9}{c}}
\toprule 
                    &\multicolumn{1}{c}{(1)}&\multicolumn{1}{c}{(2)}&\multicolumn{1}{c}{(3)}&\multicolumn{1}{c}{(4)}&\multicolumn{1}{c}{(5)}&\multicolumn{1}{c}{(6)}&\multicolumn{1}{c}{(7)}&\multicolumn{1}{c}{(8)}&\multicolumn{1}{c}{(9)}\\
\midrule 
\multicolumn{8}{l}{\textbf{Panel A. Immediate Reaction}: Buy-and-Hold Abnormal Return in Event Time 0 to 1} \\
\midrule 
Long-Term   &      0.0053\sym{***}&      0.0021\sym{***}&      0.0016\sym{*}  &      0.0045\sym{***}&      0.0015         &     -0.0002         &      0.0052\sym{***}&      0.0028\sym{***}&      0.0035\sym{***}\\
                    &    (0.0005)         &    (0.0006)         &    (0.0009)         &    (0.0010)         &    (0.0012)         &    (0.0017)         &    (0.0007)         &    (0.0008)         &    (0.0012)         \\
& & & & & & & \\[\dimexpr-\normalbaselineskip+3pt]
Constant        &     -0.0033\sym{***}&      0.0110\sym{***}&     -0.0003         &     -0.0299\sym{***}&     -0.0081\sym{*}  &     -0.0258\sym{***}&      0.0123\sym{***}&      0.0178\sym{***}&      0.0149\sym{***}\\
                    &    (0.0004)         &    (0.0025)         &    (0.0005)         &    (0.0009)         &    (0.0042)         &    (0.0009)         &    (0.0006)         &    (0.0034)         &    (0.0006)         \\
\midrule 
$R^2$                 &      0.0851         &      0.1522         &      0.2076         &      0.0083         &      0.2119         &      0.2902         &      0.0300         &      0.1464         &      0.2181         \\
\midrule 
\multicolumn{8}{l}{\textbf{Panel B. Delayed Reaction}: Buy-and-Hold Abnormal Return in Event Time 2 to 75} \\
\midrule 
Long-Term      &      0.0283\sym{***}&      0.0208\sym{***}&      0.0161\sym{***}&      0.0277\sym{***}&      0.0181\sym{***}&      0.0137\sym{***}&      0.0283\sym{***}&      0.0218\sym{***}&      0.0192\sym{***}\\
                    &    (0.0022)         &    (0.0021)         &    (0.0028)         &    (0.0034)         &    (0.0039)         &    (0.0053)         &    (0.0025)         &    (0.0027)         &    (0.0037)         \\
& & & & & & & \\[\dimexpr-\normalbaselineskip+3pt]
Constant            &     -0.0400\sym{***}&      0.0519\sym{***}&     -0.0286\sym{***}&     -0.0409\sym{***}&      0.0606\sym{***}&     -0.0284\sym{***}&     -0.0393\sym{***}&      0.0410\sym{***}&     -0.0289\sym{***}\\
                    &    (0.0022)         &    (0.0077)         &    (0.0015)         &    (0.0033)         &    (0.0128)         &    (0.0027)         &    (0.0022)         &    (0.0113)         &    (0.0019)         \\
\midrule 
$R^2$                 &      0.0043         &      0.1536         &      0.1778         &      0.0036         &      0.2411         &      0.2908         &      0.0046         &      0.1732         &      0.2200         \\
\midrule 
\multicolumn{8}{l}{\textbf{Panel C. Total Reaction}: Buy-and-Hold Abnormal Return in Event Time 0 to 75} \\
\midrule 
Long-Term             &      0.0341\sym{***}&      0.0231\sym{***}&      0.0181\sym{***}&      0.0322\sym{***}&      0.0190\sym{***}&      0.0132\sym{**} &      0.0343\sym{***}&      0.0256\sym{***}&      0.0234\sym{***}\\
                    &    (0.0023)         &    (0.0023)         &    (0.0031)         &    (0.0035)         &    (0.0042)         &    (0.0057)         &    (0.0027)         &    (0.0029)         &    (0.0040)         \\

& & & & & & & \\[\dimexpr-\normalbaselineskip+3pt]
Constant            &     -0.0436\sym{***}&      0.0643\sym{***}&     -0.0290\sym{***}&     -0.0717\sym{***}&      0.0544\sym{***}&     -0.0547\sym{***}&     -0.0268\sym{***}&      0.0586\sym{***}&     -0.0134\sym{***}\\
                    &    (0.0023)         &    (0.0085)         &    (0.0016)         &    (0.0035)         &    (0.0138)         &    (0.0030)         &    (0.0023)         &    (0.0123)         &    (0.0020)         \\
\midrule 
$R^2$                 &      0.0151         &      0.1603         &      0.1859         &      0.0039         &      0.2336         &      0.2894         &      0.0108         &      0.1776         &      0.2251         \\
\midrule 
Year-Quarter Fixed Effects&                     &           X         &           X         &                     &           X         &           X         &                     &           X         &           X         \\
Firm Fixed Effects  &                     &           X         &           X         &                     &           X         &           X         &                     &           X         &           X         \\
Controls            &                     &           X         &           X         &                     &           X         &           X         &                     &           X         &           X         \\
Observations        & 104,919         & 104,703         &  75,080         &  33,722         &  33,155         &  24,820         &  63,395         &  62,977         &  43,751         \\

\bottomrule
\end{tabular*}
\begin{tablenotes}
\item \textit{Notes}: Table \ref{table:2_stocktwits} reports the coefficients of a regression of the immediate response of securities to earning surprises defined by the following regression:
\begin{equation*}
BHAR^{w}_{t,k}=\alpha + \beta \times \mathbb{I}_{t,k, \text{Long-term} = 1} + \sum_{j=2}^{11} \Big[\delta_j \times \mathbb{I}_{t,k, \text{Earnings Quantile} = j}\Big] + \gamma^{k} \times X_{t,k}+\lambda_{t} +\eta_{k} +\epsilon_{t,k},  \label{E4}
\end{equation*}
where $BHAR^{w}_{t,k}$ is defined as in Equation~\eqref{eq:bhardef}. $\beta$ is the added response of the securities in the group followed by long-term investors; $\delta$ capture the relation between the earning surprise and the immediate response; $X_{t,k}$ is a set of control variables which include the size of the security, the dispersion of analysts forecasts, the number of analysts, the buy-and-hold abnormal returns [-30, -1], valence and standard deviation of valence over [-30, -1], institutional ownership \& turnover, volatility, and abnormal short interest. To address the concern that companies followed by long-term investors may also have unobservable features that differ from companies followed by short-term investors, we control for firm fixed effects in most regression specifications. Finally, we also include time fixed effects by year-quarter. The standard errors account for heteroskedasticity as well as correlation of errors across securities making an announcement on the same day by clustering observations by day of announcement and by firm. Table \ref{table:2_stocktwits} reports $\alpha$ and $\beta$. Columns (1), (4) and (7) have no controls or fixed effects; Columns (2), (5), and (8) have the full set of control as defined in $X_{t,k}$ along with firm and time fixed effects; Columns (3), (6), and (9) carries out matching on the control variables, followed by a regression with matching weights and time and firm fixed effects. Columns (4-6) are focusing on negative surprises, while Columns (7-9) are focusing on positive surprises, with and without control. To mitigate the impact of outliers, the dependent variables are winsorized at the 1st and 99th percentiles. \sym{*} \(p<0.10\), \sym{**} \(p<0.05\), \sym{***} \(p<0.01\). 
\end{tablenotes}
\end{threeparttable}
\end{table}
\end{landscape}

%% file: tables/table_4.tex
\begin{landscape}
\begin{table}[htbp]
\renewcommand{\arraystretch}{1.5}
\newcommand{\sym}[1]{\ifmmode^{#1}\else\(^{#1}\)\fi}
\begin{center}
\begin{threeparttable}
\caption{Regressions of Earnings Response by Earning Surprise Groups}
\label{table:3_stocktwits}
\scriptsize
\begin{tabulary}{24cm}{L C C C C C C C C C C C}
\toprule 
\textbf{Earnings Surprise Group}&          -5&          -4&          -3&          -2&          -1&     No News&           1&           2&           3&           4&        5 \\
\midrule 
\multicolumn{12}{l}{\textbf{Panel A. Immediate Reaction}: Buy-and-Hold Abnormal Return in Event Time 0 to 1} \\
Long-Term           &      0.0007         &     -0.0024         &     -0.0010         &     -0.0017         &      0.0019         &      0.0056\sym{*}  &      0.0022         &      0.0039\sym{**} &      0.0033         &      0.0053\sym{**} &      0.0043\sym{*}  \\
                    &    (0.0035)         &    (0.0034)         &    (0.0032)         &    (0.0029)         &    (0.0024)         &    (0.0028)         &    (0.0015)         &    (0.0019)         &    (0.0021)         &    (0.0022)         &    (0.0022)         \\

Constant       &     -0.0221\sym{*}  &     -0.0184\sym{*}  &      0.0008         &     -0.0063         &      0.0008         &      0.0181         &      0.0089         &      0.0156\sym{*}  &      0.0213\sym{**} &      0.0205\sym{**} &      0.0273\sym{***}\\
                    &    (0.0123)         &    (0.0111)         &    (0.0114)         &    (0.0118)         &    (0.0109)         &    (0.0121)         &    (0.0070)         &    (0.0080)         &    (0.0085)         &    (0.0084)         &    (0.0079)         \\
\midrule
$R^2$          &      0.3000         &      0.3663         &      0.3838         &      0.3838         &      0.3345         &      0.3276         &      0.2133         &      0.2584         &      0.2636         &      0.2800         &      0.2678         \\
\midrule 
\multicolumn{12}{l}{\textbf{Panel B. Delayed Reaction}: Buy-and-Hold Abnormal Return in Event Time 2 to 75} \\
Long-Term    &      0.0147         &      0.0319\sym{***}&      0.0208\sym{**} &      0.0125         &      0.0109\sym{*}  &      0.0190\sym{**} &      0.0040         &      0.0203\sym{***}&      0.0270\sym{***}&      0.0317\sym{***}&      0.0252\sym{***}\\
                    &    (0.0117)         &    (0.0115)         &    (0.0100)         &    (0.0097)         &    (0.0066)         &    (0.0083)         &    (0.0043)         &    (0.0054)         &    (0.0058)         &    (0.0065)         &    (0.0072)         \\

Constant            &      0.0214         &      0.0252         &      0.0401         &      0.0598         &      0.0641\sym{**} &      0.0352         &      0.0295         &      0.0098         &      0.0137         &     -0.0142         &      0.0593\sym{**} \\
                    &    (0.0353)         &    (0.0420)         &    (0.0411)         &    (0.0430)         &    (0.0303)         &    (0.0328)         &    (0.0219)         &    (0.0258)         &    (0.0233)         &    (0.0241)         &    (0.0272)         \\
\midrule
$R^2$          &      0.3595         &      0.3710         &      0.4041         &      0.4032         &      0.3501         &      0.3701         &      0.2492         &      0.2892         &      0.3046         &      0.3264         &      0.3077         \\
\midrule 
\multicolumn{12}{l}{\textbf{Panel C. Total Reaction}: Buy-and-Hold Abnormal Return in Event Time 0 to 75} \\
Long-Term         &      0.0140         &      0.0294\sym{**} &      0.0198\sym{*}  &      0.0071         &      0.0127\sym{*}  &      0.0253\sym{***}&      0.0066         &      0.0245\sym{***}&      0.0306\sym{***}&      0.0390\sym{***}&      0.0311\sym{***}\\
                    &    (0.0126)         &    (0.0121)         &    (0.0109)         &    (0.0102)         &    (0.0072)         &    (0.0088)         &    (0.0046)         &    (0.0060)         &    (0.0062)         &    (0.0070)         &    (0.0079)         \\
Constant                &      0.0070         &      0.0093         &      0.0378         &      0.0577         &      0.0653\sym{**} &      0.0555         &      0.0357         &      0.0256         &      0.0349         &      0.0064         &      0.0888\sym{***}\\
                    &    (0.0373)         &    (0.0450)         &    (0.0438)         &    (0.0434)         &    (0.0324)         &    (0.0353)         &    (0.0233)         &    (0.0275)         &    (0.0249)         &    (0.0265)         &    (0.0303)         \\
\midrule
$R^2$             &      0.3591         &      0.3650         &      0.3898         &      0.3855         &      0.3565         &      0.3605         &      0.2536         &      0.2950         &      0.3091         &      0.3215         &      0.3082         \\ \midrule 
Observations         &   5,923        &   5,680        &   5,551        &   5,605        &   5,837        &   6,636        &  11,977        &  11,798        &  11,737        &  11,687        &  11,801        \\
\bottomrule
\end{tabulary}

\begin{tablenotes}[flushleft]
  \scriptsize
  \item \textit{Notes}: For each of the five positive and negative surprise categories as well as the no surprise category, $k$, where $k$ is a earning surprise category, we run the following regression: 
\begin{equation*}
BHAR^{w}_{t,k}=\alpha + \beta \times [\text{Long-term} = 1] +  + \gamma^{k} \times X_{it}+\lambda_{t}+\eta_{k}+\epsilon_{it,k},
\end{equation*}

where $BHAR^{w}_{t,k}$ is defined as in Equation~\eqref{eq:bhardef}. $\beta$ is the added response of the securities in the group followed by long-term investors; $X_{t,k}$ is a set of control variables which include the size of the security, the dispersion of analysts forecasts, the number of analysts, the buy-and-hold abnormal returns [-30, -1], valence and standard deviation of valence over [-30, -1], institutional ownership \& turnover, volatility, and abnormal short interest. To address the concern that companies followed by long-term investors may also have unobservable features that differ from companies followed by short-term investors, we control include firm fixed effects. Finally, we also include time fixed effects by year-quarter. The standard errors account for heteroskedasticity as well as correlation of errors across securities making an announcement on the same day by clustering observations by day of announcement and by firm. Table \ref{table:3_stocktwits} reports $\alpha$ and $\beta$. To mitigate the impact of outliers, the dependent variables are winsorized at the 1st and 99th percentiles. \sym{*} \(p<0.10\), \sym{**} \(p<0.05\), \sym{***} \(p<0.01\). 
\end{tablenotes}
\end{threeparttable}
\end{center}
\end{table}
\end{landscape}

%% file: tables/trading_strategy.tex
\begin{landscape}
\begin{table}[htbp]
\scriptsize
\centering
\begin{threeparttable}
\caption{Factor Regressions for Long--Short Drift Portfolios by Horizon}\label{table:factor_horizon}
\def\sym#1{\ifmmode^{#1}\else\(^{#1}\)\fi}
\begin{tabular*}{\hsize}{@{\hskip\tabcolsep\extracolsep\fill}l*{13}{c}}
\toprule 
                    &\multicolumn{1}{c}{(1)}&\multicolumn{1}{c}{(2)}&\multicolumn{1}{c}{(3)}&\multicolumn{1}{c}{(4)}&\multicolumn{1}{c}{(5)}&\multicolumn{1}{c}{(6)}&\multicolumn{1}{c}{(7)}&\multicolumn{1}{c}{(8)}&\multicolumn{1}{c}{(9)}&\multicolumn{1}{c}{(10)}&\multicolumn{1}{c}{(11)}&\multicolumn{1}{c}{(12)}&\multicolumn{1}{c}{(13)}\\
                    &\multicolumn{1}{c}{Overall}&\multicolumn{1}{c}{Top 3}&\multicolumn{1}{c}{Bin 1}&\multicolumn{1}{c}{Bin 2}&\multicolumn{1}{c}{Bin 3}&\multicolumn{1}{c}{Bin 4}&\multicolumn{1}{c}{Bin 5}&\multicolumn{1}{c}{Bin 6}&\multicolumn{1}{c}{Bin 7}&\multicolumn{1}{c}{Bin 8}&\multicolumn{1}{c}{Bin 9}&\multicolumn{1}{c}{Bin 10}&\multicolumn{1}{c}{Bin 11}\\
\midrule 
Mkt-RF              &     -0.1664\sym{***}&     -0.1732\sym{***}&     -0.0399         &      0.0444         &     -0.3360\sym{***}&     -0.3427\sym{**} &     -0.1168         &     -0.1469         &     -0.1222         &     -0.1462         &     -0.1095         &     -0.2937\sym{***}&     -0.2160         \\
                    &    (0.0481)         &    (0.0545)         &    (0.1938)         &    (0.1636)         &    (0.1223)         &    (0.1573)         &    (0.1185)         &    (0.1031)         &    (0.0840)         &    (0.1024)         &    (0.0809)         &    (0.0945)         &    (0.1487)         \\
[1em]
SMB                 &      0.0561         &      0.0465         &     -0.3620         &     -0.0314         &     -0.1746         &     -0.2835         &      0.0635         &      0.1681         &      0.2597\sym{**} &      0.1866         &      0.0473         &     -0.0233         &      0.1597         \\
                    &    (0.0760)         &    (0.0864)         &    (0.2884)         &    (0.3331)         &    (0.2165)         &    (0.2272)         &    (0.2336)         &    (0.2417)         &    (0.1256)         &    (0.1479)         &    (0.1140)         &    (0.1676)         &    (0.2056)         \\
[1em]
HML                 &      0.2481\sym{***}&      0.2116\sym{***}&     -0.3989         &      0.0418         &      0.2995         &      0.0356         &      0.3336\sym{**} &      0.2942\sym{**} &      0.3099\sym{***}&      0.4116\sym{***}&      0.1806         &      0.3877\sym{***}&      0.0550         \\
                    &    (0.0478)         &    (0.0510)         &    (0.3500)         &    (0.2095)         &    (0.2739)         &    (0.2078)         &    (0.1287)         &    (0.1260)         &    (0.0889)         &    (0.1074)         &    (0.1162)         &    (0.1232)         &    (0.1224)         \\
[1em]
RMW                 &      0.3280\sym{***}&      0.3585\sym{**} &      0.3309         &      0.8090\sym{**} &      0.2556         &      0.0965         &      0.3082         &      0.3034         &      0.3016         &      0.4114         &      0.2002\sym{*}  &      0.4296\sym{***}&      0.4479         \\
                    &    (0.1126)         &    (0.1422)         &    (0.2498)         &    (0.3614)         &    (0.3141)         &    (0.2345)         &    (0.2263)         &    (0.2646)         &    (0.1827)         &    (0.2704)         &    (0.1151)         &    (0.1508)         &    (0.3644)         \\
[1em]
CMA                 &     -0.1230         &     -0.1773         &      0.6834         &      0.1818         &     -0.1997         &      0.3052         &     -0.8694\sym{***}&      0.2252         &     -0.0779         &     -0.4546\sym{**} &      0.1345         &     -0.5018\sym{**} &     -0.1275         \\
                    &    (0.1238)         &    (0.1691)         &    (0.5676)         &    (0.4531)         &    (0.3795)         &    (0.2785)         &    (0.2732)         &    (0.3205)         &    (0.1746)         &    (0.2292)         &    (0.2627)         &    (0.2064)         &    (0.3123)         \\
[1em]
$\alpha$            &      0.2948\sym{**} &      0.3810\sym{***}&     -0.1039         &     -0.3288         &      0.3422         &      0.1595         &      0.2142         &      0.8011\sym{*}  &     -0.0312         &      0.0735         &      0.4784\sym{**} &      0.3636         &      0.2998         \\
                    &    (0.1223)         &    (0.1363)         &    (0.7572)         &    (0.5837)         &    (0.4924)         &    (0.3070)         &    (0.3148)         &    (0.4579)         &    (0.1971)         &    (0.3208)         &    (0.2163)         &    (0.3424)         &    (0.3587)         \\
\midrule 
Observations        &    136       &    136       &    136       &    136       &    136       &    136       &    136       &    136       &    136       &    136       &    136       &    136       &    136       \\
\bottomrule 
\end{tabular*}
\begin{tablenotes}
\item \textit{Notes}: This table reports Fama--French Five-Factor regressions of monthly excess returns for zero-cost horizon-channel portfolios. These portfolios are constructed by going long stocks predominantly followed by Long-Term investors and shorting stocks followed by Short-Term investors, within each earnings surprise bin. Columns report results for the overall Announcement-Weighted Portfolio (Column (1)), the Top Three Bin Announcement-Weighted Portfolio (Column (2)), and the individual bin difference portfolios (Columns (3-13)). All regressions include the market factor (Mkt--RF), SMB, HML, RMW, and CMA. Standard errors (in parentheses) are heteroskedasticity- and autocorrelation-robust using Newey--West with 6 lags. Stars denote significance levels: \sym{*} \(p<0.10\), \sym{**} \(p<0.05\), \sym{***} \(p<0.01\).
\end{tablenotes}
\end{threeparttable}
\end{table}
\end{landscape}

%% file: sections/discussion.tex
\section{Discussion}

Building on our main findings that long- and short-horizon retail investors respond differently to earnings announcements, this section examines how those differences manifest in pre‐announcement behavior, sentiment dynamics, and communication patterns. Rather than treating horizon groups as mechanically distinct, we explore whether they differ in the information they emphasize, the mood they express around announcements, and the way these factors translate into return patterns. By analyzing pre‐EA momentum, testing an optimism–bias mechanism, and studying both hourly sentiment and message content, we shed light on why horizon-specific return responses arise and what they reveal about investors’ expectations and information processing.

\subsection{Momentum}

One potential non-behavioral explanation for our findings is simple momentum: if long-horizon stocks earn higher returns in the month leading up to earnings announcements, their stronger post-announcement performance could mechanically reflect continuation rather than differences in information processing. To assess this possibility, Figure~\ref{fig:figure2ab} plots cumulative buy-and-hold abnormal returns from trading day \(-30\) to \(-1\). The two series move closely together, with the long-horizon group only marginally ahead throughout the pre-event window. Because our regressions already control for past abnormal returns \([-30,-1]\), this slight lead is absorbed in the specification and does not drive our post-announcement results. The return differentials we document therefore arise after the announcement and reflect how the two investor groups process new information rather than residual momentum. This finding shifts the focus to behavioral mechanisms, which we examine next.

\input{figures/figure_3}

\subsection{Mechanism}

To evaluate whether an optimism--bias channel drives the cross-sectional differences in post-announcement returns, we augment the regression with an indicator for above-median pre-announcement valence (computed over days $[-90,-1]$) and its interaction with the long-horizon indicator (Table~\ref{table:2_stocktwits_mechanism}). Although our baseline specification already includes $[-30,-1]$ sentiment and its dispersion, this design isolates whether broader pre-EA enthusiasm shapes the price response and whether that effect differs by investor horizon. Under an optimism-bias mechanism, elevated pre-EA sentiment should (i) dampen the day~0--1 reaction because prices are partially bid up beforehand, and (ii) predict more negative post-EA drift as expectations revert—effects that should be strongest for short-horizon stocks and attenuated for long-horizon stocks. We formalize these predictions in Section~\ref{sec:theory}.

Across all specifications, the valence indicator loads negatively and significantly. In the immediate window (Panel~A), above-median sentiment is associated with $-0.31$ to $-0.73$~pp lower $BHAR$, consistent with partial pre-announcement run-up. The effect is much larger in the drift window (Panel~B): coefficients range from $-1.89$ to $-4.90$~pp ($p<0.01$ in all but one case), indicating that high pre-EA enthusiasm is followed by substantially weaker performance over days~2--75. The pattern persists in the full 0--75 window (Panel~C), where valence coefficients lie between $-2.65$ and $-5.42$~pp.

The long-term$\times$valence interaction is uniformly positive and precisely estimated in the delayed- and total-response regressions, between $+1.29$ and $+3.25$~pp in Panel~B and between $+1.06$ and $+3.25$~pp in Panel~C. These magnitudes offset a sizable share of the negative valence effect, implying that long-horizon investor presence meaningfully dampens the reversal that follows elevated sentiment. In the day~0--1 window (Panel~A), the interaction is smaller but still positive and significant in most specifications, offsetting $0.24$--$0.54$~pp of the immediate overreaction.

High pre-EA sentiment reliably predicts subsequent giveback, especially in stocks followed predominantly by short-horizon investors, while long-horizon presence attenuates that reversal. The cross-sectional pattern fits an optimism-bias (over-extrapolation) mechanism rather than a pure risk-based or mechanical momentum explanation.

\input{tables/table_5}

\subsection{Mood Around Announcements}

We next examine how investor sentiment evolves around earnings announcements. Specifically, we track hourly sentiment (extracted via \ \cite{vamossy2023emtract}) in the 24 hours before and after each release. To prevent firms with exceptionally heavy message traffic from dominating the series, we first average posts at the ticker--hour level and then take the cross-sectional mean.\footnote{Averaging at the ticker--hour level first prevents firms with very high post counts from dominating the sample.} Figure~\ref{fig:sentiment_ann} plots the resulting sentiment paths by news type: panel (a) negative, panel (b) neutral, and panel (c) positive.

Two patterns stand out. First, stocks followed predominantly by short-horizon investors display higher pre-announcement sentiment, consistent with the optimism-bias mechanism tested above. Second, sentiment for these stocks reacts more sharply to the news itself: it rises more after good news and falls more after bad news than sentiment for stocks followed by long-horizon investors. By contrast, long-horizon sentiment shows only muted movements and reverts quickly.

These mood dynamics mirror our return results. Short-horizon investors exhibit more volatile, expectation-driven swings in sentiment, while long-horizon investors maintain a steadier outlook anchored in fundamentals. The evidence underscores how horizon-specific expectations shape both the tone of investor discourse and the corresponding price response around earnings announcements.

\subsection{Information Content}

Finally, to understand the distinct information sets used by these groups, we examine posts made from day $-90$ to day $-1$ before each earnings announcement. Word clouds built from term frequencies provide a first look. Panel (a) of Figure~\ref{fig:user_characteristics} highlights words such as \textit{financial}, \textit{earnings}, \textit{EPS}, \textit{rating}, and \textit{report} for the long-term group, pointing to a focus on fundamentals. Panel (b) for the short-term group features \textit{support}, \textit{volume}, \textit{break}, \textit{target}, and \textit{stop}, language typical of technical trading.

We quantify these lexical differences using a BERT-based sentence transformer model (\textit{bert-base-nli-mean-tokens}). From each group, we randomly draw 20,000 posts and split them into two subsets of 10,000 messages to compute pairwise cosine similarities. Panel (c) of Figure~\ref{fig:user_characteristics} plots the similarity distributions for within-group and between-group comparisons. Two-sample Kolmogorov–Smirnov tests reject equality of the distributions at the 1 percent level: within-group similarities are significantly higher than between-group similarities. Thus, messages by long-term investors are cohesive among themselves, as are those by short-term investors, but the two groups rely on noticeably different distinct vocabularies. This linguistic divergence supports our central claim that short-horizon investors operate on technical signals and sentiment, while long-horizon investors process fundamental information.

\input{figures/figure_6}
\input{figures/figure_5}

The results in this section provide a coherent behavioral explanation for the horizon-based return patterns we document. Pre-announcement returns do not differ meaningfully across groups, ruling out mechanical momentum. Instead, the cross-sectional evidence points to an optimism–bias mechanism: high pre-EA sentiment leads to weaker subsequent performance, and this reversal is materially attenuated in stocks followed by long-horizon investors. Consistent with this mechanism, short-horizon investors display more elevated and more reactive sentiment around announcements, and their discussions emphasize technical trading cues rather than fundamentals. Long-horizon investors, by contrast, communicate in a more fundamentals-oriented and semantically cohesive manner and exhibit steadier sentiment dynamics. 

%% file: figures/figure_3.tex
\begin{figure*}[t!]
    \centering
    \includegraphics[scale=0.32]{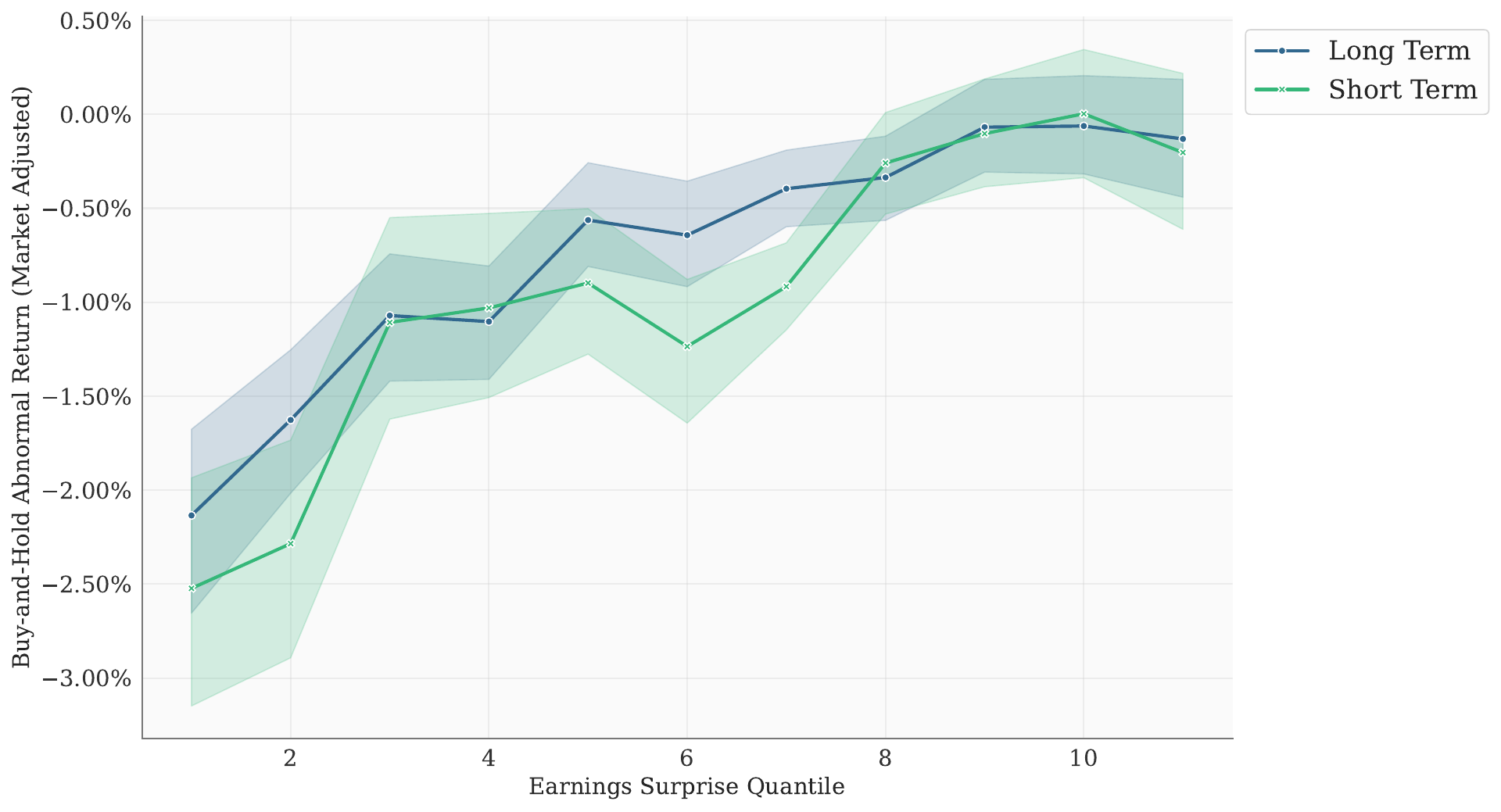}
    \caption*{(a) Response to Earnings Surprise from -30 to -1}
    \label{fig:dellavigna_1c}

    \caption{Pre-announcement behavior before earnings surprises}\label{fig:figure2ab}
    \captionsetup{justification=justified, font=scriptsize}
    \caption*{Notes: Figure \ref{fig:figure2ab} illustrates the mean buy-and-hold abnormal return reactions. From January 2010 until June 2021, stocks in CRSP are matched with quarterly earnings declarations in I/B/E/S. In the timeline of the event, the day of the announcement is designated as day 0. The buy-and-hold abnormal return for each stock is the raw buy-and-hold return adjusted based on the estimated beta from the market model. The earnings surprise for an announcement is defined as the difference between the actual earnings for the quarter, as recorded by I/B/E/S, and the median analyst forecast featured in the I/B/E/S detail file within the 30 days preceding the quarterly earnings announcement. This is then scaled by the stock price 5 trading days before the announcement. Earnings announcements are divided into 11 groups: Quintiles 1 to 5 consist of five quintiles of negative earnings surprises, while quintiles 7 to 11 comprise five quintiles of positive earnings surprises. Quintile 6 includes all announcements where the earnings surprise is zero. The breakpoints are determined quarterly.}
\end{figure*}

%% file: tables/table_5.tex
\begin{landscape}
\begin{table}[htbp]
\scriptsize
\centering
\begin{threeparttable}
\caption{Response to Earnings Surprise: Mechanism}\label{table:2_stocktwits_mechanism}
\def\sym#1{\ifmmode^{#1}\else\(^{#1}\)\fi}
\begin{tabular*}{\hsize}{@{\hskip\tabcolsep\extracolsep\fill}l*{9}{c}}
\toprule 
                    &\multicolumn{1}{c}{(1)}&\multicolumn{1}{c}{(2)}&\multicolumn{1}{c}{(3)}&\multicolumn{1}{c}{(4)}&\multicolumn{1}{c}{(5)}&\multicolumn{1}{c}{(6)}&\multicolumn{1}{c}{(7)}&\multicolumn{1}{c}{(8)}&\multicolumn{1}{c}{(9)}\\
\midrule 
\multicolumn{8}{l}{\textbf{Panel A. Immediate Reaction}: Buy-and-Hold Abnormal Return in Event Time 0 to 1} \\
\midrule 
Long-Term     &      0.0028\sym{***}&      0.0001         &     -0.0001         &      0.0015         &     -0.0014         &     -0.0032         &      0.0032\sym{***}&      0.0008         &      0.0017         \\
                    &    (0.0007)         &    (0.0008)         &    (0.0012)         &    (0.0014)         &    (0.0015)         &    (0.0021)         &    (0.0009)         &    (0.0010)         &    (0.0015)         \\
Valence &     -0.0047\sym{***}&     -0.0036\sym{***}&     -0.0044\sym{***}&     -0.0057\sym{***}&     -0.0064\sym{***}&     -0.0073\sym{***}&     -0.0036\sym{***}&     -0.0031\sym{***}&     -0.0044\sym{**} \\
                    &    (0.0008)         &    (0.0008)         &    (0.0013)         &    (0.0014)         &    (0.0015)         &    (0.0023)         &    (0.0009)         &    (0.0010)         &    (0.0018)         \\
Long-Term $\times$ Valence  0.0030\sym{***}&      0.0034\sym{***}&      0.0030\sym{**} &      0.0039\sym{**} &      0.0049\sym{***}&      0.0054\sym{**} &      0.0024\sym{*}  &      0.0034\sym{***}&      0.0034\sym{*}  \\
                    &    (0.0010)         &    (0.0010)         &    (0.0014)         &    (0.0018)         &    (0.0019)         &    (0.0026)         &    (0.0012)         &    (0.0013)         &    (0.0020)         \\
\midrule 
$R^2$                 &      0.0856         &      0.1524         &      0.2080         &      0.0090         &      0.2125         &      0.2910         &      0.0303         &      0.1465         &      0.2184         \\
\midrule 
\multicolumn{8}{l}{\textbf{Panel B. Delayed Reaction}: Buy-and-Hold Abnormal Return in Event Time 2 to 75} \\
\midrule 
Long-Term                 &      0.0062\sym{**} &      0.0083\sym{***}&      0.0100\sym{***}&      0.0030         &      0.0082\sym{*}  &      0.0101         &      0.0079\sym{***}&      0.0074\sym{**} &      0.0102\sym{**} \\
                    &    (0.0026)         &    (0.0025)         &    (0.0036)         &    (0.0041)         &    (0.0047)         &    (0.0068)         &    (0.0030)         &    (0.0033)         &    (0.0048)         \\
Valence     &     -0.0444\sym{***}&     -0.0347\sym{***}&     -0.0275\sym{***}&     -0.0490\sym{***}&     -0.0284\sym{***}&     -0.0189\sym{**} &     -0.0419\sym{***}&     -0.0388\sym{***}&     -0.0336\sym{***}\\
                    &    (0.0024)         &    (0.0024)         &    (0.0041)         &    (0.0046)         &    (0.0048)         &    (0.0075)         &    (0.0028)         &    (0.0030)         &    (0.0054)         \\
Long-Term $\times$ Valence&      0.0236\sym{***}&      0.0187\sym{***}&      0.0078\sym{*}  &      0.0293\sym{***}&      0.0151\sym{**} &      0.0039         &      0.0207\sym{***}&      0.0219\sym{***}&      0.0129\sym{**} \\
                    &    (0.0032)         &    (0.0031)         &    (0.0047)         &    (0.0059)         &    (0.0060)         &    (0.0086)         &    (0.0038)         &    (0.0039)         &    (0.0061)         \\
\midrule 
$R^2$                 &      0.0094         &      0.1558         &      0.1801         &      0.0086         &      0.2423         &      0.2917         &      0.0097         &      0.1762         &      0.2229         \\
\midrule 
\multicolumn{8}{l}{\textbf{Panel C. Total Reaction}: Buy-and-Hold Abnormal Return in Event Time 0 to 75} \\
\midrule 
Long-Term       &      0.0097\sym{***}&      0.0088\sym{***}&      0.0104\sym{***}&      0.0049         &      0.0062         &      0.0066         &      0.0120\sym{***}&      0.0094\sym{***}&      0.0129\sym{**} \\
                    &    (0.0027)         &    (0.0027)         &    (0.0038)         &    (0.0043)         &    (0.0050)         &    (0.0074)         &    (0.0033)         &    (0.0035)         &    (0.0051)         \\
Valence        &     -0.0490\sym{***}&     -0.0381\sym{***}&     -0.0318\sym{***}&     -0.0542\sym{***}&     -0.0345\sym{***}&     -0.0265\sym{***}&     -0.0454\sym{***}&     -0.0416\sym{***}&     -0.0376\sym{***}\\
                    &    (0.0026)         &    (0.0025)         &    (0.0043)         &    (0.0047)         &    (0.0050)         &    (0.0080)         &    (0.0031)         &    (0.0031)         &    (0.0057)         \\
Long-Term $\times$ Valence &     0.0263\sym{***}&      0.0221\sym{***}&      0.0106\sym{**} &      0.0325\sym{***}&      0.0199\sym{***}&      0.0095         &      0.0228\sym{***}&      0.0250\sym{***}&      0.0155\sym{**} \\
                    &    (0.0034)         &    (0.0033)         &    (0.0049)         &    (0.0062)         &    (0.0063)         &    (0.0092)         &    (0.0041)         &    (0.0042)         &    (0.0065)         \\
\midrule 
$R^2$                 &      0.0204         &      0.1626         &      0.1883         &      0.0095         &      0.2352         &      0.2908         &      0.0158         &      0.1804         &      0.2282         \\
\midrule 
Year-Quarter Fixed Effects&                     &           X         &           X         &                     &           X         &           X         &                     &           X         &           X         \\
Firm Fixed Effects  &                     &           X         &           X         &                     &           X         &           X         &                     &           X         &           X         \\
Controls            &                     &           X         &           X         &                     &           X         &           X         &                     &           X         &           X         \\
Observations       & 104,919        & 104,703        &  75,080        &  33,722        &  33,155        &  24,820        &  63,395        &  62,977        &  43,751        \\
\bottomrule
\end{tabular*}
\begin{tablenotes}
\item \textit{Notes}: Table \ref{table:2_stocktwits_mechanism} reports the coefficients of a regression of the immediate response of securities to earning surprises defined by the following regression:
\begin{equation*}
\mathrm{BHAR}^{w}_{t,k}
=\alpha
+ \beta\,\mathbb{I}_{t,k,\text{Long-term}=1}
+ \theta\,\mathbb{I}_{t,k,\text{Valence}>=0.5}
+ \phi\,\Big(\mathbb{I}_{t,k,\text{Long-term}=1}\times \mathbb{I}_{t,k,\text{Valence}>=0.5}\Big)
+ \sum_{j=2}^{11}\delta_j\,\mathbb{I}_{t,k,\text{Earnings Quantile}=j}
+ \gamma^{k} X_{t,k}
+ \lambda_t + \eta_k + \epsilon_{t,k}.
\label{E4}
\end{equation*}
where $BHAR^{w}_{t,k}$ is defined as in Equation~\eqref{eq:bhardef}. \(\beta\) is the added response of the securities in the group followed by long-term investors; \(\theta\) is the added response to above-median pre-announcement sentiment (\(\mathbb{I}_{\text{Valence}}=1\)) for the baseline (short-horizon) group; and \(\phi\) is the additional differential response for the long-term group, that is, how the sentiment effect changes when \(\mathbb{I}_{\text{Long-term}}=1\). $\delta$ capture the relation between the earning surprise and the immediate response;  $X_{t,k}$ is a set of control variables which include the size of the security, the dispersion of analysts forecasts, the number of analysts, the buy-and-hold abnormal returns [-30, -1], valence and standard deviation of valence over [-30, -1], institutional ownership \& turnover, volatility, and abnormal short interest. To address the concern that companies followed by long-term investors may also have unobservable features that differ from companies followed by short-term investors, we control for firm fixed effects in most regression specifications. Finally, we also include time fixed effects by year-quarter. The standard errors account for heteroskedasticity as well as correlation of errors across securities making an announcement on the same day by clustering observations by day of announcement and by firm. Table \ref{table:2_stocktwits_mechanism} reports $\beta$, $\theta$ and $\phi$. Columns (1), (4) and (7) have no controls or fixed effects; Columns (2), (5), and (8) have the full set of control as defined in $X_{it}$ along with firm and time fixed effects; Columns (3), (6), and (9) carries out matching on the control variables, followed by a regression with matching weights and time and firm fixed effects. Columns (4-6) are focusing on negative surprises, while Columns (7-9) are focusing on positive surprises, with and without control. To mitigate the impact of outliers, the dependent variables are winsorized at the 1st and 99th percentiles. \sym{*} \(p<0.10\), \sym{**} \(p<0.05\), \sym{***} \(p<0.01\). 
\end{tablenotes}
\end{threeparttable}
\end{table}
\end{landscape}

%% file: figures/figure_6.tex
\begin{figure*}[t!]
    \centering
    \begin{minipage}[b]{.9\linewidth}
        \centering
        \includegraphics[scale=0.425]{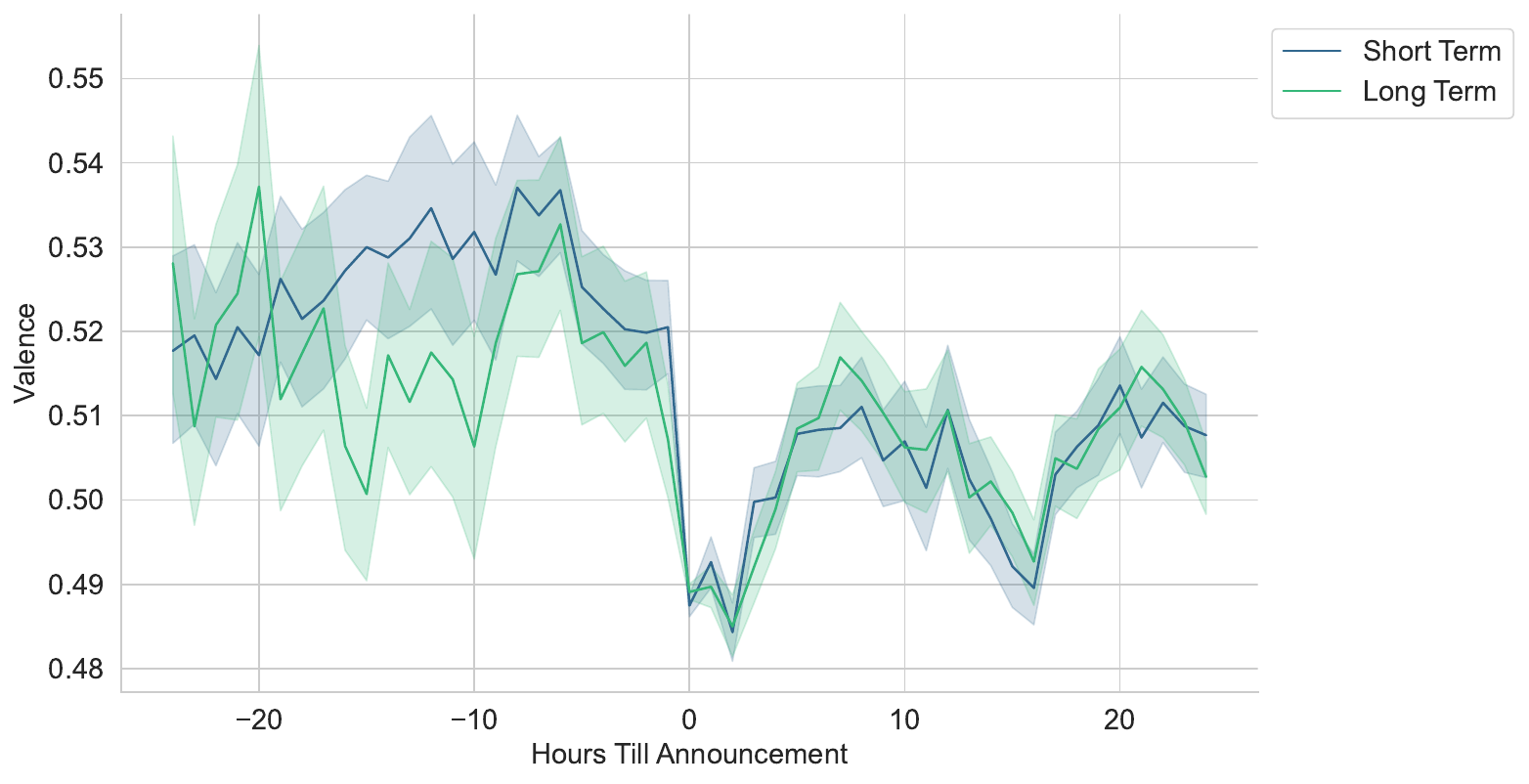}
        \caption*{(a) Negative News}
    \end{minipage}
    \begin{minipage}[b]{.9\linewidth}
        \centering
        \includegraphics[scale=0.425]{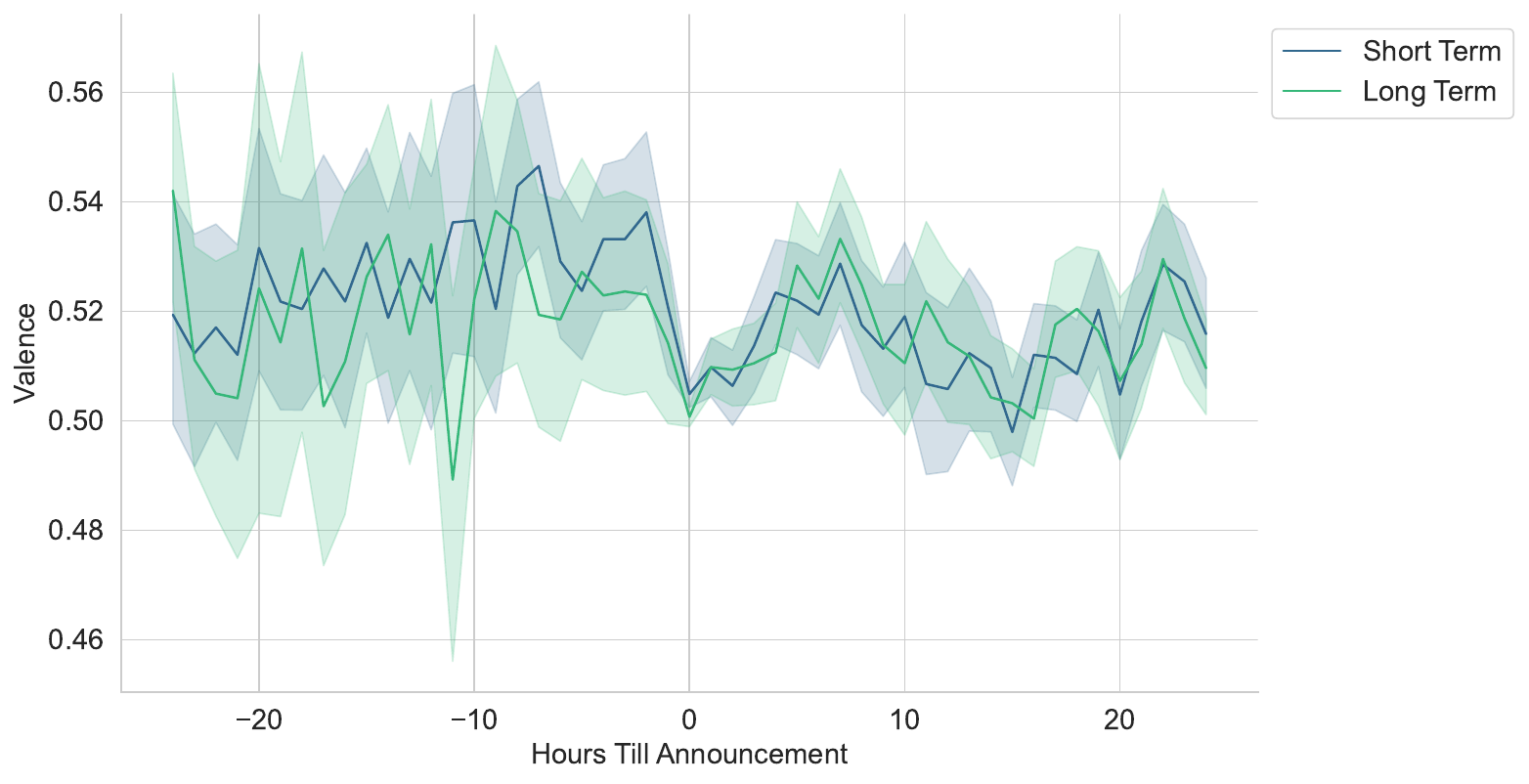}
        \caption*{(b) No News}
    \end{minipage}
    \begin{minipage}[b]{.9\linewidth}
        \centering
        \includegraphics[scale=0.425]{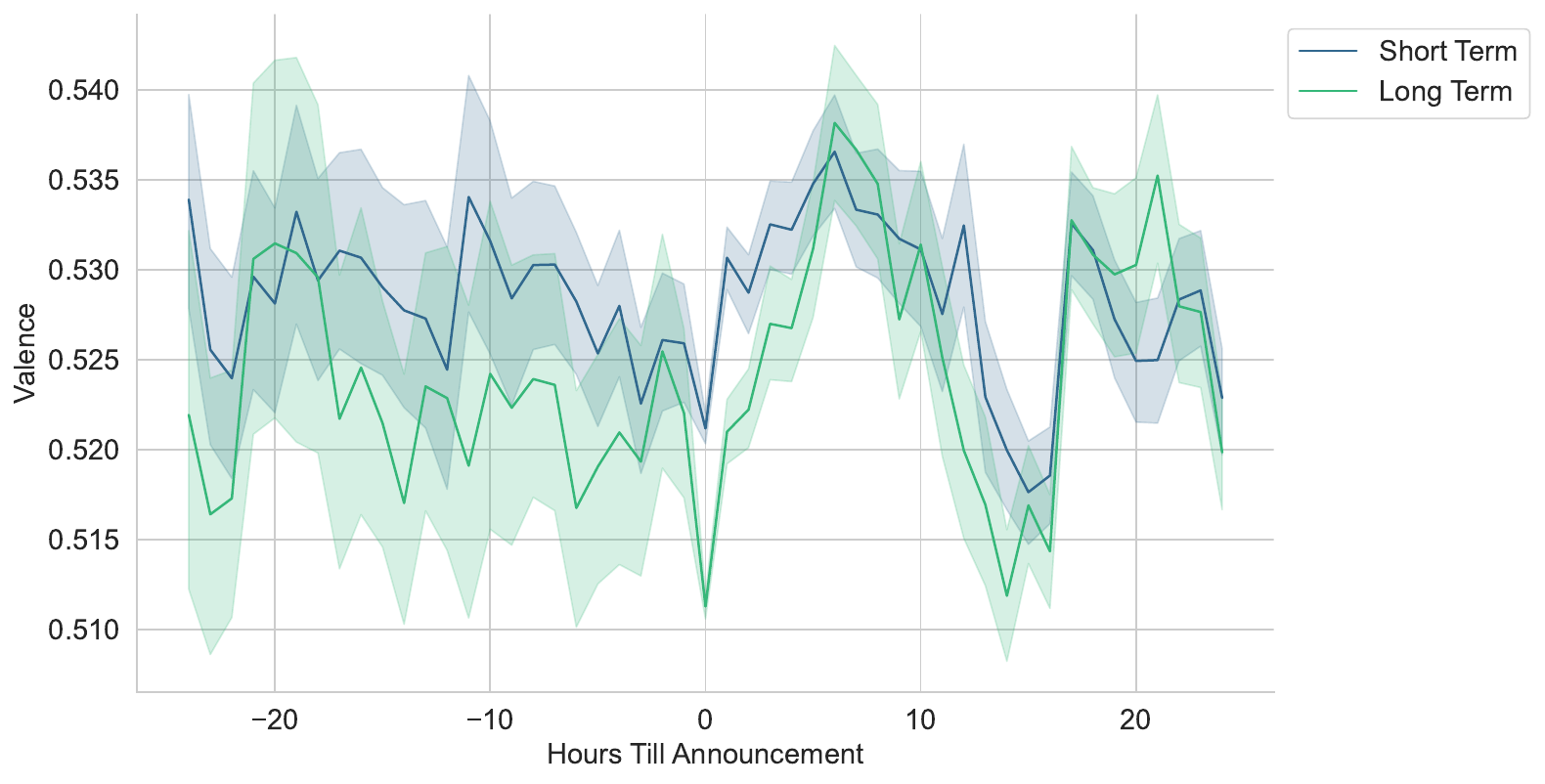}
        \caption*{(c) Positive News}
    \end{minipage}
    \caption{Mood Surrounding Earnings Announcements by Investor Horizon}\label{fig:sentiment_ann}
    \captionsetup{justification=justified, font=scriptsize}
    \caption*{Notes: Hourly average valence for stocks associated with short-term and long-term investors, 24 hours before and after earnings announcements.}
\end{figure*}

%% file: figures/figure_5.tex
\begin{figure*}[ht!]
    \centering
    
    \begin{minipage}[b]{0.5\linewidth}
        \centering
        \includegraphics[scale=0.5]{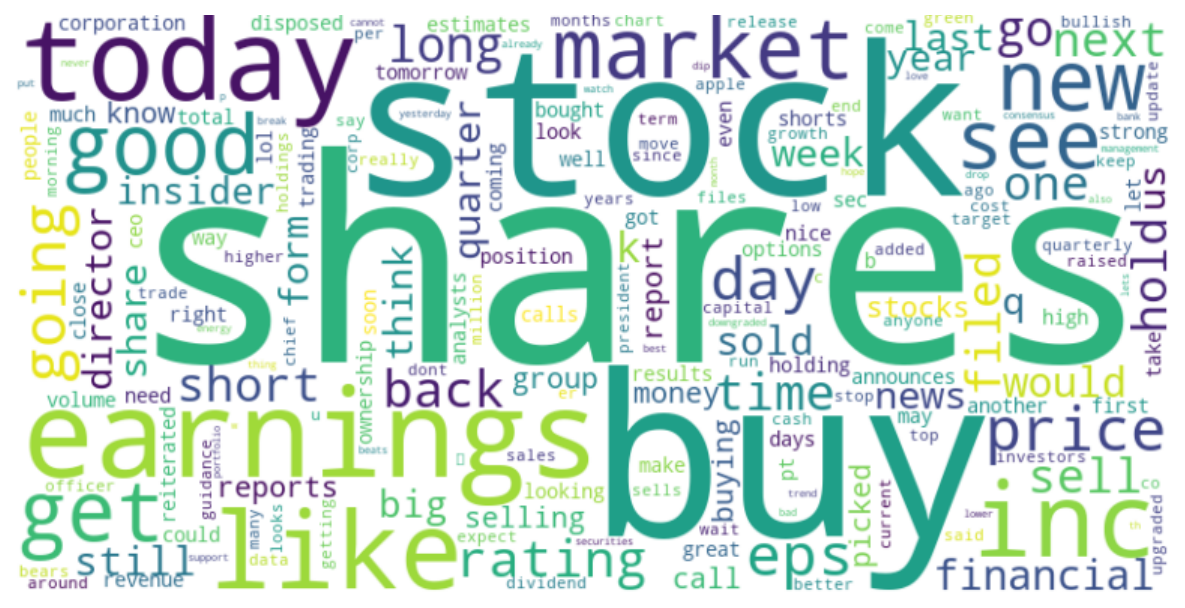}
        \caption*{(a) ``Long-Term" Users: Word Frequencies}
    \end{minipage}%
    \hfill 
    \begin{minipage}[b]{0.5\linewidth}
        \centering
        \includegraphics[scale=0.5]{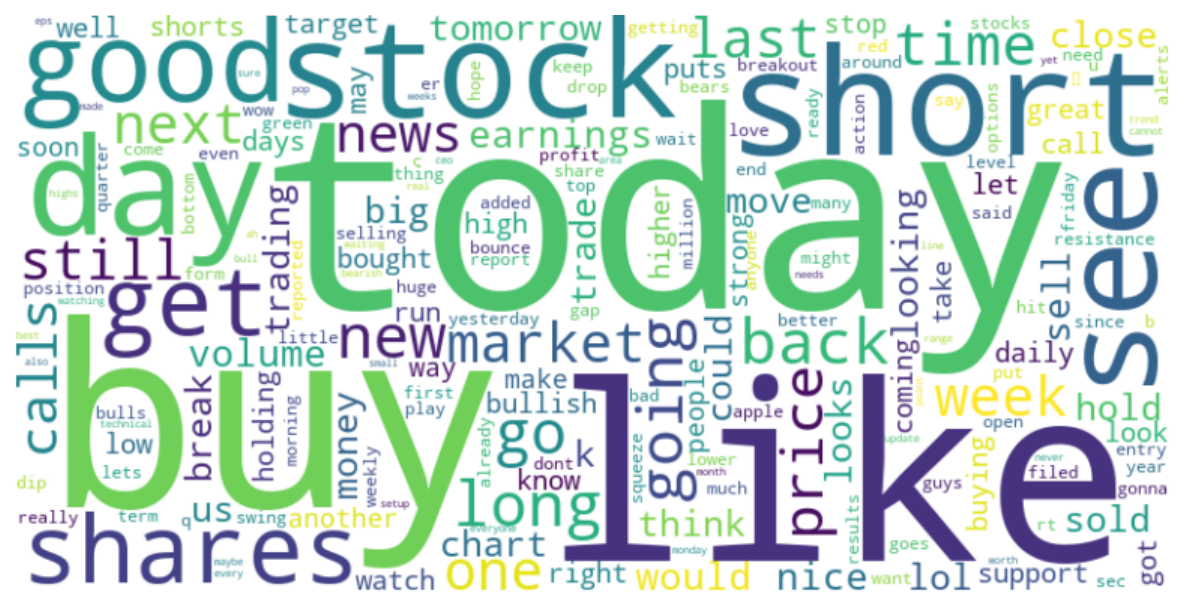}
        \caption*{(b) ``Short-Term" Users: Word Frequencies}
    \end{minipage}

    \vspace{0.5cm} 

    \begin{minipage}[b]{\linewidth}
        \centering
        \includegraphics[scale=0.4]{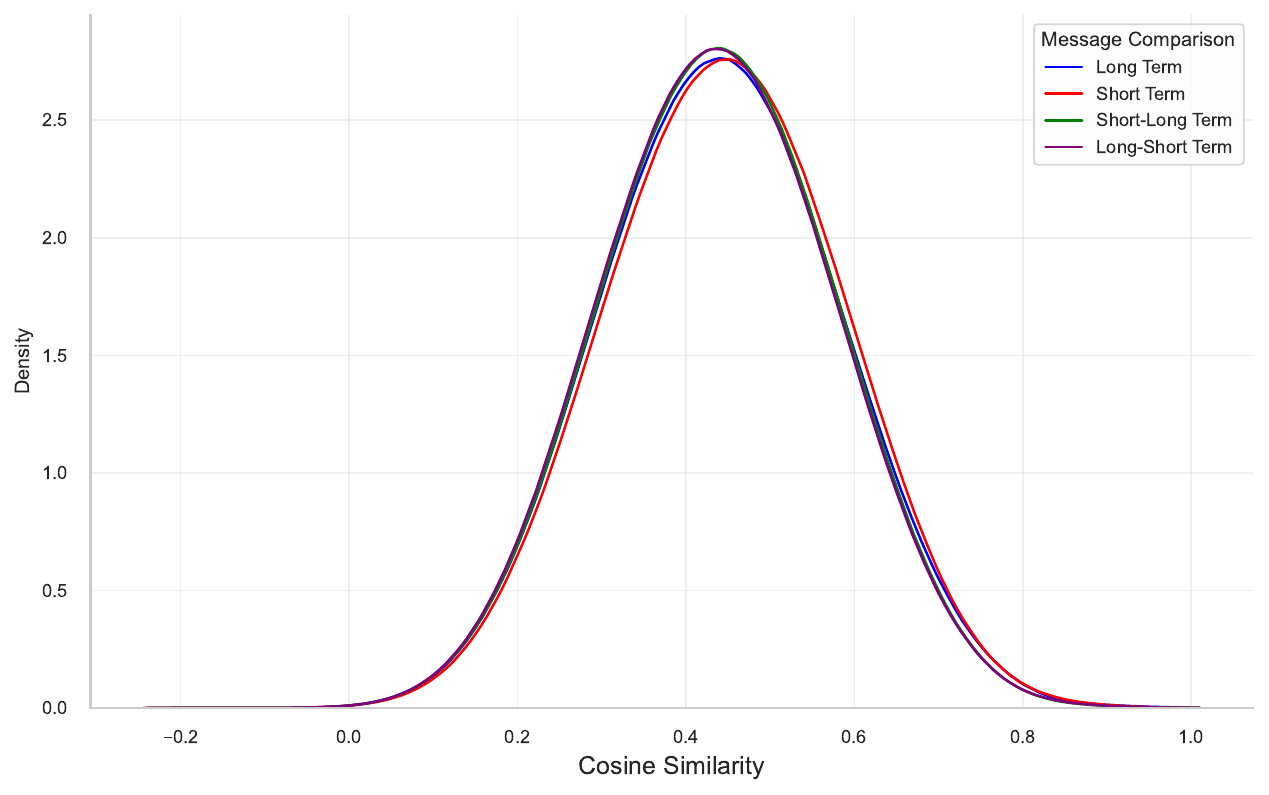}
        \caption*{(c) Post Similarities}
    \end{minipage}

    \caption{Information Content by Investor Horizon}\label{fig:user_characteristics}
    \captionsetup{justification=justified, font=scriptsize}
    
    \caption*{Notes: Panels (a) and (b) display word frequencies from posts authored by long-term versus short-term users. Panel (c) shows the distribution of post similarities, illustrating differences in user communication content. 
    \begin{minipage}[t]{\linewidth} 
    The Two-Sample Kolgomorov-Smirnov statistics for testing whether the post similarity distributions are the same are: 0.0173 (Long vs. Short), 0.0096 (Long vs. Short-Long), 0.0246 (Short vs. Short-Long), 0.0140 (Long vs. Long-Short), and 0.0305 (Short vs. Long-Short). Each test rejects the null hypothesis with a $p$-value $<$ 0.01.
    \end{minipage}}
\end{figure*}

%% file: sections/conclusion.tex
\section{Conclusion}

This paper investigates whether the investment horizon of retail investors relates to how stock prices incorporate earnings news. By leveraging self-reported holding periods from StockTwits to classify stocks as predominantly long- or short-horizon, we document a systematic link between investor composition, information processing, and return dynamics around earnings announcements.

Our findings reveal that investor horizon is a significant predictor of post-announcement returns. Relative to stocks dominated by short-horizon traders, those followed by long-horizon investors exhibit a significantly stronger immediate price reaction—particularly to positive news—and a larger, more persistent post-announcement drift. In our fully controlled specifications, long-horizon stocks earn an additional 0.21 percentage points over the first two days and a further 2.08 percentage points over the subsequent quarter. These results challenge the view that high-frequency retail activity necessarily improves price efficiency; instead, we find that short-horizon environments are associated with sluggish information incorporation and subsequent corrections.

We propose a behavioral explanation for these patterns rooted in optimism bias and information sets. Short-horizon investors exhibit elevated pre-announcement sentiment and communicate using technical trading vocabulary (e.g., \textit{break}, \textit{target}, \textit{volume}), behaviors that correlate with pre-event run-ups and subsequent reversals. In contrast, long-horizon investors focus on fundamental concepts (e.g., \textit{EPS}, \textit{financial}, \textit{report}) and display steadier sentiment dynamics, a pattern consistent with dampened mispricing associated with pre-announcement euphoria. The data rule out simple momentum or flow-based explanations, suggesting instead distinct differences in how these groups form expectations and process fundamental news.

The economic magnitude of these differentials is substantial. A simple, monthly rebalanced trading strategy that goes long long-horizon stocks and shorts short-horizon stocks within earnings-surprise bins generates statistically significant risk-adjusted alphas of approximately 30 to 40 basis points per month. The strategy exhibits a defensive profile, loading negatively on market risk while tilting toward value and profitability, yet the alpha remains unexplained by standard risk factors.

Our study is subject to the caveat that identifying investor horizons relies on observational data from a single social media platform, which may proxy for correlated firm characteristics such as attention or growth narratives. However, the robustness of our results to firm fixed effects and extensive controls suggests that the horizon channel is distinct. Ultimately, these findings highlight the heterogeneity within the retail sector: "retail" is not a monolith, and the distinction between short-term speculation and long-term conviction is central to understanding the efficiency of asset prices.

%% file: sections/appendix.tex
\appendix
\addcontentsline{toc}{section}{Appendices}
\section*{Appendix}

\section{Robustness Checks}

\subsection{Three-, Four-, Five-Factor Models}

We repeat the analysis of Table~\ref{table:2_stocktwits} after redefining buy-and-hold abnormal returns with progressively richer risk adjustments: the Fama–French three-factor model (\cite{FamaFrench1993}) in Table~\ref{table:2_stocktwits_3f}, the Carhart four-factor model (\cite{Carhart1997}) in Table~\ref{table:2_stocktwits_4f}, and the Fama–French five-factor model (\cite{FamaFrench2015}) in Table~\ref{table:2_stocktwits_5f}.  The results are little changed.

\subsection{Alternative Event Window}
Table~\ref{table:2_stocktwits_event_windows} recalculates the post-announcement drift using two shorter horizons, \([2,40]\) and \([2,60]\) trading days.  Point estimates and significance levels remain in line with the baseline \([2,75]\) specification, confirming that our results are robust to the choice of event window.

\subsection{Matching Sensitivity}

To assess the robustness of our results to matching choices, Table~\ref{table:2_stocktwits_sensitivity} reports estimates using nearest-neighbor matching with varying numbers of neighbors (5 or 10) and calipers (0.001, 0.01, 0.1), as well as alternative estimators (IPW-ATT and AIPW). The main specification in other tables employs neighbor(1) with caliper(0.01).

Comparing the main winsorized specification in Column 3 of Table~\ref{table:2_stocktwits_w} to the analogous sensitivity specification with neighbor(5), caliper(0.01) in Column 2 of Table~\ref{table:2_stocktwits_sensitivity} shows that the long-term coefficient changes only slightly for the immediate reaction (0.0034 vs. 0.0032, around 6\% lower) but is around 15\% lower for the delayed (0.0314 vs. 0.0268) and total (0.0360 vs. 0.0305) reaction windows. While the point estimates are somewhat sensitive to matching parameters and outlier treatment, all effects remain positive, statistically significant, and similar in magnitude, leaving the qualitative conclusions unchanged.

\subsection{Winsorization}

Tables~\ref{table:2_stocktwits_w} removes the winsorization to the dependent variables, and repeats Table \ref{table:2_stocktwits}. Coefficient magnitudes and significance levels remain virtually unchanged, confirming that our results are not driven by extreme observations.

\input{tables/table_6}
\input{tables/table_7}
\input{tables/table_8}
\input{tables/table_9}
\input{tables/table_10}
\input{tables/table_11}

\section{Model Simulation}

\subsection*{Setup}
We simulate a market with \( N = 100,000 \) investors, each linked to a unique stock to approximate aggregate behavior. Parameters:

\begin{itemize}
    \item \textbf{Optimism Bias} (\( b \)): Reflects the optimism of short-term investors about earnings announcements, set to \( b = 0.5 \).
    \item \textbf{Risk Aversion} (\( \gamma \)): Indicates the degree of risk aversion in investors' utility functions, set to \( \gamma = 2.0 \).
    \item \textbf{Fundamental Values} (\( \theta \)) and \textbf{Earnings Noise} (\( \eta \)): Drawn from \( N(0, \sigma_\theta^2) \) with \( \sigma_\theta = 0.25 \), and \( N(0, \sigma_\eta^2) \) with \( \sigma_\eta = 0.5 \).
    \item \textbf{Earnings Announcements} (\( e = \theta + \eta \)): Combining fundamental value and earnings noise. \( \sigma_e = \sqrt{\sigma_\theta^2 + \sigma_\eta^2} \approx 0.559 \): Standard deviation of earnings announcement.
    \item \textbf{Signals} for short-term (\( z_S = e + \epsilon_S \)) and long-term investors (\( z_L = \theta + \epsilon_L \)), where \( \epsilon_S \sim N(0, \sigma_S^2) \) and \( \epsilon_L \sim N(0, \sigma_L^2) \), with \( \sigma_S = 1.0 \) and \( \sigma_L = 1.0 \).
\end{itemize}

\subsection*{Simulation Procedure}

\begin{algorithm}[H]
    \caption{Model Simulation Algorithm}
    \label{alg:model_simulation_clean}
    
    \begin{algorithmic}[1] 
        
        \STATE \textbf{Data:} Set parameters $b$, $\gamma$, $\sigma_\theta$, $\sigma_\eta$, $\sigma_S$, and $\sigma_L$.
        \STATE \textbf{Result:} Abnormal returns for short-term and long-term investors.

        \STATE \textbf{Generate Random Variables:}  Simulate $\theta \sim N(0, \sigma_\theta^2)$ and $\eta \sim N(0, \sigma_\eta^2)$. \\
        Calculate $e = \theta + \eta$. \\
        Generate signals $z_S = e + \epsilon_S$ and $z_L = \theta + \epsilon_L$, where $\epsilon_S \sim N(0, \sigma_S^2)$ and $\epsilon_L \sim N(0, \sigma_L^2)$.
        
        \STATE \textbf{Update Expectations} \\
        Short-term investors: $ E_e^S = b + \left( \frac{\sigma_e^2}{\sigma_e^2 + \sigma_S^2} \right)(z_S - b)$ \\ 
        Long-term investors:  $ E_\theta^L = \frac{\sigma_\theta^2}{\sigma_\theta^2 + \sigma_L^2} \cdot z_L$ \\
        
        \STATE \textbf{Calculate Initial Demand (Assume $p_1 = 0$)} \\
        Short-term demand: $\displaystyle k_S = \frac{E_e^S}{\gamma \sigma_S^2}$. \\
        Long-term demand: $\displaystyle k_L = \frac{E_\theta^L}{\gamma \sigma_L^2}$.

        \STATE \textbf{Update After Earnings Announcement ($t=2$)} \\ 
        Short-term investors update expectations: $ E_\theta^S = \left( \frac{\sigma_\theta^2}{\sigma_\theta^2 + \sigma_\eta^2} \right) e$ \\
        Update short-term holdings:  $ k_S^{\text{new}} = \frac{E_\theta^S - e}{\gamma \sigma_S^2}$ \\ 

        \STATE \textbf{Calculate Abnormal Returns} \\
        Short-term abnormal return: $\text{Abnormal Return}_S = p_2 - E_e^S = e - E_e^S$. \\
        Long-term abnormal return: $\text{Abnormal Return}_L = p_2 - p_1 = e$.
        
    \end{algorithmic}
\end{algorithm}

\section{Comparison of I/B/E/S-Only and Matched Samples}

This appendix compares the full set of earnings announcements available in I/B/E/S with the subset matched to StockTwits horizon classifications. Table~\ref{app_1} reports differences in firm and announcement characteristics across the two samples. Although the matched sample is smaller, the two groups are similar along most observable dimensions. Differences in institutional ownership, analyst coverage, volatility, and year are statistically significant but economically modest. These comparisons indicate that the matched sample remains broadly representative of the I/B/E/S universe, mitigating concerns that our results are driven by selection into the StockTwits-linked subsample.

\input{tables/app_1}

%% file: tables/table_6.tex
\begin{landscape}
\begin{table}[htbp]
\scriptsize
\centering
\begin{threeparttable}
\caption{Response to Earnings Surprise: Three-Factor Model}\label{table:2_stocktwits_3f}
\def\sym#1{\ifmmode^{#1}\else\(^{#1}\)\fi}
\begin{tabular*}{\hsize}{@{\hskip\tabcolsep\extracolsep\fill}l*{9}{c}}
\toprule 
                    &\multicolumn{1}{c}{(1)}&\multicolumn{1}{c}{(2)}&\multicolumn{1}{c}{(3)}&\multicolumn{1}{c}{(4)}&\multicolumn{1}{c}{(5)}&\multicolumn{1}{c}{(6)}&\multicolumn{1}{c}{(7)}&\multicolumn{1}{c}{(8)}&\multicolumn{1}{c}{(9)}\\
\midrule 
\multicolumn{8}{l}{\textbf{Panel A. Immediate Reaction}: Buy-and-Hold Abnormal Return in Event Time 0 to 1} \\
\midrule 
Long-Term   &      0.0052\sym{***}&      0.0021\sym{***}&      0.0019\sym{**} &      0.0045\sym{***}&      0.0015         &     -0.0000         &      0.0051\sym{***}&      0.0027\sym{***}&      0.0037\sym{***}\\
                    &    (0.0005)         &    (0.0006)         &    (0.0009)         &    (0.0010)         &    (0.0012)         &    (0.0017)         &    (0.0006)         &    (0.0008)         &    (0.0011)         \\
& & & & & & & \\[\dimexpr-\normalbaselineskip+3pt]
Constant       &     -0.0034\sym{***}&      0.0109\sym{***}&     -0.0007         &     -0.0299\sym{***}&     -0.0083\sym{**} &     -0.0261\sym{***}&      0.0122\sym{***}&      0.0182\sym{***}&      0.0145\sym{***}\\
                    &    (0.0004)         &    (0.0024)         &    (0.0005)         &    (0.0008)         &    (0.0042)         &    (0.0009)         &    (0.0005)         &    (0.0033)         &    (0.0006)         \\
\midrule 
$R^2$                 &      0.0852         &      0.1525         &      0.2078         &      0.0086         &      0.2123         &      0.2920         &      0.0293         &      0.1471         &      0.2187         \\
\midrule 
\multicolumn{8}{l}{\textbf{Panel B. Delayed Reaction}: Buy-and-Hold Abnormal Return in Event Time 2 to 75} \\
\midrule 
Long-Term       &      0.0242\sym{***}&      0.0199\sym{***}&      0.0160\sym{***}&      0.0266\sym{***}&      0.0180\sym{***}&      0.0151\sym{***}&      0.0219\sym{***}&      0.0183\sym{***}&      0.0185\sym{***}\\
                    &    (0.0019)         &    (0.0021)         &    (0.0028)         &    (0.0032)         &    (0.0039)         &    (0.0052)         &    (0.0021)         &    (0.0025)         &    (0.0037)         \\
& & & & & & & \\[\dimexpr-\normalbaselineskip+3pt]
Constant    &     -0.0445\sym{***}&      0.0414\sym{***}&     -0.0355\sym{***}&     -0.0447\sym{***}&      0.0496\sym{***}&     -0.0337\sym{***}&     -0.0446\sym{***}&      0.0479\sym{***}&     -0.0380\sym{***}\\
                    &    (0.0017)         &    (0.0070)         &    (0.0014)         &    (0.0028)         &    (0.0124)         &    (0.0026)         &    (0.0017)         &    (0.0096)         &    (0.0018)         \\
\midrule 
$R^2$                 &      0.0032         &      0.1265         &      0.1530         &      0.0033         &      0.2153         &      0.2689         &      0.0027         &      0.1466         &      0.2010         \\
\midrule 
\multicolumn{8}{l}{\textbf{Panel C. Total Reaction}: Buy-and-Hold Abnormal Return in Event Time 0 to 75} \\
\midrule 
Long-Term            &      0.0299\sym{***}&      0.0225\sym{***}&      0.0183\sym{***}&      0.0312\sym{***}&      0.0191\sym{***}&      0.0148\sym{***}&      0.0278\sym{***}&      0.0219\sym{***}&      0.0227\sym{***}\\
                    &    (0.0020)         &    (0.0022)         &    (0.0030)         &    (0.0033)         &    (0.0041)         &    (0.0056)         &    (0.0023)         &    (0.0028)         &    (0.0039)         \\
& & & & & & & \\[\dimexpr-\normalbaselineskip+3pt]
Constant            &     -0.0484\sym{***}&      0.0529\sym{***}&     -0.0364\sym{***}&     -0.0757\sym{***}&      0.0416\sym{***}&     -0.0602\sym{***}&     -0.0326\sym{***}&      0.0660\sym{***}&     -0.0230\sym{***}\\
                    &    (0.0018)         &    (0.0076)         &    (0.0015)         &    (0.0029)         &    (0.0132)         &    (0.0028)         &    (0.0019)         &    (0.0105)         &    (0.0019)         \\
\midrule 
$R^2$                 &      0.0116         &      0.1339         &      0.1617         &      0.0043         &      0.2081         &      0.2685         &      0.0064         &      0.1531         &      0.2076         \\
\midrule 
Year-Quarter Fixed Effects&                     &           X         &           X         &                     &           X         &           X         &                     &           X         &           X         \\
Firm Fixed Effects  &                     &           X         &           X         &                     &           X         &           X         &                     &           X         &           X         \\
Controls            &                     &           X         &           X         &                     &           X         &           X         &                     &           X         &           X         \\
Observations        & 104,919         & 104,703         &  75,080         &  33,722         &  33,155         &  24,820         &  63,395         &  62,977         &  43,751         \\

\bottomrule
\end{tabular*}
\begin{tablenotes}
\item \textit{Notes}: Table \ref{table:2_stocktwits_3f} reports the coefficients of a regression of the immediate response of securities to earning surprises defined by the following regression:
\begin{equation*}
BHAR^{w}_{t,k}=\alpha + \beta \times \mathbb{I}_{t,k, \text{Long-term} = 1} + \sum_{j=2}^{11} \Big[\delta_j \times \mathbb{I}_{t,k, \text{Earnings Quantile} = j}\Big] + \gamma^{k} \times X_{t,k}+\lambda_{t} +\eta_{k} +\epsilon_{t,k},  \label{E4}
\end{equation*}
where $BHAR^{w}_{t,k}$ is defined as in Equation~\eqref{eq:bhardef}. $\beta$ is the added response of the securities in the group followed by long-term investors; $\delta$ capture the relation between the earning surprise and the immediate response; $X_{t,k}$ is a set of control variables which include the size of the security, the dispersion of analysts forecasts, the number of analysts, the buy-and-hold abnormal returns [-30, -1], valence and standard deviation of valence over [-30, -1], institutional ownership \& turnover, volatility, and abnormal short interest. To address the concern that companies followed by long-term investors may also have unobservable features that differ from companies followed by short-term investors, we control for firm fixed effects in most regression specifications. Finally, we also include time fixed effects by year-quarter. The standard errors account for heteroskedasticity as well as correlation of errors across securities making an announcement on the same day by clustering observations by day of announcement and by firm. Table \ref{table:2_stocktwits_3f} reports $\alpha$ and $\beta$. Columns (1), (4) and (7) have no controls or fixed effects; Columns (2), (5), and (8) have the full set of control as defined in $X_{t,k}$ along with firm and time fixed effects; Columns (3), (6), and (9) carries out matching on the control variables, followed by a regression with matching weights and time and firm fixed effects. Columns (4-6) are focusing on negative surprises, while Columns (7-9) are focusing on positive surprises, with and without control. To mitigate the impact of outliers, the dependent variables are winsorized at the 1st and 99th percentiles. \sym{*} \(p<0.10\), \sym{**} \(p<0.05\), \sym{***} \(p<0.01\). 
\end{tablenotes}
\end{threeparttable}
\end{table}
\end{landscape}

%% file: tables/table_7.tex
\begin{landscape}
\begin{table}[htbp]
\scriptsize
\centering
\begin{threeparttable}
\caption{Response to Earnings Surprise: Four-Factor Model}\label{table:2_stocktwits_4f}
\def\sym#1{\ifmmode^{#1}\else\(^{#1}\)\fi}
\begin{tabular*}{\hsize}{@{\hskip\tabcolsep\extracolsep\fill}l*{9}{c}}
\toprule 
                    &\multicolumn{1}{c}{(1)}&\multicolumn{1}{c}{(2)}&\multicolumn{1}{c}{(3)}&\multicolumn{1}{c}{(4)}&\multicolumn{1}{c}{(5)}&\multicolumn{1}{c}{(6)}&\multicolumn{1}{c}{(7)}&\multicolumn{1}{c}{(8)}&\multicolumn{1}{c}{(9)}\\
\midrule 
\multicolumn{8}{l}{\textbf{Panel A. Immediate Reaction}: Buy-and-Hold Abnormal Return in Event Time 0 to 1} \\
\midrule 
Long-Term    &      0.0053\sym{***}&      0.0021\sym{***}&      0.0019\sym{**} &      0.0046\sym{***}&      0.0015         &     -0.0000         &      0.0051\sym{***}&      0.0028\sym{***}&      0.0037\sym{***}\\
                    &    (0.0005)         &    (0.0006)         &    (0.0009)         &    (0.0010)         &    (0.0012)         &    (0.0017)         &    (0.0006)         &    (0.0008)         &    (0.0012)         \\
& & & & & & & \\[\dimexpr-\normalbaselineskip+3pt]
Constant      &     -0.0034\sym{***}&      0.0114\sym{***}&     -0.0006         &     -0.0298\sym{***}&     -0.0080\sym{*}  &     -0.0259\sym{***}&      0.0122\sym{***}&      0.0185\sym{***}&      0.0145\sym{***}\\
                    &    (0.0004)         &    (0.0024)         &    (0.0005)         &    (0.0008)         &    (0.0042)         &    (0.0009)         &    (0.0005)         &    (0.0034)         &    (0.0006)         \\
\midrule 
$R^2$                 &      0.0847         &      0.1521         &      0.2071         &      0.0087         &      0.2118         &      0.2915         &      0.0293         &      0.1473         &      0.2190         \\
\midrule 
\multicolumn{8}{l}{\textbf{Panel B. Delayed Reaction}: Buy-and-Hold Abnormal Return in Event Time 2 to 75} \\
\midrule 
Long-Term     &      0.0231\sym{***}&      0.0198\sym{***}&      0.0158\sym{***}&      0.0244\sym{***}&      0.0179\sym{***}&      0.0143\sym{***}&      0.0214\sym{***}&      0.0185\sym{***}&      0.0186\sym{***}\\
                    &    (0.0019)         &    (0.0021)         &    (0.0028)         &    (0.0032)         &    (0.0039)         &    (0.0052)         &    (0.0020)         &    (0.0026)         &    (0.0037)         \\
& & & & & & & \\[\dimexpr-\normalbaselineskip+3pt]
Constant      &     -0.0429\sym{***}&      0.0421\sym{***}&     -0.0344\sym{***}&     -0.0420\sym{***}&      0.0489\sym{***}&     -0.0317\sym{***}&     -0.0434\sym{***}&      0.0463\sym{***}&     -0.0373\sym{***}\\
                    &    (0.0017)         &    (0.0071)         &    (0.0014)         &    (0.0029)         &    (0.0126)         &    (0.0026)         &    (0.0017)         &    (0.0099)         &    (0.0018)         \\
\midrule 
$R^2$                 &      0.0029         &      0.1284         &      0.1545         &      0.0028         &      0.2186         &      0.2703         &      0.0025         &      0.1480         &      0.2027         \\
\midrule 
\multicolumn{8}{l}{\textbf{Panel C. Total Reaction}: Buy-and-Hold Abnormal Return in Event Time 0 to 75} \\
\midrule 
Long-Term                   &      0.0288\sym{***}&      0.0223\sym{***}&      0.0180\sym{***}&      0.0290\sym{***}&      0.0189\sym{***}&      0.0138\sym{**} &      0.0273\sym{***}&      0.0223\sym{***}&      0.0228\sym{***}\\
                    &    (0.0020)         &    (0.0022)         &    (0.0030)         &    (0.0033)         &    (0.0041)         &    (0.0056)         &    (0.0022)         &    (0.0028)         &    (0.0040)         \\
& & & & & & & \\[\dimexpr-\normalbaselineskip+3pt]
Constant          &     -0.0467\sym{***}&      0.0542\sym{***}&     -0.0352\sym{***}&     -0.0729\sym{***}&      0.0417\sym{***}&     -0.0580\sym{***}&     -0.0313\sym{***}&      0.0647\sym{***}&     -0.0223\sym{***}\\
                    &    (0.0018)         &    (0.0077)         &    (0.0015)         &    (0.0029)         &    (0.0134)         &    (0.0028)         &    (0.0020)         &    (0.0108)         &    (0.0019)         \\
\midrule 
$R^2$                 &      0.0111         &      0.1354         &      0.1626         &      0.0036         &      0.2111         &      0.2701         &      0.0064         &      0.1542         &      0.2092         \\
\midrule 
Year-Quarter Fixed Effects&                     &           X         &           X         &                     &           X         &           X         &                     &           X         &           X         \\
Firm Fixed Effects  &                     &           X         &           X         &                     &           X         &           X         &                     &           X         &           X         \\
Controls            &                     &           X         &           X         &                     &           X         &           X         &                     &           X         &           X         \\
Observations        & 104,919         & 104,703         &  75,080         &  33,722         &  33,155         &  24,820         &  63,395         &  62,977         &  43,751         \\

\bottomrule
\end{tabular*}
\begin{tablenotes}
\item \textit{Notes}: Table \ref{table:2_stocktwits_4f} reports the coefficients of a regression of the immediate response of securities to earning surprises defined by the following regression:
\begin{equation*}
BHAR^{w}_{t,k}=\alpha + \beta \times \mathbb{I}_{t,k, \text{Long-term} = 1} + \sum_{j=2}^{11} \Big[\delta_j \times \mathbb{I}_{t,k, \text{Earnings Quantile} = j}\Big] + \gamma^{k} \times X_{t,k}+\lambda_{t} +\eta_{k} +\epsilon_{t,k},  \label{E4}
\end{equation*}
where $BHAR^{w}_{t,k}$ is defined as in Equation~\eqref{eq:bhardef}. $\beta$ is the added response of the securities in the group followed by long-term investors; $\delta$ capture the relation between the earning surprise and the immediate response; $X_{t,k}$ is a set of control variables which include the size of the security, the dispersion of analysts forecasts, the number of analysts, the buy-and-hold abnormal returns [-30, -1], valence and standard deviation of valence over [-30, -1], institutional ownership \& turnover, volatility, and abnormal short interest. To address the concern that companies followed by long-term investors may also have unobservable features that differ from companies followed by short-term investors, we control for firm fixed effects in most regression specifications. Finally, we also include time fixed effects by year-quarter. The standard errors account for heteroskedasticity as well as correlation of errors across securities making an announcement on the same day by clustering observations by day of announcement and by firm. Table \ref{table:2_stocktwits_4f} reports $\alpha$ and $\beta$. Columns (1), (4) and (7) have no controls or fixed effects; Columns (2), (5), and (8) have the full set of control as defined in $X_{t,k}$ along with firm and time fixed effects; Columns (3), (6), and (9) carries out matching on the control variables, followed by a regression with matching weights and time and firm fixed effects. Columns (4-6) are focusing on negative surprises, while Columns (7-9) are focusing on positive surprises, with and without control. To mitigate the impact of outliers, the dependent variables are winsorized at the 1st and 99th percentiles. \sym{*} \(p<0.10\), \sym{**} \(p<0.05\), \sym{***} \(p<0.01\). 
\end{tablenotes}
\end{threeparttable}
\end{table}
\end{landscape}

%% file: tables/table_8.tex
\begin{landscape}
\begin{table}[htbp]
\scriptsize
\centering
\begin{threeparttable}
\caption{Response to Earnings Surprise: Five-Factor Model}\label{table:2_stocktwits_5f}
\def\sym#1{\ifmmode^{#1}\else\(^{#1}\)\fi}
\begin{tabular*}{\hsize}{@{\hskip\tabcolsep\extracolsep\fill}l*{9}{c}}
\toprule 
                    &\multicolumn{1}{c}{(1)}&\multicolumn{1}{c}{(2)}&\multicolumn{1}{c}{(3)}&\multicolumn{1}{c}{(4)}&\multicolumn{1}{c}{(5)}&\multicolumn{1}{c}{(6)}&\multicolumn{1}{c}{(7)}&\multicolumn{1}{c}{(8)}&\multicolumn{1}{c}{(9)}\\
\midrule 
\multicolumn{8}{l}{\textbf{Panel A. Immediate Reaction}: Buy-and-Hold Abnormal Return in Event Time 0 to 1} \\
\midrule 
Long-Term       &      0.0053\sym{***}&      0.0022\sym{***}&      0.0019\sym{**} &      0.0047\sym{***}&      0.0018         &      0.0004         &      0.0051\sym{***}&      0.0027\sym{***}&      0.0036\sym{***}\\
                    &    (0.0005)         &    (0.0006)         &    (0.0009)         &    (0.0010)         &    (0.0012)         &    (0.0017)         &    (0.0006)         &    (0.0008)         &    (0.0011)         \\
& & & & & & & \\[\dimexpr-\normalbaselineskip+3pt]
Constant  &     -0.0034\sym{***}&      0.0106\sym{***}&     -0.0007         &     -0.0300\sym{***}&     -0.0089\sym{**} &     -0.0264\sym{***}&      0.0122\sym{***}&      0.0178\sym{***}&      0.0146\sym{***}\\
                    &    (0.0004)         &    (0.0024)         &    (0.0005)         &    (0.0008)         &    (0.0042)         &    (0.0009)         &    (0.0005)         &    (0.0034)         &    (0.0006)         \\
\midrule 
$R^2$                 &      0.0852         &      0.1522         &      0.2079         &      0.0089         &      0.2124         &      0.2925         &      0.0291         &      0.1464         &      0.2188         \\
\midrule 
\multicolumn{8}{l}{\textbf{Panel B. Delayed Reaction}: Buy-and-Hold Abnormal Return in Event Time 2 to 75} \\
\midrule 
Long-Term     &      0.0243\sym{***}&      0.0198\sym{***}&      0.0158         &      0.0272\sym{***}&      0.0177\sym{***}&      0.0160\sym{***}&      0.0217\sym{***}&      0.0182\sym{***}&      0.0182\sym{***}\\
                    &    (0.0019)         &    (0.0021)         &         (.)         &    (0.0031)         &    (0.0040)         &    (0.0052)         &    (0.0021)         &    (0.0025)         &    (0.0037)         \\
& & & & & & & \\[\dimexpr-\normalbaselineskip+3pt]
Constant     &     -0.0446\sym{***}&      0.0374\sym{***}&     -0.0355         &     -0.0457\sym{***}&      0.0425\sym{***}&     -0.0349\sym{***}&     -0.0443\sym{***}&      0.0431\sym{***}&     -0.0379\sym{***}\\
                    &    (0.0016)         &    (0.0069)         &         (.)         &    (0.0028)         &    (0.0125)         &    (0.0026)         &    (0.0017)         &    (0.0098)         &    (0.0018)         \\
\midrule 
$R^2$                 &      0.0031         &      0.1238         &      0.1522         &      0.0033         &      0.2119         &      0.2693         &      0.0026         &      0.1432         &      0.1993         \\
\midrule 
\multicolumn{8}{l}{\textbf{Panel C. Total Reaction}: Buy-and-Hold Abnormal Return in Event Time 0 to 75} \\
\midrule 
Long-Term             &      0.0301\sym{***}&      0.0223\sym{***}&      0.0181\sym{***}&      0.0318\sym{***}&      0.0189\sym{***}&      0.0157\sym{***}&      0.0276\sym{***}&      0.0218\sym{***}&      0.0222\sym{***}\\
                    &    (0.0020)         &    (0.0022)         &    (0.0030)         &    (0.0032)         &    (0.0042)         &    (0.0057)         &    (0.0023)         &    (0.0028)         &    (0.0040)         \\
& & & & & & & \\[\dimexpr-\normalbaselineskip+3pt]
Constant           &     -0.0485\sym{***}&      0.0488\sym{***}&     -0.0364\sym{***}&     -0.0767\sym{***}&      0.0344\sym{***}&     -0.0615\sym{***}&     -0.0322\sym{***}&      0.0608\sym{***}&     -0.0228\sym{***}\\
                    &    (0.0017)         &    (0.0076)         &    (0.0015)         &    (0.0028)         &    (0.0133)         &    (0.0029)         &    (0.0019)         &    (0.0107)         &    (0.0019)         \\
\midrule 
$R^2$                 &      0.0117         &      0.1316         &      0.1609         &      0.0045         &      0.2050         &      0.2691         &      0.0061         &      0.1503         &      0.2063         \\
\midrule 
Year-Quarter Fixed Effects&                     &           X         &           X         &                     &           X         &           X         &                     &           X         &           X         \\
Firm Fixed Effects  &                     &           X         &           X         &                     &           X         &           X         &                     &           X         &           X         \\
Controls            &                     &           X         &           X         &                     &           X         &           X         &                     &           X         &           X         \\
Observations        & 104,919         & 104,703         &  75,080         &  33,722         &  33,155         &  24,820         &  63,395         &  62,977         &  43,751         \\

\bottomrule
\end{tabular*}
\begin{tablenotes}
\item \textit{Notes}: Table \ref{table:2_stocktwits_5f} reports the coefficients of a regression of the immediate response of securities to earning surprises defined by the following regression:
\begin{equation*}
BHAR^{w}_{t,k}=\alpha + \beta \times \mathbb{I}_{t,k, \text{Long-term} = 1} + \sum_{j=2}^{11} \Big[\delta_j \times \mathbb{I}_{t,k, \text{Earnings Quantile} = j}\Big] + \gamma^{k} \times X_{t,k}+\lambda_{t} +\eta_{k} +\epsilon_{t,k},  \label{E4}
\end{equation*}
where $BHAR^{w}_{t,k}$ is defined as in Equation~\eqref{eq:bhardef}. $\beta$ is the added response of the securities in the group followed by long-term investors; $\delta$ capture the relation between the earning surprise and the immediate response; $X_{t,k}$ is a set of control variables which include the size of the security, the dispersion of analysts forecasts, the number of analysts, the buy-and-hold abnormal returns [-30, -1], valence and standard deviation of valence over [-30, -1], institutional ownership \& turnover, volatility, and abnormal short interest. To address the concern that companies followed by long-term investors may also have unobservable features that differ from companies followed by short-term investors, we control for firm fixed effects in most regression specifications. Finally, we also include time fixed effects by year-quarter. The standard errors account for heteroskedasticity as well as correlation of errors across securities making an announcement on the same day by clustering observations by day of announcement and by firm. Table \ref{table:2_stocktwits_5f} reports $\alpha$ and $\beta$. Columns (1), (4) and (7) have no controls or fixed effects; Columns (2), (5), and (8) have the full set of control as defined in $X_{t,k}$ along with firm and time fixed effects; Columns (3), (6), and (9) carries out matching on the control variables, followed by a regression with matching weights and time and firm fixed effects. Columns (4-6) are focusing on negative surprises, while Columns (7-9) are focusing on positive surprises, with and without control. To mitigate the impact of outliers, the dependent variables are winsorized at the 1st and 99th percentiles. \sym{*} \(p<0.10\), \sym{**} \(p<0.05\), \sym{***} \(p<0.01\). 
\end{tablenotes}
\end{threeparttable}
\end{table}
\end{landscape}

%% file: tables/table_9.tex
\begin{landscape}
\begin{table}[htbp]
\scriptsize
\centering
\begin{threeparttable}
\caption{Response to Earnings Surprise: Alternative Event Windows}\label{table:2_stocktwits_event_windows}
\def\sym#1{\ifmmode^{#1}\else\(^{#1}\)\fi}
\begin{tabular*}{\hsize}{@{\hskip\tabcolsep\extracolsep\fill}l*{9}{c}}
\toprule 
                    &\multicolumn{1}{c}{(1)}&\multicolumn{1}{c}{(2)}&\multicolumn{1}{c}{(3)}&\multicolumn{1}{c}{(4)}&\multicolumn{1}{c}{(5)}&\multicolumn{1}{c}{(6)}&\multicolumn{1}{c}{(7)}&\multicolumn{1}{c}{(8)}&\multicolumn{1}{c}{(9)}\\
\midrule 
\multicolumn{8}{l}{\textbf{Panel A. Delayed Reaction}: Buy-and-Hold Abnormal Return in Event Time 2 to 40} \\
\midrule 
Long-Term        &      0.0120\sym{***}&      0.0086\sym{***}&      0.0074\sym{***}&      0.0117\sym{***}&      0.0097\sym{***}&      0.0049         &      0.0123\sym{***}&      0.0082\sym{***}&      0.0067\sym{***}\\
                    &    (0.0015)         &    (0.0013)         &    (0.0018)         &    (0.0022)         &    (0.0026)         &    (0.0033)         &    (0.0017)         &    (0.0016)         &    (0.0022)         \\
& & & & & & & \\[\dimexpr-\normalbaselineskip+3pt]
Constant     &     -0.0173\sym{***}&      0.0187\sym{***}&     -0.0127\sym{***}&     -0.0177\sym{***}&      0.0266\sym{***}&     -0.0107\sym{***}&     -0.0171\sym{***}&      0.0108         &     -0.0106\sym{***}\\
                    &    (0.0013)         &    (0.0050)         &    (0.0009)         &    (0.0020)         &    (0.0085)         &    (0.0017)         &    (0.0013)         &    (0.0072)         &    (0.0011)         \\
\midrule 
$R^2$                 &      0.0028         &      0.1290         &      0.1728         &      0.0026         &      0.2080         &      0.2832         &      0.0030         &      0.1598         &      0.2221         \\
\midrule 
\multicolumn{8}{l}{\textbf{Panel B. Delayed Reaction}: Buy-and-Hold Abnormal Return in Event Time 2 to 60} \\
\midrule 
Long-Term           &      0.0201\sym{***}&      0.0146\sym{***}&      0.0111\sym{***}&      0.0197\sym{***}&      0.0122\sym{***}&      0.0062         &      0.0199\sym{***}&      0.0150\sym{***}&      0.0128\sym{***}\\
                    &    (0.0019)         &    (0.0017)         &    (0.0023)         &    (0.0029)         &    (0.0033)         &    (0.0045)         &    (0.0021)         &    (0.0022)         &    (0.0030)         \\
& & & & & & & \\[\dimexpr-\normalbaselineskip+3pt]
Constant     &     -0.0324\sym{***}&      0.0327\sym{***}&     -0.0230\sym{***}&     -0.0337\sym{***}&      0.0386\sym{***}&     -0.0202\sym{***}&     -0.0315\sym{***}&      0.0266\sym{***}&     -0.0226\sym{***}\\
                    &    (0.0018)         &    (0.0063)         &    (0.0012)         &    (0.0028)         &    (0.0106)         &    (0.0023)         &    (0.0018)         &    (0.0091)         &    (0.0015)         \\
\midrule 
$R^2$                 &      0.0033         &      0.1471         &      0.1758         &      0.0029         &      0.2293         &      0.2892         &      0.0035         &      0.1731         &      0.2234         \\
\midrule 
Year-Quarter Fixed Effects&                     &           X         &           X         &                     &           X         &           X         &                     &           X         &           X         \\
Firm Fixed Effects  &                     &           X         &           X         &                     &           X         &           X         &                     &           X         &           X         \\
Controls            &                     &           X         &           X         &                     &           X         &           X         &                     &           X         &           X         \\
Observations        & 104,919         & 104,703         &  75,080         &  33,722         &  33,155         &  24,820         &  63,395         &  62,977         &  43,751         \\

\bottomrule
\end{tabular*}
\begin{tablenotes}
\item \textit{Notes}: Table \ref{table:2_stocktwits_event_windows} reports the coefficients of a regression of the immediate response of securities to earning surprises defined by the following regression:
\begin{equation*}
BHAR^{w}_{t,k}=\alpha + \beta \times \mathbb{I}_{t,k, \text{Long-term} = 1} + \sum_{j=2}^{11} \Big[\delta_j \times \mathbb{I}_{t,k, \text{Earnings Quantile} = j}\Big] + \gamma^{k} \times X_{t,k}+\lambda_{t} +\eta_{k} +\epsilon_{t,k},  \label{E4}
\end{equation*}
where $BHAR^{w}_{t,k}$ is defined as in Equation~\eqref{eq:bhardef}. $\beta$ is the added response of the securities in the group followed by long-term investors; $\delta$ capture the relation between the earning surprise and the immediate response; $X_{t,k}$ is a set of control variables which include the size of the security, the dispersion of analysts forecasts, the number of analysts, the buy-and-hold abnormal returns [-30, -1], valence and standard deviation of valence over [-30, -1], institutional ownership \& turnover, volatility, and abnormal short interest. To address the concern that companies followed by long-term investors may also have unobservable features that differ from companies followed by short-term investors, we control for firm fixed effects in most regression specifications. Finally, we also include time fixed effects by year-quarter. The standard errors account for heteroskedasticity as well as correlation of errors across securities making an announcement on the same day by clustering observations by day of announcement and by firm. Table \ref{table:2_stocktwits_event_windows} reports $\alpha$ and $\beta$. Columns (1), (4) and (7) have no controls or fixed effects; Columns (2), (5), and (8) have the full set of control as defined in $X_{t,k}$ along with firm and time fixed effects; Columns (3), (6), and (9) carries out matching on the control variables, followed by a regression with matching weights and time and firm fixed effects. Columns (4-6) are focusing on negative surprises, while Columns (7-9) are focusing on positive surprises, with and without control. To mitigate the impact of outliers, the dependent variables are winsorized at the 1st and 99th percentiles. \sym{*} \(p<0.10\), \sym{**} \(p<0.05\), \sym{***} \(p<0.01\). 
\end{tablenotes}
\end{threeparttable}
\end{table}
\end{landscape}

%% file: tables/table_10.tex
\begin{landscape}
\begin{table}[htbp]
\scriptsize
\centering
\begin{threeparttable}
\caption{Response to Earnings Surprise: Sensitivity}\label{table:2_stocktwits_sensitivity}
\def\sym#1{\ifmmode^{#1}\else\(^{#1}\)\fi}
\begin{tabular*}{\hsize}{@{\hskip\tabcolsep\extracolsep\fill}l*{8}{c}}
\toprule 
                    &\multicolumn{1}{c}{(1)}&\multicolumn{1}{c}{(2)}&\multicolumn{1}{c}{(3)}&\multicolumn{1}{c}{(4)}&\multicolumn{1}{c}{(5)}&\multicolumn{1}{c}{(6)}&\multicolumn{1}{c}{(7)}&\multicolumn{1}{c}{(8)}\\
\midrule 
\multicolumn{8}{l}{\textbf{Panel A. Immediate Reaction}: Buy-and-Hold Abnormal Return in Event Time 0 to 1} \\
\midrule 
Long-Term     &      0.0030\sym{***}&      0.0030\sym{***}&      0.0030\sym{***}&      0.0030\sym{***}&      0.0030\sym{***}&      0.0030\sym{***}&      0.0016\sym{**} & 0.0026\sym{***}                    \\
                    &    (0.0010)         &    (0.0010)         &    (0.0010)         &    (0.0009)         &    (0.0009)         &    (0.0009)         &    (0.0007)         &                 (0.0007)    \\
& & & & & & & \\[\dimexpr-\normalbaselineskip+3pt]
  &      0.0153\sym{***}&      0.0153\sym{***}&      0.0153\sym{***}&      0.0153\sym{***}&      0.0153\sym{***}&      0.0153\sym{***}&     -0.0005         &      \\
                    &    (0.0005)         &    (0.0005)         &    (0.0005)         &    (0.0005)         &    (0.0005)         &    (0.0005)         &    (0.0004)         &             \\
\midrule 
$R^2$                 &      0.1930         &      0.1928         &      0.1929         &      0.1868         &      0.1867         &      0.1867         &      0.1783         &      0.1839         \\
\midrule 
\multicolumn{8}{l}{\textbf{Panel B. Delayed Reaction}: Buy-and-Hold Abnormal Return in Event Time 2 to 75} \\
\midrule 
Long-Term     &      0.0184\sym{***}&      0.0184\sym{***}&      0.0184\sym{***}&      0.0186\sym{***}&      0.0187\sym{***}&      0.0187\sym{***}&      0.0155\sym{***}&      0.0245\sym{***}                \\
                    &    (0.0031)         &    (0.0031)         &    (0.0031)         &    (0.0030)         &    (0.0030)         &    (0.0030)         &    (0.0023)         &    (0.0021)                   \\
& & & & & & & \\[\dimexpr-\normalbaselineskip+3pt]
Constant      &     -0.0299\sym{***}&     -0.0299\sym{***}&     -0.0300\sym{***}&     -0.0305\sym{***}&     -0.0305\sym{***}&     -0.0306\sym{***}&     -0.0296\sym{***}&     \\
                    &    (0.0016)         &    (0.0016)         &    (0.0016)         &    (0.0015)         &    (0.0015)         &    (0.0015)         &    (0.0012)         &           \\
\midrule 
$R^2$                 &      0.1863         &      0.1862         &      0.1862         &      0.1804         &      0.1803         &      0.1804         &      0.1402         &      0.1726         \\
\midrule 
\multicolumn{8}{l}{\textbf{Panel C. Total Reaction}: Buy-and-Hold Abnormal Return in Event Time 0 to 75} \\
\midrule 
Long-Term                &      0.0220\sym{***}&      0.0220\sym{***}&      0.0220\sym{***}&      0.0223\sym{***}&      0.0223\sym{***}&      0.0224\sym{***}&      0.0173\sym{***}&     0.0273\sym{***}                  \\
                    &    (0.0034)         &    (0.0034)         &    (0.0034)         &    (0.0034)         &    (0.0034)         &    (0.0034)         &    (0.0024)         &      (0.0023)                 \\
& & & & & & & \\[\dimexpr-\normalbaselineskip+3pt]
Constant            &     -0.0141\sym{***}&     -0.0142\sym{***}&     -0.0141\sym{***}&     -0.0147\sym{***}&     -0.0147\sym{***}&     -0.0148\sym{***}&     -0.0299\sym{***}&    \\
                    &    (0.0017)         &    (0.0017)         &    (0.0017)         &    (0.0017)         &    (0.0017)         &    (0.0017)         &    (0.0013)         &           \\
\midrule 
$R^2$                 &      0.1934         &      0.1932         &      0.1932         &      0.1874         &      0.1873         &      0.1873         &      0.1499         &      0.1769         \\
\midrule 
Method              &NN=5,cal=.001         &NN=5,cal=0.01         &NN=5,cal=0.1         &NN=10,cal=.001         &NN=10,cal=0.01         &NN=10,cal=0.1         &     IPW-ATT         &        AIPW         \\
Year-Quarter Fixed Effects             &           X         &           X         &           X         &           X         &           X         &           X         &           X         &           X         \\
Firm Fixed Effects             &           X         &           X         &           X         &           X         &           X         &           X         &           X         &           X         \\
Observations          &  56,179       &  56,194       &  56,196       &  59,737       &  59,752       &  59,754       & 104,703       &  52,895       \\

\bottomrule
\end{tabular*}
\begin{tablenotes}
\item \textit{Notes}: Table \ref{table:2_stocktwits} reports the coefficients of a regression of the immediate response of securities to earning surprises defined by the following regression:
\begin{equation*}
BHAR^{w}_{t,k}=\alpha + \beta \times \mathbb{I}_{t,k, \text{Long-term} = 1} + \sum_{j=2}^{11} \Big[\delta_j \times \mathbb{I}_{t,k, \text{Earnings Quantile} = j}\Big] + \gamma^{k} \times X_{t,k}+\lambda_{t} +\eta_{k} +\epsilon_{t,k},  \label{E4}
\end{equation*}
where $BHAR^{w}_{t,k}$ is defined as in Equation~\eqref{eq:bhardef}. $\beta$ is the added response of the securities in the group followed by long-term investors; $\delta$ capture the relation between the earning surprise and the immediate response; $X_{t,k}$ is a set of control variables which include the size of the security, the dispersion of analysts forecasts, the number of analysts, the buy-and-hold abnormal returns [-30, -1], valence and standard deviation of valence over [-30, -1], institutional ownership \& turnover, volatility, and abnormal short interest. To address the concern that companies followed by long-term investors may also have unobservable features that differ from companies followed by short-term investors, we control for firm fixed effects in most regression specifications. Finally, we also include time fixed effects by year-quarter. The standard errors account for heteroskedasticity as well as correlation of errors across securities making an announcement on the same day by clustering observations by day of announcement and by firm. Table \ref{table:2_stocktwits} reports $\alpha$ and $\beta$. Columns (1)–(6) apply nearest-neighbor matching estimators with varying numbers of neighbors (5 or 10) and calipers (0.001, 0.01, or 0.1) to assess robustness to matching parameters. Column (7) uses an Inverse Probability Weighting estimator for the Average Treatment effect on the Treated (IPW-ATT), while Column (8) implements a manual Augmented Inverse Probability Weighting (AIPW) estimator combining outcome regression and inverse-probability weighting to improve efficiency and robustness.  To mitigate the impact of outliers, the dependent variables are winsorized at the 1st and 99th percentiles.  \sym{*} \(p<0.10\), \sym{**} \(p<0.05\), \sym{***} \(p<0.01\). 
\end{tablenotes}
\end{threeparttable}
\end{table}
\end{landscape}

%% file: tables/table_11.tex
\begin{landscape}
\begin{table}[htbp]
\scriptsize
\centering
\begin{threeparttable}
\caption{Response to Earnings Surprise: No Winsorization}\label{table:2_stocktwits_w}
\def\sym#1{\ifmmode^{#1}\else\(^{#1}\)\fi}
\begin{tabular*}{\hsize}{@{\hskip\tabcolsep\extracolsep\fill}l*{9}{c}}
\toprule 
                    &\multicolumn{1}{c}{(1)}&\multicolumn{1}{c}{(2)}&\multicolumn{1}{c}{(3)}&\multicolumn{1}{c}{(4)}&\multicolumn{1}{c}{(5)}&\multicolumn{1}{c}{(6)}&\multicolumn{1}{c}{(7)}&\multicolumn{1}{c}{(8)}&\multicolumn{1}{c}{(9)}\\
\midrule 
\multicolumn{8}{l}{\textbf{Panel A. Immediate Reaction}: Buy-and-Hold Abnormal Return in Event Time 0 to 1} \\
\midrule 
Long-Term   &      0.0051\sym{***}&      0.0018\sym{***}&      0.0012         &      0.0046\sym{***}&      0.0012         &     -0.0007         &      0.0049\sym{***}&      0.0028\sym{***}&      0.0033\sym{***}\\
                    &    (0.0006)         &    (0.0007)         &    (0.0010)         &    (0.0011)         &    (0.0014)         &    (0.0019)         &    (0.0007)         &    (0.0009)         &    (0.0012)         \\
& & & & & & & \\[\dimexpr-\normalbaselineskip+3pt]
Constant       &     -0.0031\sym{***}&      0.0122\sym{***}&      0.0001         &     -0.0308\sym{***}&     -0.0079\sym{*}  &     -0.0258\sym{***}&      0.0133\sym{***}&      0.0177\sym{***}&      0.0158\sym{***}\\
                    &    (0.0005)         &    (0.0028)         &    (0.0006)         &    (0.0010)         &    (0.0046)         &    (0.0010)         &    (0.0006)         &    (0.0038)         &    (0.0006)         \\
\midrule 
$R^2$                 &      0.0778         &      0.1467         &      0.2042         &      0.0083         &      0.2176         &      0.3933         &      0.0288         &      0.1480         &      0.2192         \\
\midrule 
\multicolumn{8}{l}{\textbf{Panel B. Delayed Reaction}: Buy-and-Hold Abnormal Return in Event Time 2 to 75} \\
\midrule 
Long-Term        &      0.0295\sym{***}&      0.0191\sym{***}&      0.0148\sym{***}&      0.0295\sym{***}&      0.0170\sym{***}&      0.0131\sym{**} &      0.0290\sym{***}&      0.0181\sym{***}&      0.0205\sym{***}\\
                    &    (0.0026)         &    (0.0032)         &    (0.0036)         &    (0.0042)         &    (0.0047)         &    (0.0060)         &    (0.0028)         &    (0.0043)         &    (0.0040)         \\
& & & & & & & \\[\dimexpr-\normalbaselineskip+3pt]
Constant        &     -0.0396\sym{***}&      0.0695\sym{***}&     -0.0267\sym{***}&     -0.0402\sym{***}&      0.0920\sym{***}&     -0.0257\sym{***}&     -0.0388\sym{***}&      0.0672\sym{***}&     -0.0279\sym{***}\\
                    &    (0.0026)         &    (0.0096)         &    (0.0018)         &    (0.0042)         &    (0.0169)         &    (0.0030)         &    (0.0026)         &    (0.0175)         &    (0.0020)         \\
\midrule 
$R^2$                 &      0.0035         &      0.1690         &      0.1980         &      0.0035         &      0.2359         &      0.2751         &      0.0034         &      0.2017         &      0.2308         \\
\midrule 
\multicolumn{8}{l}{\textbf{Panel C. Total Reaction}: Buy-and-Hold Abnormal Return in Event Time 0 to 75} \\
\midrule 
Long-Term        &      0.0357\sym{***}&      0.0215\sym{***}&      0.0167\sym{***}&      0.0353\sym{***}&      0.0188\sym{***}&      0.0131\sym{**} &      0.0349\sym{***}&      0.0214\sym{***}&      0.0250\sym{***}\\
                    &    (0.0027)         &    (0.0034)         &    (0.0039)         &    (0.0043)         &    (0.0049)         &    (0.0063)         &    (0.0031)         &    (0.0047)         &    (0.0043)         \\
& & & & & & & \\[\dimexpr-\normalbaselineskip+3pt]
Constant      &     -0.0434\sym{***}&      0.0820\sym{***}&     -0.0270\sym{***}&     -0.0729\sym{***}&      0.0807\sym{***}&     -0.0528\sym{***}&     -0.0257\sym{***}&      0.0879\sym{***}&     -0.0122\sym{***}\\
                    &    (0.0027)         &    (0.0103)         &    (0.0020)         &    (0.0043)         &    (0.0178)         &    (0.0032)         &    (0.0027)         &    (0.0187)         &    (0.0021)         \\
\midrule 
$R^2$                 &      0.0111         &      0.1777         &      0.2032         &      0.0033         &      0.2371         &      0.2795         &      0.0081         &      0.2083         &      0.2358         \\
\midrule 
Year-Quarter Fixed Effects&                     &           X         &           X         &                     &           X         &           X         &                     &           X         &           X         \\
Firm Fixed Effects  &                     &           X         &           X         &                     &           X         &           X         &                     &           X         &           X         \\
Controls            &                     &           X         &           X         &                     &           X         &           X         &                     &           X         &           X         \\
Observations        & 104,919         & 104,703         &  75,080         &  33,722         &  33,155         &  24,820         &  63,395         &  62,977         &  43,751         \\

\bottomrule
\end{tabular*}
\begin{tablenotes}
\item \textit{Notes}: Table \ref{table:2_stocktwits_w} reports the coefficients of a regression of the immediate response of securities to earning surprises defined by the following regression:
\begin{equation*}
BHAR^{w}_{t,k}=\alpha + \beta \times \mathbb{I}_{t,k, \text{Long-term} = 1} + \sum_{j=2}^{11} \Big[\delta_j \times \mathbb{I}_{t,k, \text{Earnings Quantile} = j}\Big] + \gamma^{k} \times X_{t,k}+\lambda_{t} +\eta_{k} +\epsilon_{t,k},  \label{E4}
\end{equation*}
where $BHAR^{w}_{t,k}$ is defined as in Equation~\eqref{eq:bhardef}. $\beta$ is the added response of the securities in the group followed by long-term investors; $\delta$ capture the relation between the earning surprise and the immediate response; $X_{t,k}$ is a set of control variables which include the size of the security, the dispersion of analysts forecasts, the number of analysts, the buy-and-hold abnormal returns [-30, -1], valence and standard deviation of valence over [-30, -1], institutional ownership \& turnover, volatility, and abnormal short interest. To address the concern that companies followed by long-term investors may also have unobservable features that differ from companies followed by short-term investors, we control for firm fixed effects in most regression specifications. Finally, we also include time fixed effects by year-quarter. The standard errors account for heteroskedasticity as well as correlation of errors across securities making an announcement on the same day by clustering observations by day of announcement and by firm. Table \ref{table:2_stocktwits_w} reports $\alpha$ and $\beta$. Columns (1), (4) and (7) have no controls or fixed effects; Columns (2), (5), and (8) have the full set of control as defined in $X_{t,k}$ along with firm and time fixed effects; Columns (3), (6), and (9) carries out matching on the control variables, followed by a regression with matching weights and time and firm fixed effects. Columns (4-6) are focusing on negative surprises, while Columns (7-9) are focusing on positive surprises, with and without control. \sym{*} \(p<0.10\), \sym{**} \(p<0.05\), \sym{***} \(p<0.01\). 
\end{tablenotes}
\end{threeparttable}
\end{table}
\end{landscape}

%% file: tables/app_1.tex
\begin{table}[htbp]
\centering
\scriptsize 
\caption{Differences Between I/B/E/S-Only and Matched Sample}\label{app_1}
\begin{threeparttable}
\begin{tabular*}{\hsize}{@{\hskip\tabcolsep\extracolsep\fill}lccccc}
\toprule
& Matched Sample & Original Sample & & Norm.\ diff. & t-stat \\
& (N = 104{,}919) & (N = 123{,}880) & & & \\
\midrule
Market Cap              & 12{,}965.8   & 12{,}902.8   & &  0.00 &   0.15 \\
Volatility              & 0.0260       & 0.0250       & &  0.01 &   2.53 \\
Turnover                & 0.2330       & 0.2330       & &  0.00 &   0.42 \\
Institutional Ownership & 0.7000       & 0.6880       & &  0.04 &  10.40 \\
Past 3-Month Return     & 0.0270       & 0.0280       & & -0.00 &  -1.17 \\
Abnormal Short Interest & 0.0000       & 0.0000       & &  0.00 &   0.18 \\
Year                    & 2015.799     & 2015.460     & &  0.10 &  25.00 \\
Earnings Surprise (SUE) & 0.0000       & 0.0000       & &  0.00 &   0.82 \\
Number of Analysts      & 8.958        & 8.494        & &  0.07 &  15.84 \\
Forecast Dispersion     & 0.0820       & 0.0800       & &  0.01 &   1.30 \\
\bottomrule
\end{tabular*}
\end{threeparttable}
\end{table}

%% file: bib.bib
@article{FamaFrench1993,
  author  = {Fama, Eugene F. and French, Kenneth R.},
  title   = {Common Risk Factors in the Returns on Stocks and Bonds},
  journal = {Journal of Financial Economics},
  year    = {1993},
  volume  = {33},
  number  = {1},
  pages   = {3--56}
}

@article{Carhart1997,
  author  = {Carhart, Mark M.},
  title   = {On Persistence in Mutual Fund Performance},
  journal = {The Journal of Finance},
  year    = {1997},
  volume  = {52},
  number  = {1},
  pages   = {57--82}
}

@article{FamaFrench2015,
  author  = {Fama, Eugene F. and French, Kenneth R.},
  title   = {A Five-Factor Asset Pricing Model},
  journal = {Journal of Financial Economics},
  year    = {2015},
  volume  = {116},
  number  = {1},
  pages   = {1--22}
}

@article{friedman2024retail,
  title={Retail investor trading and market reactions to earnings announcements},
  author={Friedman, Henry L and Zeng, Zitong},
  journal={Available at SSRN 3817979},
  year={2024}
}

@article{barber2008retail,
  title={Do retail trades move markets?},
  author={Barber, Brad M and Odean, Terrance and Zhu, Ning},
  journal={The Review of Financial Studies},
  volume={22},
  number={1},
  pages={151--186},
  year={2008},
  publisher={Society for Financial Studies}
}

@article{barber2023resolving,
  title={Resolving a paradox: Retail trades positively predict returns but are not profitable},
  author={Barber, Brad M and Lin, Shengle and Odean, Terrance},
  journal={Journal of Financial and Quantitative Analysis},
  pages={1--35},
  year={2023}
}

@article{vamossy2021investor,
title = {Investor emotions and earnings announcements},
journal = {Journal of Behavioral and Experimental Finance},
volume = {30},
pages = {100474},
year = {2021},
issn = {2214-6350},
doi = {https://doi.org/10.1016/j.jbef.2021.100474},
author = {Domonkos F. Vamossy},
}

@article{vamossy2023emtract,
title = {EmTract: Extracting emotions from social media},
journal = {International Review of Financial Analysis},
volume = {97},
pages = {103769},
year = {2025},
issn = {1057-5219},
doi = {https://doi.org/10.1016/j.irfa.2024.103769},
author = {Domonkos F. Vamossy and Rolf P. Skog},
}

@article{vamossy2023social,
author = {Domonkos F. Vamossy},
title = {Social Media Emotions and IPO Returns},
journal = {Journal of Money, Credit and Banking},
volume = {57},
number = {1},
pages = {31-67},
doi = {https://doi.org/10.1111/jmcb.13152},
year = {2025}
}

@article{cella:2013,
  title={Investors' horizons and the amplification of market shocks},
  author={Cella, Cristina and Ellul, Andrew and Giannetti, Mariassunta},
  journal={Review of Financial Studies},
  pages={hht023},
  year={2013}
  }

@article{cohen:2002,
  title={Who underreacts to cash-flow news? Evidence from trading between individuals and institutions},
  author={Cohen, Randolph B and Gompers, Paul A and Vuolteenaho, Tuomo},
  journal={Journal of financial Economics},
  volume={66},
  number={2},
  pages={409--462},
  year={2002}
}

@article{dellavigna:2009,
  title={Investor inattention and Friday earnings announcements},
  author={DellaVigna, Stefano and Pollet, Joshua M},
  journal={The Journal of Finance},
  volume={64},
  number={2},
  pages={709--749},
  year={2009}
  }

@article{gaspar:2005,
  title={Shareholder investment horizons and the market for corporate control},
  author={Gaspar, Jos{\'e}-Miguel and Massa, Massimo and Matos, Pedro},
  journal={Journal of Financial Economics},
  volume={76},
  number={1},
  pages={135--165},
  year={2005}
}

@article{gompers:2001,
author = {Gompers, Paul A. and Metrick, Andrew}, 
title = {Institutional Investors and Equity Prices},
volume = {116}, 
number = {1}, 
pages = {229-259}, 
year = {2001}, 
journal = {The Quarterly Journal of Economics} 
}

@article{grinblatt:2000,
  title={The investment behavior and performance of various investor types: a study of Finland's unique data set},
  author={Grinblatt, Mark and Keloharju, Matti},
  journal={Journal of financial economics},
  volume={55},
  number={1},
  pages={43--67},
  year={2000}
  }

@article{grinblatt:1995,
  title={Momentum investment strategies, portfolio performance, and herding: A study of mutual fund behavior},
  author={Grinblatt, Mark and Titman, Sheridan and Wermers, Russ},
  journal={The American economic review},
  pages={1088--1105},
  year={1995}
  }

@article{hotchkiss:2003,
  title={Does shareholder composition matter? Evidence from the market reaction to corporate earnings announcements},
  author={Hotchkiss, Edith S and Strickland, Deon},
  journal={The Journal of Finance},
  volume={58},
  number={4},
  pages={1469--1498},
  year={2003}
  }

@article{ke:2006,
  title={The effect of investment horizon on institutional investors' incentives to acquire private information on long-term earnings},
  author={Ke, Bin and Ramalingegowda, Santhosh and Yu, Yong},
  journal={Available at SSRN 673881},
  year={2006}
}

@article{lakonishok:1992,
  title={The impact of institutional trading on stock prices},
  author={Lakonishok, Josef and Shleifer, Andrei and Vishny, Robert W},
  journal={Journal of financial economics},
  volume={32},
  number={1},
  pages={23--43},
  year={1992},
  publisher={Elsevier}
}

@article{nofsinger:1999,
  title={Herding and feedback trading by institutional and individual investors},
  author={Nofsinger, John R and Sias, Richard W},
  journal={The Journal of Finance},
  volume={54},
  number={6},
  pages={2263--2295},
  year={1999}
  }

@article{wermers:1999,
  title={Mutual fund herding and the impact on stock prices},
  author={Wermers, Russ},
  journal={The Journal of Finance},
  volume={54},
  number={2},
  pages={581--622},
  year={1999}
  }

@article{yan:2009,
  title={Institutional investors and equity returns: are short-term institutions better informed?},
  author={Yan, Xuemin and Zhang, Zhe},
  journal={The Review of Financial Studies},
  volume={22},
  number={2},
  pages={893--924},
  year={2009},
  publisher={Oxford University Press}
}
